\documentclass{aa}
\usepackage{graphicx}
\usepackage{multirow}
\usepackage{multicol}
\usepackage{txfonts}
\usepackage{upgreek}    
\begin{document}

   \title{Comprehensive analysis of southern eclipsing systems with pulsating components: The cases of HM~Pup, V632~Sco and TT~Vel}
\titlerunning{Comprehensive analysis of HM~Pup, V632~Sco and TT~Vel}

   \author{A. Liakos\inst{1}
             \and
              D. J. W. Moriarty\inst{2}
              \and
              M. G. Blackford\inst{3}
             \and
              J. F. West\inst{4}
             \and
             P. Evans\inst{5}
             \and
             C.~M. Moriarty\inst{4}
             \and
             S.~M. Sweet\inst{2, 6}
            }
   \institute{Institute for Astronomy, Astrophysics, Space Applications and Remote Sensing, National Observatory of Athens,\\
              Metaxa \& Vas. Pavlou St., GR-15236, Penteli, Athens, Greece \\
              \email{alliakos@noa.gr}
             \and
              School of Mathematics and Physics, The University of Queensland, Queensland, 4072, Australia
              \and
              Variable Stars South (VSS), Congarinni Observatory, Congarinni, NSW, 2447, Australia
              \and
              Astronomical Association of Queensland, St. Lucia, Queensland, 4067, Australia
              \and
              El Sauce Observatory, Coquimbo Province, Chile
              \and
              ARC Centre of Excellence for All Sky Astrophysics in 3 Dimensions (ASTRO 3D), Australia
             }

           \date{Received September XX, 2021; accepted March XX, 2021}


\abstract
{This work presents an extensive analysis of the properties of three southern semi-detached eclipsing binaries hosting a pulsating component, namely HM~Pup, V632~Sco, and TT~Vel. Systematic multi-filtered photometric observations were obtained using telescopes located in Australia and Chile mostly between 2018-2021. These observations were combined with data from the Transiting Exoplanet Survey Satellite ($TESS$) mission for a detailed analysis of pulsations. Spectral types and radial velocities were determined from spectra obtained with the Australian National University's 2.3~m telescope and Wide Field Spectrograph. The data are modelled and the absolute parameters of all components are derived. The light curve residuals are further analysed using Fourier transformation techniques for the determination of the pulsation frequencies. Using theoretical models, the most probable modes of the principal oscillations are also identified. Eclipse-timing variation analysis is also made for all systems and the most likely mechanisms modulating the orbital period are proposed. The physical properties of these systems are compared with other similar cases and the locations of their components are plotted in the M-R and HR diagrams. Finally, the pulsational properties of the oscillating components are compared with currently known systems of this type within the orbital-pulsation period and $\log g$-pulsation period diagrams. These systems are identified as oEA stars by definition, with the primaries to be pulsating stars of $\delta$~Scuti type, while evidence of mass flow from the evolved secondary components is present in their Na~I~D spectra.}


   \maketitle
%

\section{Introduction}
\label{sec:intro}

A goal of asteroseismology is to improve the physics of stellar structure and evolution models by requiring such models to fit observed oscillation frequencies and modes of $\delta$~Scuti stars \citep{ANT14}. The single pulsating stars of $\delta$~Scuti type exhibit radial and non-radial oscillation modes triggered mostly by the $\kappa$-mechanism \citep[c.f.][]{AER10, BAL15}. Their pulsations are, in general, relatively fast (20~min-8~h), and typically show multiperiodic behaviour. The masses of these stars range between 1.5-2.5~$M_{\sun}$ \citep{AER10}, their temperatures between 6100-9800~K (A-F spectral classes), their luminosity classes between III-V. They are located inside the classical instability strip and share a border in the Colour-Magnitude diagram (i.e. red-edge of the classical instability strip) with the $\gamma$~Doradus pulsating stars, which oscillate in lower frequencies up to 4~cycle~d$^{-1}$ \citep{HAN99}. In an analysis of 750 A-F type stars, observed by $Kepler$ mission \citep{BOR10, KOC10}, \citet{UYT11} discovered that $63\%$ are of $\delta$~Scuti or $\gamma$~Doradus type or present hybrid behaviour. The latter result regarding the hybrid nature of these stars is an open question to date.


The eclipsing binaries (EBs) are the ultimate astrophysical tools for determining the absolute parameters of stellar components. Their light curves (LCs) provide information about the luminosity and radius ratio of their components, as well as the orbital period, eccentricity and inclination of the system. The radial velocities (RVs) of both components are required for calculating the mass ratio of the system. That is extremely critical for modelling eclipsing systems accurately, compared to models using only photometric data (i.e. degeneracy of solutions). Thus, combining photometry and spectroscopy of the EBs and using the Kepler's third law, we are able to calculate stellar physical parameters with high accuracy. However, although the aforementioned methods concern the current status of an EB, using past eclipse timings and by applying the `Eclipse Timing Variation' (ETV) method (also called as `O$-$C analysis'), it is feasible to identify short-, mid- or long-term orbital period changes (such as cyclic or secular). These changes can be attributed to standard orbital period ($P_{\rm orb}$) modulation mechanisms, such as mass transfer/loss, additional components orbiting the EB or even magnetic braking. Semi-detached Algol-type EBs have cool secondary stars that have filled their inner critical Lagrangian surface and are transferring mass to the hotter star, which was originally the less massive component. Emission in the H$_\alpha$ spectra of these systems has provided evidence of the mass flow and presence of circumstellar gas \citep{RIC99}. In systems where the cool component is very faint, H$_\alpha$ emission would only be detected during primary eclipses if the secondary component was larger than the hotter star. However, the Na~I~D lines of late K stars are relatively much stronger than those of their hot companions, thus they can reveal the presence of circumstellar gas at quadrature orbital phases, especially if the systemic velocity is large \citep{MOR19}.

The high accuracy (order of $10^{-4}$~mag) and the continuity in time of the photometric data of $Kepler$ \citep{KOC10, BOR10}, $K2$ \citep{HOW14} and Transiting Exoplanet Survey Satellite \citep[$TESS$;][]{RIC09, RIC15} missions provide the means for detailed studies of stellar oscillations, especially in the case they are combined with multi-band photometry and high-resolution spectroscopy. The time resolution of the data of these missions vary from 2 up to 30 min and they are quite useful for the determination of short-period frequencies, such as those met in $\delta$~Scuti stars. Given their accuracy, they also provide the opportunity to detect low-amplitude pulsations of the order of a few $\upmu$mag \citep[c.f.][]{MUR13, BOW18, LIA20b, KUR20, LIAN20, KIM21}. Moreover, these missions provide long time-coverage (some times of the order of years) that is very useful for the calculation of many minima timings of EBs, which in turn, allows for a better study of their orbital period changes. 

Eclipsing binary systems in which one component is a pulsating star are especially important in modern astrophysics as they provide the means to calculate the absolute properties of the pulsating star. The latter is quite significant and allows for the checking of stellar evolution theory and for the asteroseismology as well. Moreover, EBs hosting oscillating components and exhibiting mass transfer are very important in the field of asteroseismology, as they show different pulsational behaviour in terms of initiation, preservation and evolution in comparison with those that are members of detached systems or are single stars \citep{LIAN17, LIA17, MKR18a, BOW19}. EBs of the Algol type, designated as oscillating eclipsing Algol ($oEA$) systems \citep{MKR02}, contain an active cool G-K giant or subgiant star that transfers mass to a hot (A-F type) mass-accreting primary star of the $\delta$~Scuti type.

Currently, more than 330 binary systems with a $\delta$~Scuti component are known \citep[][and personal collection from literature from 2020 and later - available online\footnote{\url{http://alexiosliakos.weebly.com/catalogue.html}}]{LIAN17, LIA18, LIA20a}. However, only 233 of these systems are EBs, while only 48 of these 233 are also double-line spectroscopic binaries (SB2), i.e. they can provide accurate absolute stellar properties. From these 48 EB+SB2 systems, only 21 are oEA stars, while the rest are in detached configurations. The aforementioned numbers indicate the difficulty and the importance in determining the absolute properties of pulsating stars in EBs.   

This work is a continuation of the detailed study on individual oEA stars with a $\delta$~Scuti component \citep[see also][]{LIA12, LIA13, SOY13, LIA14, LIA17, LIA18, ULA20} and the first paper of a series dedicated to southern systems. We report analyses of  three EBs that were apparently single-lined: HM~Puppis, V632~Scorpii and TT~Velorum. These systems are previously known EBs exhibiting pulsations \citep[][respectively]{MOR13, STR16, MKR18b} but there is no detailed study for any of them nor any classification for their Roche geometry based on models to date. Therefore, this work enriches significantly the sample of oEAs with well determined absolute stellar properties by $\sim14\%$ and the total number of known oEAs by $\sim2.4\%$.  For the selected systems, multi-band ground-based and space-borne photometric and low- and high-resolution spectroscopic data (see Section~\ref{sec:OBS}) have been collected in order to: a) perform a detailed modelling (Section~\ref{sec:LCRVMDL}), b) calculate the stellar parameters with high accuracy (Section~\ref{sec:LCRVMDL}), c) perform ETV analysis for the determination of the most possible orbital period modulating mechanisms (Section~\ref{sec:OCMDL}), d) detect the most powerful pulsation frequencies (Section~\ref{sec:PULMDL}), and e) estimate the oscillation modes (Section~\ref{sec:PULMDL}). The properties of the systems as well as the pulsation properties of their oscillating stars are compared with others of similar type (Section~\ref{sec:EVOL}). Finally, summary and conclusions are given in Section~\ref{sec:DIS}. In all tables of the present study the errors of the parameters are given in parentheses alongside values and correspond to the last digit(s).

\section{Observations, data reduction and analysis}
\label{sec:OBS}

In this study, three different data sources are used: a) Ground-based multi-band photometric time series, b) $TESS$ data obtained with long- (30~min), mid- (10~min) and short-cadence (2~min) modes, and c) high and low resolution spectroscopic data for the RVs calculation and spectral type determination.

\subsection{Ground-based photometry}
\label{sec:GBPHOT}

The ground-based data were obtained from three different observatories located at the southern hemisphere. Small-size telescopes in Australia and Chile, equipped with CCD cameras and photometric filters, were employed for the observations. The majority of the data were collected between 2018-2021, however, for one system, data from 2013 and 2014 were also used in our study. The detailed observations log is given in Table~\ref{tab:GBPhotlog}, where we list: the name of the system, the observing site (inc. longitude and latitude), the telescope (diameter, focal ratio, and optical design), the camera (model, chip model, pixel size, and pixels array), the filter used ($Fil.$), the exposure time ($Exp.$) for the images in each band, the number of nights ($Nig.$) with data obtained for each filter, the date range of the observations (in DD/MM/YY format), and the total number of points for each filter. It should be noted, that in one night more than one filter might have been used. After the completion of the LC of the EB, the $B$ filter observations continued around the quadratures of the systems for a longer time coverage of the pulsations.

The data reduction was made with various photometric packages, i.e. MaximDL (Diffraction Limited 2012); AstroImageJ \citep{COL17}, using the differential aperture photometry technique. The magnitude calibration of the variables was based on the respective apparent magnitudes of the comparison stars taken from the AAVSO Photometric All-Sky Survey (APASS) DR9 \citep{HEN15} and the differential magnitude \citep[for details see][]{MOR19}. The phase-folded LCs of the systems are given in Fig.~\ref{fig:LCSRVS}.

\begin{table*}
\centering
\caption{Ground-based photometric observations log.}
\label{tab:GBPhotlog}
\scalebox{0.95}{
\begin{tabular}{l cc cc cc cc}
\hline																	
System	&	Site	&	Telescope	&	Camera	&	Fil.	&	Exp.	&	Nig.	&	Date range	&	Points	\\
	&		&		&		&		&	(s)	&		&		&		\\
\hline																	
\multirow{3}{*}{HM Pup}	&	El Sauce,	&	35.6 cm, f/7.2	&	SBIG STT 1603-3,	&	$B$	&	80-90	&	17	&	\multirow{ 3}{*}{26/11/20-17/1/21}	&	5225	\\
	&	Chile	&	CDK$^a$	&	KAF-1603, 9~$\mu$m,	&	$V$	&	60	&	15	&		&	5078	\\
	&	70.76$\degr$W, 30.47$\degr$S	&		&	1536$\times$1024	&	$I$	&	90	&	11	&		&	2840	\\
\hline																	
\multirow{ 6}{*}{V632~Sco}	&	Glen Aplin,	&	35.6 cm, f/11	&	Moravian G3-6303	&	$B$	&	140	&	10	&	31/7/13,	&	772	\\
	&	Australia	&	S-C$^b$	&	KAF-6303E, 9~$\mu$m,	&	$V$	&	60	&	7	&	28/5/14,	&	670	\\
	&	151.85$\degr$E, 28.75$\degr$S	&		&	3072$\times$2048	&	$I$	&	70	&	7	&	13-21/6/18	&	614	\\
\cline{2-9}																	
	&	Congarinni,	&	35.5 cm, f/8	&	SBIG STT 3200-ME, 	&	$B$	&	40-55	&	8	&	\multirow{ 3}{*}{23/7-21/8/20}	&	3106	\\
	&	Australia 	&	RC$^c$	&	KAF-3200ME, 6.8~$\mu$m,	&	$V$	&	40	&	2	&		&	926	\\
	&	152.86$\degr$E, 30.73$\degr$S	&		&	2184$\times$1472	&	$I$	&	40	&	1	&		&	296	\\
\hline																	
\multirow{ 3}{*}{TT~Vel}	&	Congarinni,	&	35.5 cm, f/8	&	SBIG STT 3200-ME, 	&	$B$	&	45	&	26	&	\multirow{ 3}{*}{20/4-20/6/21}	&	2621	\\
	&	Australia 	&	RC$^c$	&	KAF-3200ME, 6.8~$\mu$m,	&	$V$	&	30	&	19	&		&	507	\\
	&	152.86$\degr$E, 30.73$\degr$S	&		&	2184$\times$1472	&	$I$	&	45	&	19	&		&	491	\\
\hline
\end{tabular}}
\tablefoot{$^a$Corrected Dall-Kirkham, $^b$Schmidt-Cassegrain, $^c$Ritchey-Chr\'{e}tien}
\end{table*}

\begin{figure*}[h!]
\centering
\begin{tabular}{c}
\includegraphics[width=17.2cm]{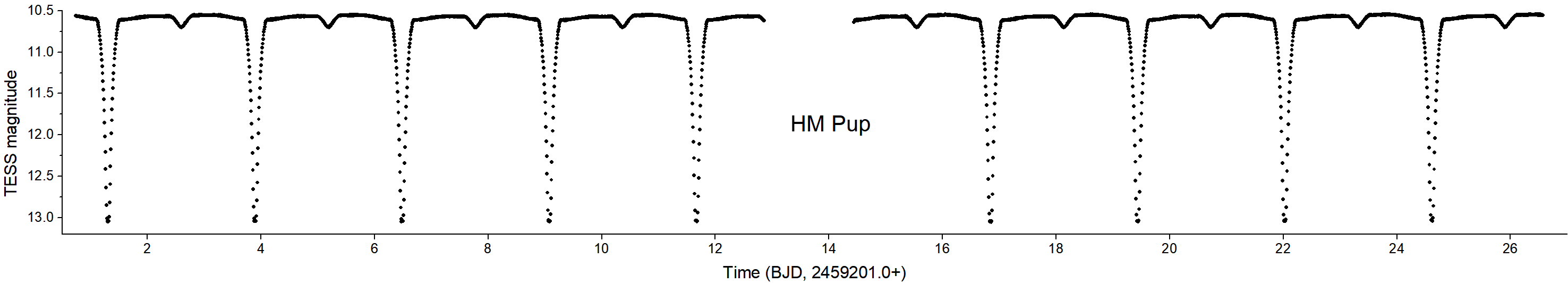}\\
\includegraphics[width=17.2cm]{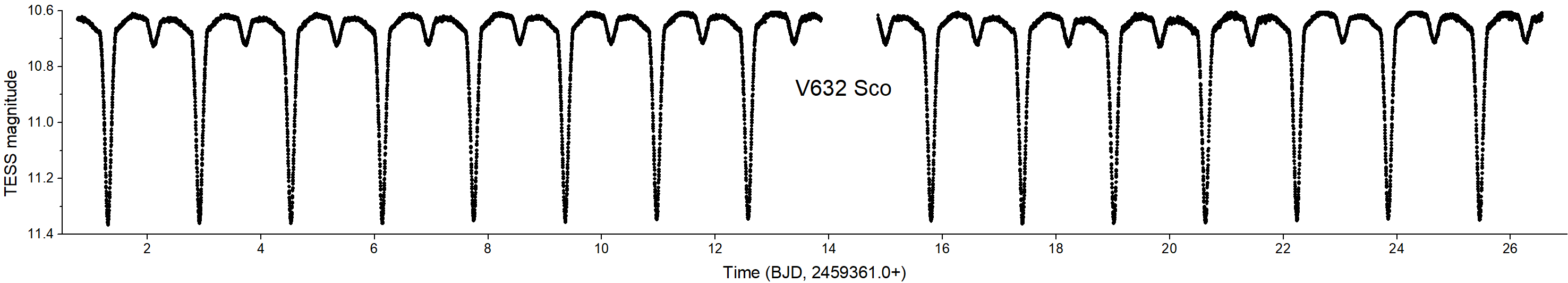}\\
\includegraphics[width=17.2cm]{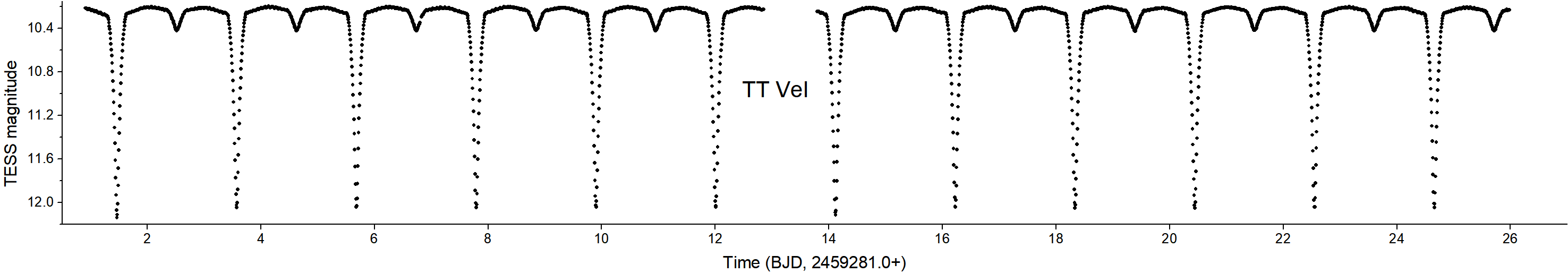}\\
\end{tabular}
\caption{$TESS$ LCs that cover approximately 26~days of continuous monitoring. The rest data are not shown for scaling reasons. The time resolution of the data sets is 10, 2, and 10~min for HM~Pup (top), V632~Sco (middle), and TT~Vel (bottom), respectively.}
\label{fig:TESSLCs}
\end{figure*}

\subsection{TESS photometry}
\label{sec:TESSPHOT}

\begin{table}
\centering
\caption{$TESS$ observations log.}
\label{tab:TESSlog}
\scalebox{0.9}{
\begin{tabular}{l ccc c cc}
\hline						
System	&	$m_{TESS}$	&	BJD begin	&	Days	&	$Sr$	&	Res. 	&	Points	\\
	&	(mag)	&		&		&		&	(min)	&		\\
\hline													
\multirow{5}{*}{HM Pup}	&	\multirow{5}{*}{10.55}	&	2458491.85	&	24.16	&	7	&	30	&	1000	\\
	&		&	2458518.10	&	23.83	&	8	&	30	&	682	\\
	&		&	2459201.73	&	25.83	&	33	&	10	&	3495	\\
	&		&	2459228.76	&	25.31	&	34	&	10	&	3645	\\
	&		&	2459255.00	&	24.98	&	35	&	10	&	3597	\\
\hline													
\multirow{2}{*}{V632 Sco}	&	\multirow{2}{*}{10.62}	&	2458624.97	&	27.90	&	12	&	30	&	1244	\\
	&		&	2459361.78	&	27.84	&	39	&	2	&	18330	\\
\hline													
\multirow{3}{*}{TT~Vel}	&	\multirow{3}{*}{10.21}	&	2458544.56	&	23.85	&	9	&	30	&	940	\\
	&		&	2458571.72	&	23.92	&	10	&	30	&	878	\\
	&		&	2459280.91	&	25.08	&	36	&	10	&	3612	\\

\hline												
\end{tabular}}
\end{table}

\begin{table*}[h!]
\centering
\caption{Spectroscopic observations log. All data were obtained with the WiFeS instrument on the 2.3~m ANU telescope.}
\label{tab:SPEClog}
\scalebox{0.97}{
\begin{tabular}{l cc cc cc}
\hline													
	System&	Date range	&	Phase	&	Spectra	&	Expos.	&	\multicolumn{2}{c}{Grating}			\\
	&		&		&		&	(s)	&		&		\\
\hline													
\multirow{4}{*}{HM Pup}	&	30/4/18	&	0.00	&	2	&	1100	&	B3000 (S/N=150)	&	R7000 (S/N=140)	\\
	&	28/5/21	&	0.00	&	2	&	1000	&B3000 (S/N=150)	&	R7000 (S/N=140)	\\
	&	28/3/21	&	0.50	&	3	&	100	&	\multicolumn{2}{c}{B3000 (S/N=210)}			\\
	&	14/2/17-27/5/21	&	0.20-0.36; 0.62-0.88	&	41	&	180-420	&	B7000 (190$<$S/N$<$250)	&	R7000 (200$<$S/N$<$250)	\\
\hline													
\multirow{2}{*}{V632~Sco}	&	12/6/17	&	0.50	&	3	&	60	&	\multicolumn{2}{c}{B3000 (S/N=220)}			\\
	&	2/5/18-8/7/20	&	0.17-0.32; 0.65-0.75	&	23	&	100-420	&	B7000 (340$<$S/N$<$370)	&	R7000 (270$<$S/N$<$360)	\\
\hline													
\multirow{2}{*}{TT~Vel}	&	26/4/21	&	0.50	&	3	&	60	&	\multicolumn{2}{c}{B3000 (S/N=130)}			\\
	&	1/12/20 - 28/5/21	&	0.16-0.32; 0.66-0.85	&	35	&	180-420	&	B7000 (350$<$S/N$<$160)	&	R7000 (340$<$S/N$<$500)	\\
\hline																																											
\end{tabular}}
\end{table*}

The $TESS$ data were downloaded from the Mikulski Archive for Space Telescopes (MAST\footnote{\url{https://mast.stsci.edu}}) archive. The observations log for each system is given in Table~\ref{tab:TESSlog}, where the following are listed: the BJD of the beginning, the number of continuous days of observations, the sector ($Sr$) of $TESS$, the time resolution ($Res.$), and the number of points of each data set.

For all systems, the Pre-search Data Conditioning Simple Aperture Photometry (PDCSAP) flux, which typically is corrected for long-term trends caused by instrumental effects, was used, when available. However, in the data sets where the PDC pipeline produced distorted LCs, the SAP flux values were used instead. The magnitudes of the systems (assumed as the maximum brightness values) were taken from the TESS Input Catalog v8.0 \citep{STA19} and were used for the flux-to-magnitude conversion (Table~\ref{tab:TESSlog}).

For HM~Pup, there are five data sets, two in long- and three in mid-cadence modes. For the latter ones (i.e. data sets of sectors 33-35), although PDCSAP flux values are available, they present obvious distortions, therefore the SAP fluxes were used. V632~Sco was observed in long- and short-cadence modes. For the long-cadence data set, only the SAP flux is given, while the PDCSAP flux values are available for the short-cadence set. The time-series data sets for TT~Vel were made in long- and mid-cadence modes. The PDCSAP flux was available, but as in the case of HM~Pup, there are lots of distortions, thus, the SAP flux values were again used. Samples of the $TESS$ LCs for all systems are plotted in Fig.~\ref{fig:TESSLCs}, while the folded data in the orbital period are given in Fig. \ref{fig:LCSRVS}.



\subsection{Spectroscopy}
\label{sec:SPEC}

Spectra were observed with the wide field spectrograph (WiFeS) mounted at the Nasmyth focal position of the Australian National University's 2.3~m telescope at Siding Spring Observatory, Australia \citep{DOP07, DOP10}. The spectrograph is equipped with two Fairchild Imaging CCDs with a resolution of 4096$\times$4096 pixels and a pixel scale of 0.5~arcsec. The 560~nm dichroic beam splitter was used. The detailed spectroscopic observations log is given in Table~\ref{tab:SPEClog} and includes for each system: the date range (in DD/MM/YY format), the orbital phase of the system at the time of observations, the number of the obtained spectra and the corresponding exposure time ($Expos.$), and the grating used along with the resulting signal-to-noise (S/N) ratio. See \citet{MOR19} for details of the procedures used to determine RVs and spectral types. In general, for RV determinations, sets of three or four spectra were observed at each phase with the B7000 grating, together with a spectrum of the Ne-Ar arc lamp before and after each set or closely grouped sets. Exposure times with the B7000 grating were long enough to allow detection of the faint secondary stars' metal lines with the broadening function in RAVESPAN \citep{PIL17}. On several occasions, at the same time as long exposure times were used with the B7000 grating, shorter exposure times were used with the R7000 grating to enhance the separation of Na~I~D lines of the primary stars from the lines due to circumbinary gas. As HM~Pup is a totally eclipsing system, spectra of its secondary component can be observed during the 42~minute period of totality. We observed it twice with the B3000 grating for spectral classification and with the R7000 grating to seek evidence of chromospherical activity and mass flow.

\subsubsection{Spectral classification}
\label{sec:SPECCLASS}

\begin{figure*}
\centering
\includegraphics[width=18.2cm]{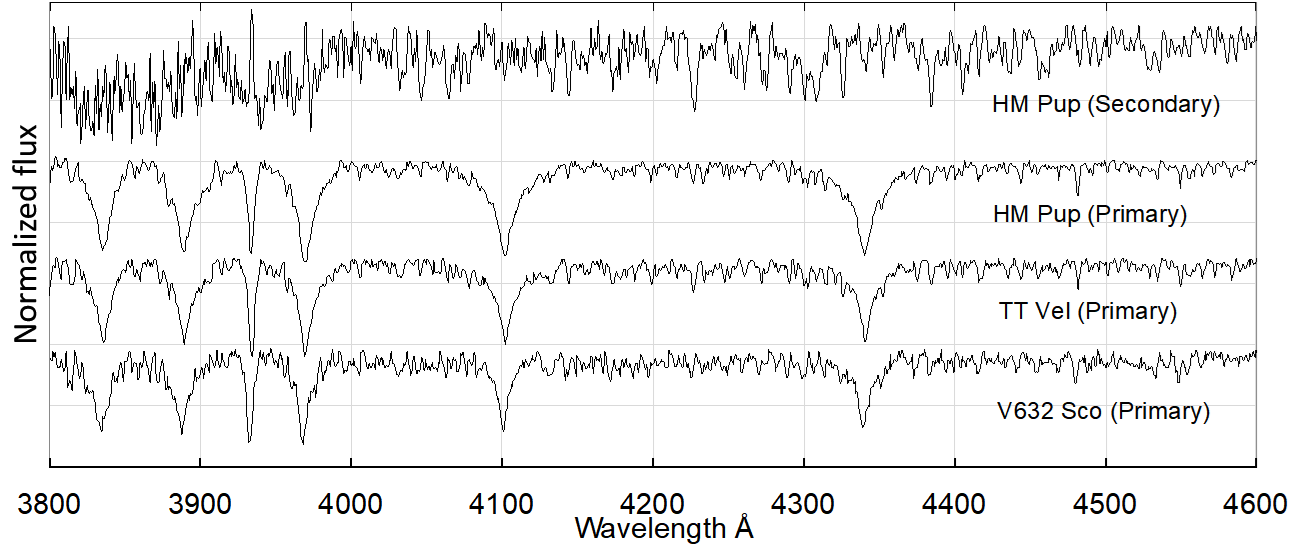}
\caption{Rectified spectra of the primary components of each system and the secondary component of HM~Pup.}
\label{fig:CLASS}
\end{figure*}
\begin{figure*}
\centering
\includegraphics[width=17.5cm]{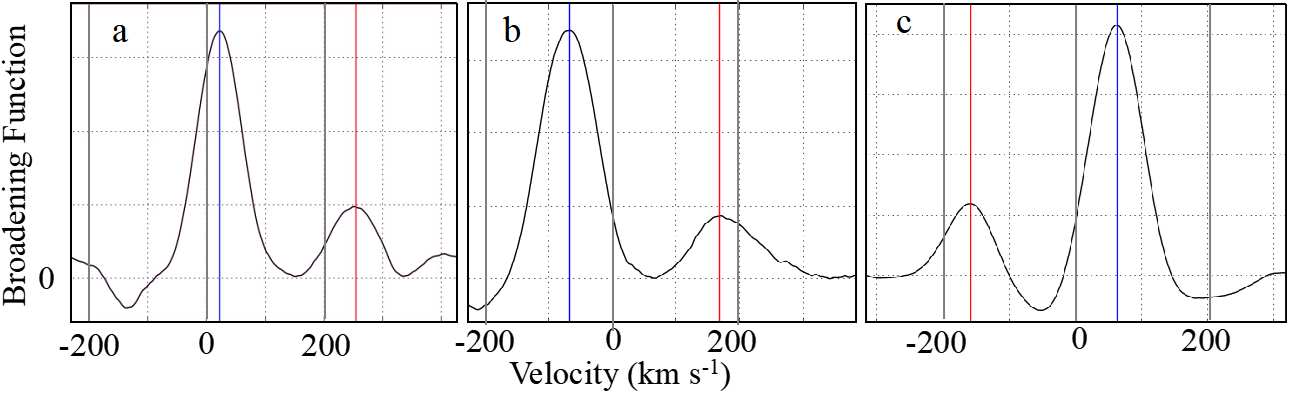}
\caption{Examples of Broadening Functions method on the spectra of (a) HM~Pup at phase 0.25, (b) V632~Sco at phase 0.25, and (c) TT~Vel at phase 0.75. Blue lines denote the RVs of the primary and red lines those of the secondary components.}
\label{fig:BFs}
\end{figure*}

Spectra for classification were observed with the low-resolution B3000 grating (708~lines~mm$^{-1}$, R=3000). Its wavelength range of 3200 to 5900~{\AA} includes the Balmer lines from H$_{\beta}$-H$_{\eta}$, and the important Ca~I~K line (the Ca~II~H line is blended with H$_{\epsilon}$ in the spectra of the A and F-type stars in the systems studied here). The spectra of the primary components of HM~Pup, V632~Sco, and TT~Vel were observed during the secondary eclipse (phase 0.5) with short exposures to exclude interference by their faint secondary components. The secondary component of HM~Pup was observed at phase 0.0. All these spectra were corrected to heliocentric values by subtracting the systemic radial velocities of 58, -17, and 24~km~s$^{-1}$ respectively (see Table~\ref{tab:LCRV}). Spectral types were determined using `XCLASS' and `MKCLASS' programmes \citep{GRA14}. XCLASS provides direct comparison of the target spectrum with reference spectra from the MKCLASS\footnote{\url{http://www.appstate.edu/~grayro/mkclass/}} libraries, which encompass integer spectral temperature types from O6 to M5, and a wide range of luminosity types between V and Ia. The target spectra were normalised to unity at 4503~{\AA} to match the library spectra.


We classified the spectra initially by visual comparisons with selected reference spectra across the wavelength range of 3800-4600~{\AA}. The best fits of the primary stars were HM~Pup: A6V, V632~Sco: F2V and TT~Vel: A8V. We classified the HM~Pup secondary star as K8IV/III. The spectra were rectified to deliver a linear continuum baseline using the `Autorectify' function in the XMK25 display routine in XCLASS (Fig.~\ref{fig:CLASS}). These were then compared with other reference spectra, including the examples provided in the work of \citet{GRA09}.

\subsubsection{Radial Velocities}
\label{sec:SPECRVS}

In general for radial velocity determinations, sets of three or four spectra were observed at each phase with the high-resolution B7000 grating (1530~lines~mm$^{-1}$, R=7000) that provides a velocity resolution of 45~km~s$^{-1}$ and wavelength range between 4180-5580~{\AA}. Exposure times were long enough to allow detection of the metal lines of the faint secondary stars with the broadening function method in RAVESPAN software \citep{RUC92, RUC02, PIL17}. Spectra in the phase ranges shown in Table~\ref{tab:SPEClog} were analysed. The optimum settings used for the three systems studied here were as follows. Spectra were normalised and analysed in the wavelength range 4355-4840~{\AA} and 4880-5548~{\AA} with the Balmer line masked. A template with an effective temperature of 4000~K, gravity coefficient set at 2.5 and metallicity at 0.0, was selected from the synthetic spectra supplied with the RAVESPAN software \citep{COE05}. A 4th-order polynomial was used with resolution set at 2.0~km~s$^{-1}$ and v-range 1.8. With these settings, the broadening function velocity peak of the secondary components was clearly evident (Fig.~\ref{fig:BFs}). The phased RVs curves are plotted in Fig.~\ref{fig:LCSRVS}, while the complete list of RV values is given in Table~\ref{tab:RVs} in App.~\ref{sec:AppRVs}.



\subsubsection{Chromospherical activity and gas streaming}
\label{sec:SPECEMIT}
\begin{figure}
\begin{tabular}{cc}
\includegraphics[width=4.05cm]{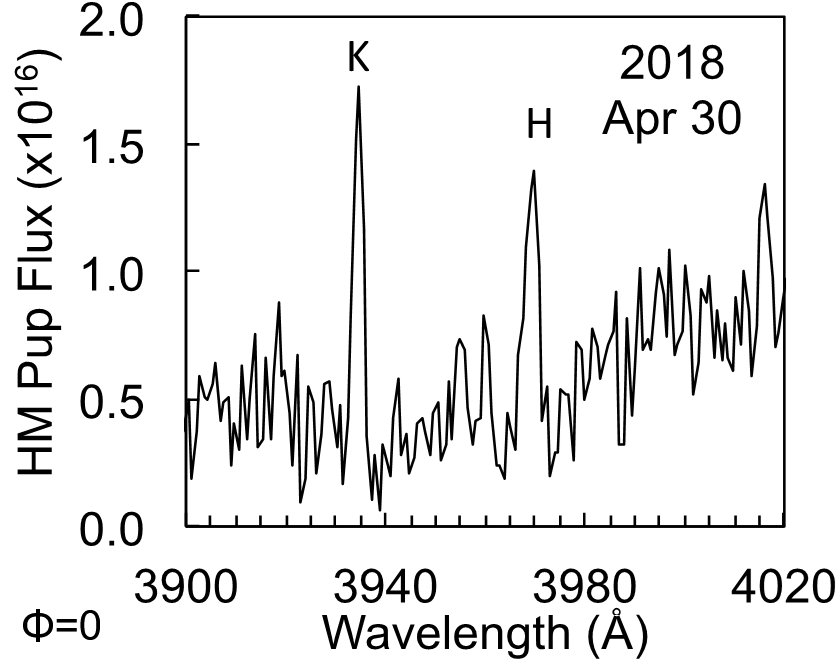}&\includegraphics[width=4.05cm]{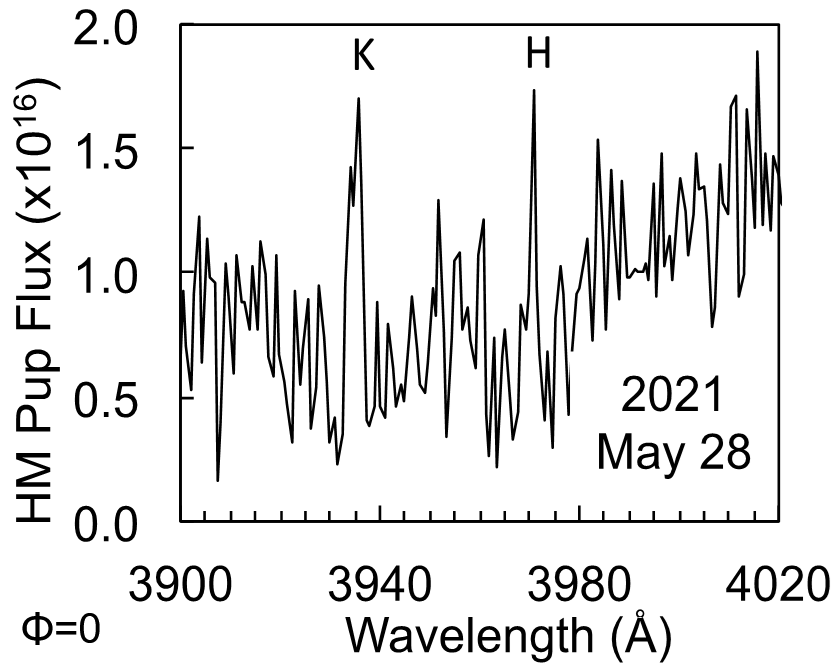}\\
\end{tabular}
\caption{Spectra of the HM~Pup secondary star in the range 3900-4020~{\AA} (with R=3000) during totality on two occasions. The K and H emission peaks have a red shift of 58~km~s$^{-1}$.}
\label{fig:PupCa}
\end{figure}

\begin{figure*}
\centering
\begin{tabular}{cccc}
\includegraphics[width=4.1cm]{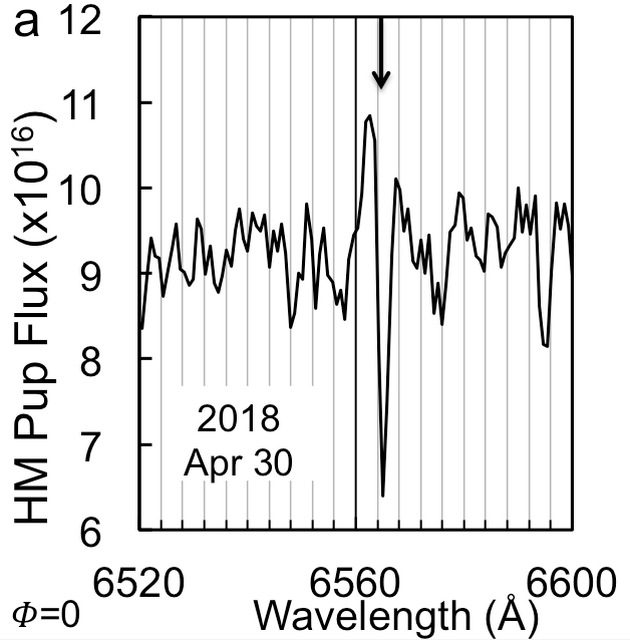}&\includegraphics[width=4.1cm]{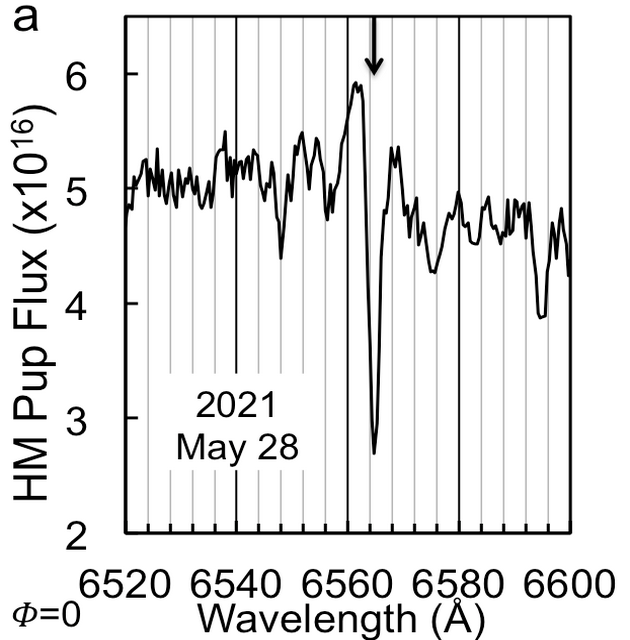}&\includegraphics[width=4.1cm]{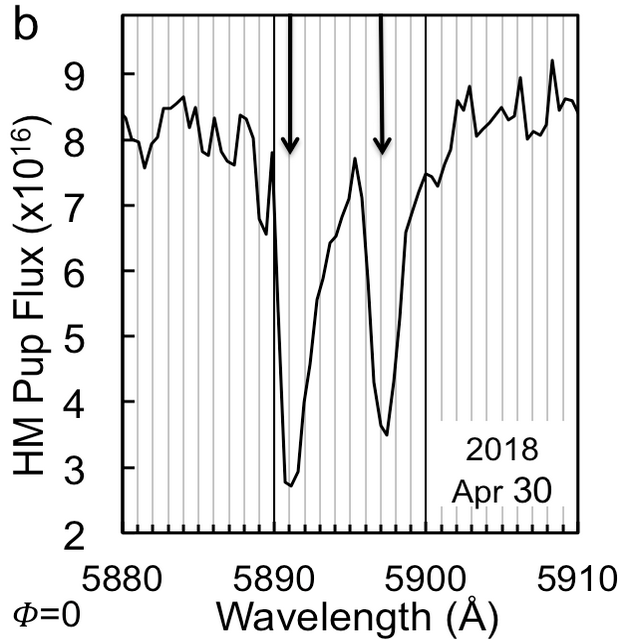}&\includegraphics[width=4.1cm]{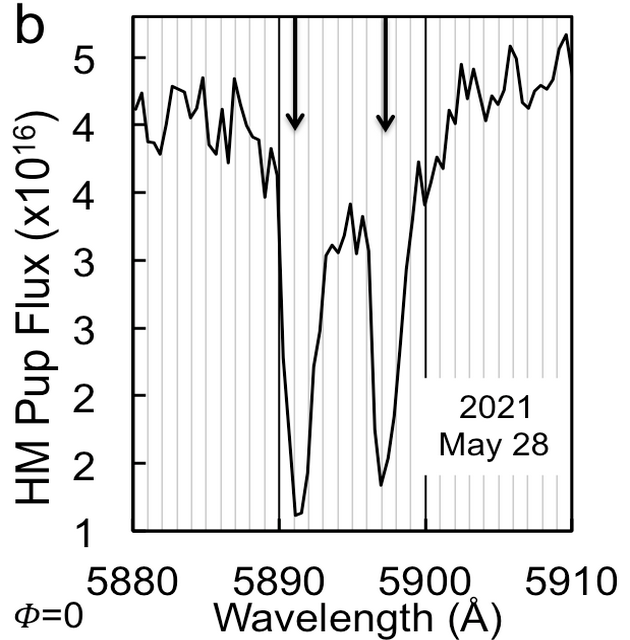}\\
\end{tabular}
\caption{The H$_{\alpha}$ (a) and Na~I~D (b) spectra of the HM~Pup secondary star observed at the same time as the Ca~II spectra in Fig.~\ref{fig:PupCa}, but with R=7000. Arrows mark the calculated wavelength positions of the spectral line centres at the systemic velocity.}
\label{fig:PupHaNaD}
\begin{tabular}{cccc}
\includegraphics[width=4.1cm]{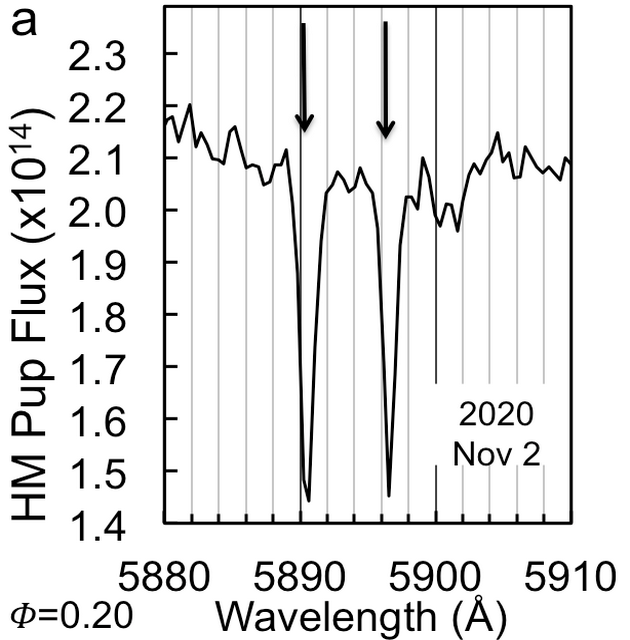}&\includegraphics[width=4.1cm]{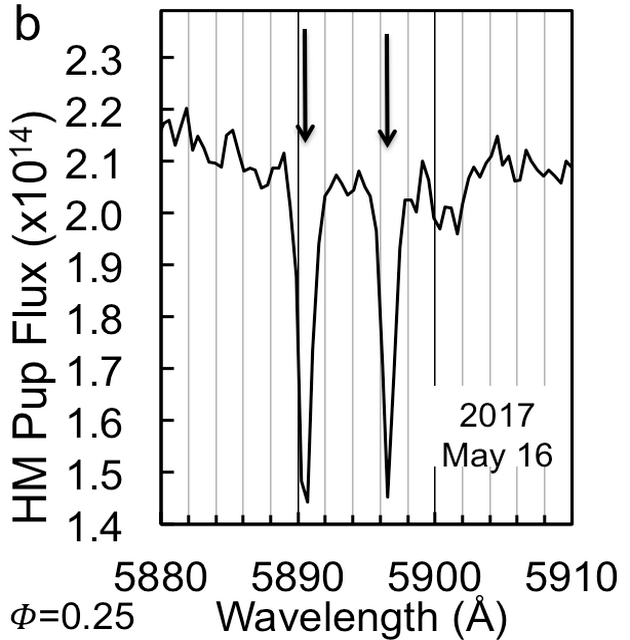}&\includegraphics[width=4.1cm]{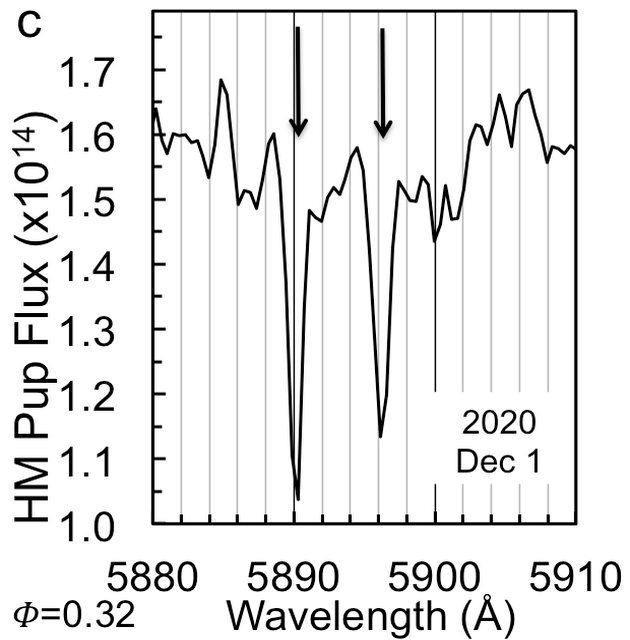}&\includegraphics[width=4.1cm]{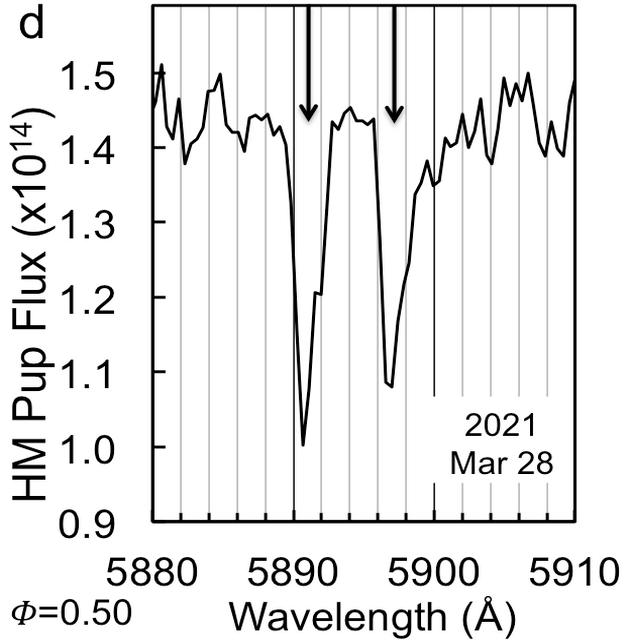}\\
\includegraphics[width=4.1cm]{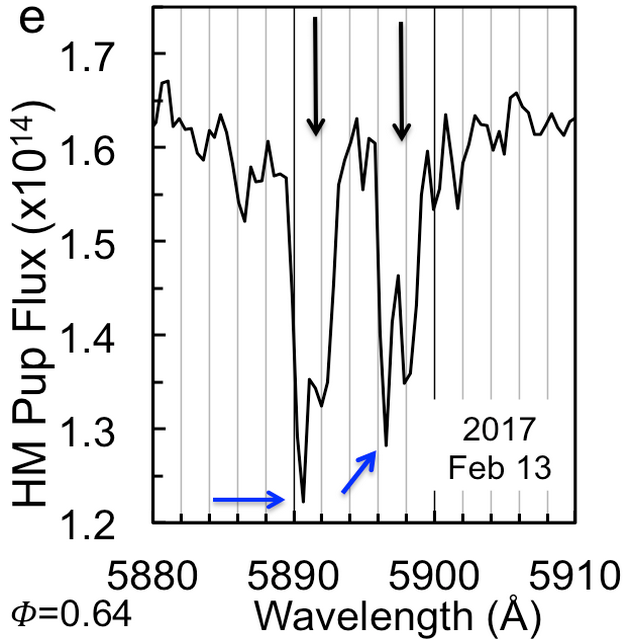}&\includegraphics[width=4.1cm]{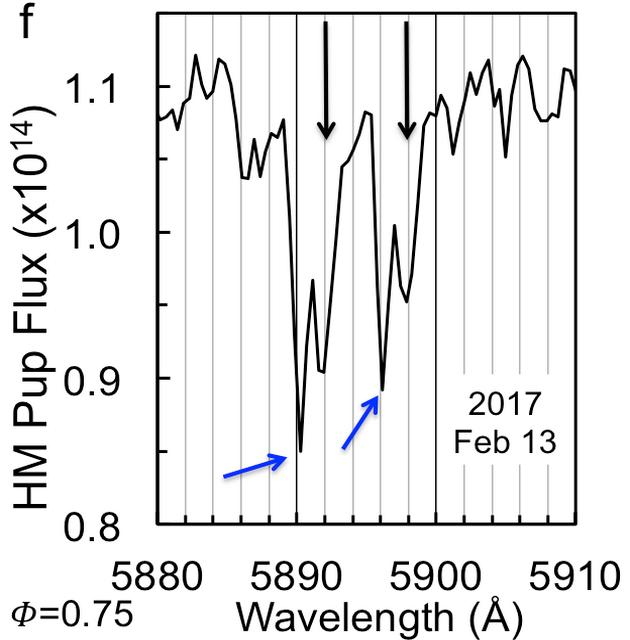}&\includegraphics[width=4.1cm]{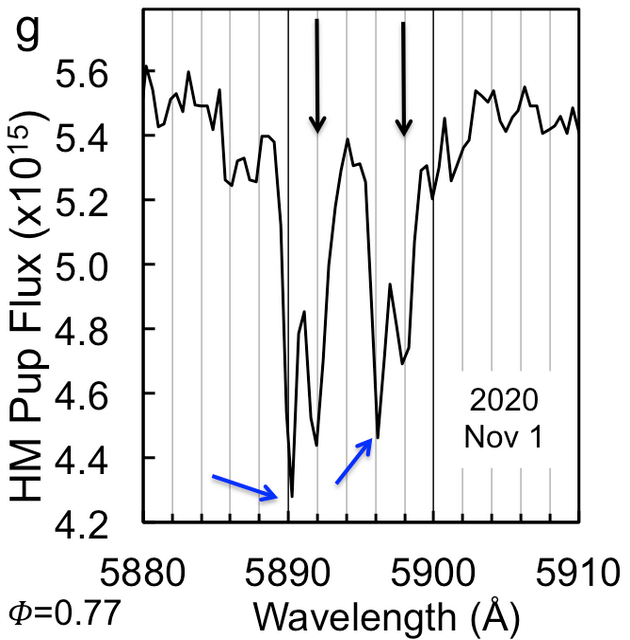}&\includegraphics[width=4.1cm]{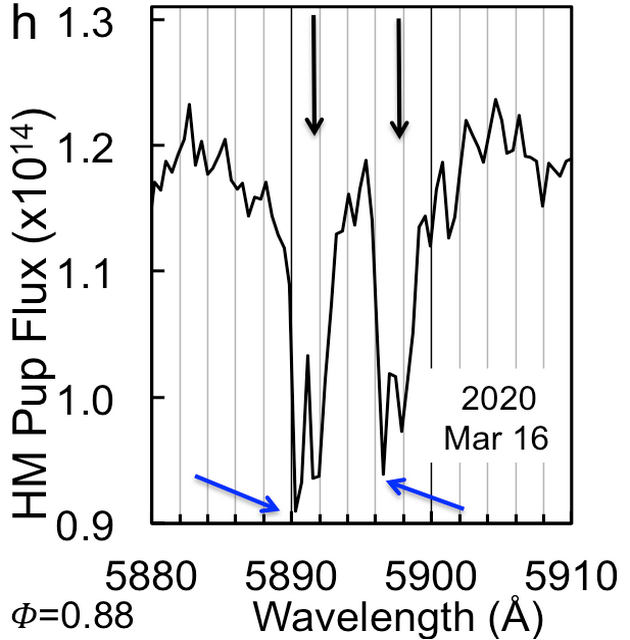}\\
\end{tabular}
\caption{Spectra of HM~Pup Na~I~D lines at several orbital phases ($\Phi$) with R=7000. Downward arrows mark the calculated position of the line centres for the primary star that are the sum of the systemic and orbital velocities. Lines of circumstellar gas are marked with blue arrows.}
\label{fig:PupNa}
\begin{tabular}{cccc}
\includegraphics[width=4.1cm]{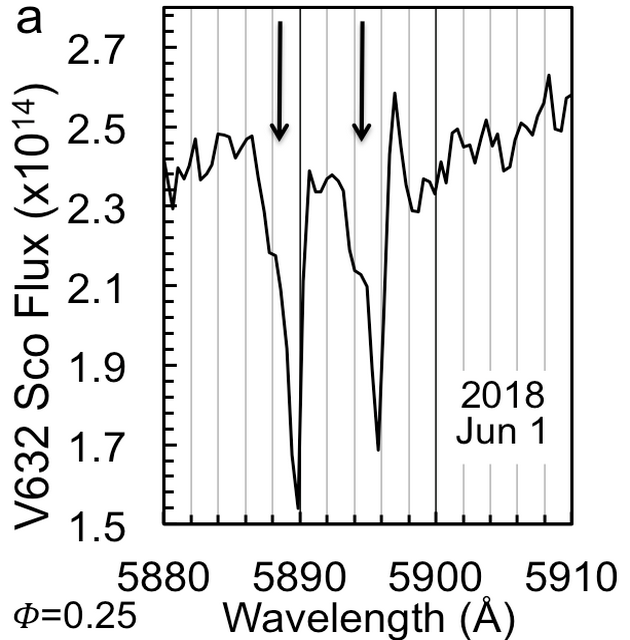}&\includegraphics[width=4.1cm]{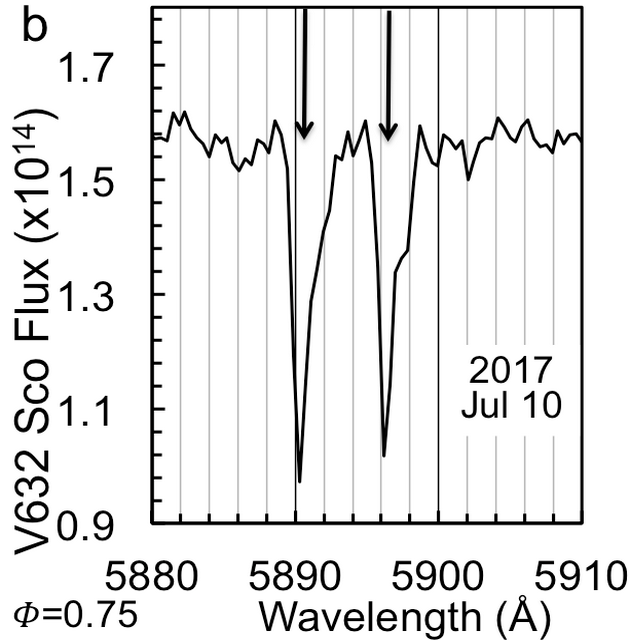}&\includegraphics[width=4.1cm]{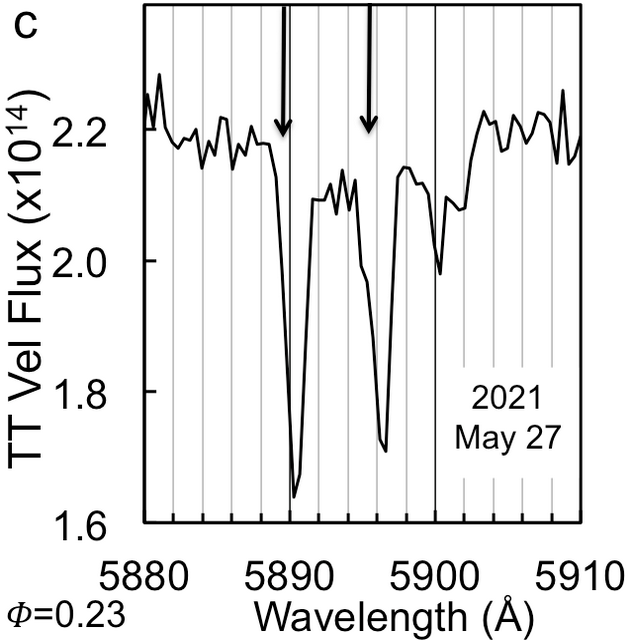}&\includegraphics[width=4.1cm]{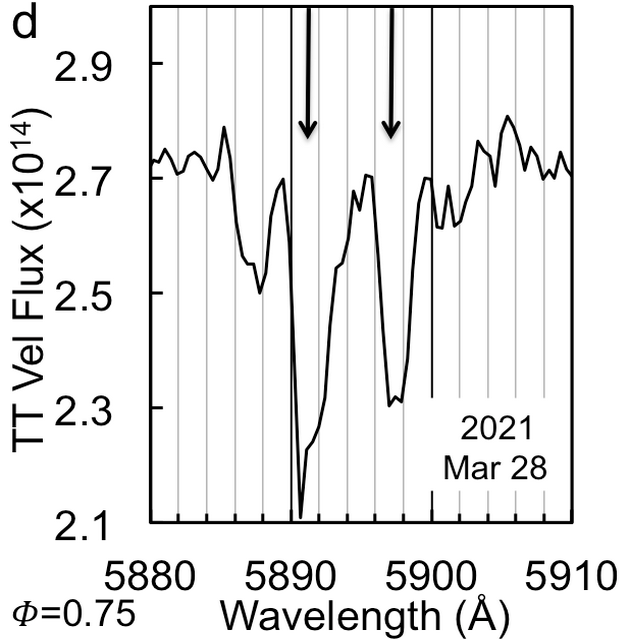}\\
\end{tabular}
\caption{Spectra of Na~I~D lines at quadrature phases (R = 7000): (a, b) V632~Sco, (c, d) TT~Vel. Downward arrows mark the calculated position of the line centres for primary components that are the sum of their orbital and systemic velocities.}
\label{fig:ScoVelNaD}
\end{figure*}

Emission was observed in the centres of the Ca~II H and K lines of the secondary component of HM~Pup during the total primary eclipse. The emission peaks had a red shift of 58~km~s$^{-1}$ from the systemic velocity, indicating they were due to chromospherical activity and not circumbinary gas (Fig.~\ref{fig:PupCa}). On the same nights, the H$_{\alpha}$ spectra had an emission peak with a blue shift of 120~km~s$^{-1}$ and the absorption line was broadened with a red shift from 33 to 110~km~s$^{-1}$ (Fig.~\ref{fig:PupHaNaD}). The Na~I D lines of the secondary component on those nights were broadened towards red at velocities of 110-170~km~s$^{-1}$ (Fig.~\ref{fig:PupHaNaD}). Sky background flux values were subtracted from all the spectra in Figs~\ref{fig:PupCa} and \ref{fig:PupHaNaD}, as the exposures used were long. The sky background flux values in all spectra in Figs~\ref{fig:PupNa} and \ref{fig:ScoVelNaD} were 2-3 orders of magnitude lower than the stellar flux values; the difference was similar to that found for ST~Cen and V775~Cen \citep{MOR19}. Flux units are ergs~cm$^{-2}$~s$^{-1}$~arcsec$^{-2}$~{\AA}$^{-1}$.

Broadening or splitting in the spectra of the sodium D doublet lines at different phases indicates evidence of gas streaming from the secondary components of Algol systems \citep[see][]{MOR19}. Circumstellar gas was particularly evident around HM~Pup. The primary component's sodium~D line centres at phase 0.75 are red-shifted with a velocity of 103~km~s$^{-1}$ (the sum of the orbital and systemic velocities of 45 and 58~km~s$^{-1}$ respectively). A clear separation of the sodium lines of the slower moving gas from the photospheric Na~I~D absorption lines of the receding primary star itself is clearly evident. Conversely, at phase 0.25, the sodium~D lines of the approaching primary component were blended with the gas lines, which are evident as a broadening of the lines to the red (compare Fig.~\ref{fig:PupNa}a-d with Fig.~\ref{fig:PupNa}e-h).

In contrast to HM~Pup, the sodium~D line centres of V632~Sco, which has a negative systemic velocity, are blue shifted at phase 0.25 with a velocity of -69~km~s$^{-1}$, (the sum of the orbital and systemic velocities of -52 and -17~km~s$^{-1}$ respectively). The narrow lines to the red side are evidence of circumstellar gas (Fig.~\ref{fig:ScoVelNaD}a). At phase 0.75 the circumstellar gas lines are blended with those of the primary component, which are rotationally broadened (Fig.~\ref{fig:ScoVelNaD}b). There was also evidence of circumbinary gas in the Na~I~D spectra of TT~Vel, but its stellar and gas sodium lines were similarly blended at the resolution available to us (Fig.~\ref{fig:ScoVelNaD}c, d). The sodium lines of the primary components of each system were strong in comparison to those of their secondary stars, which have low luminosities.

\begin{figure}
\centering
\includegraphics[width=5.5cm]{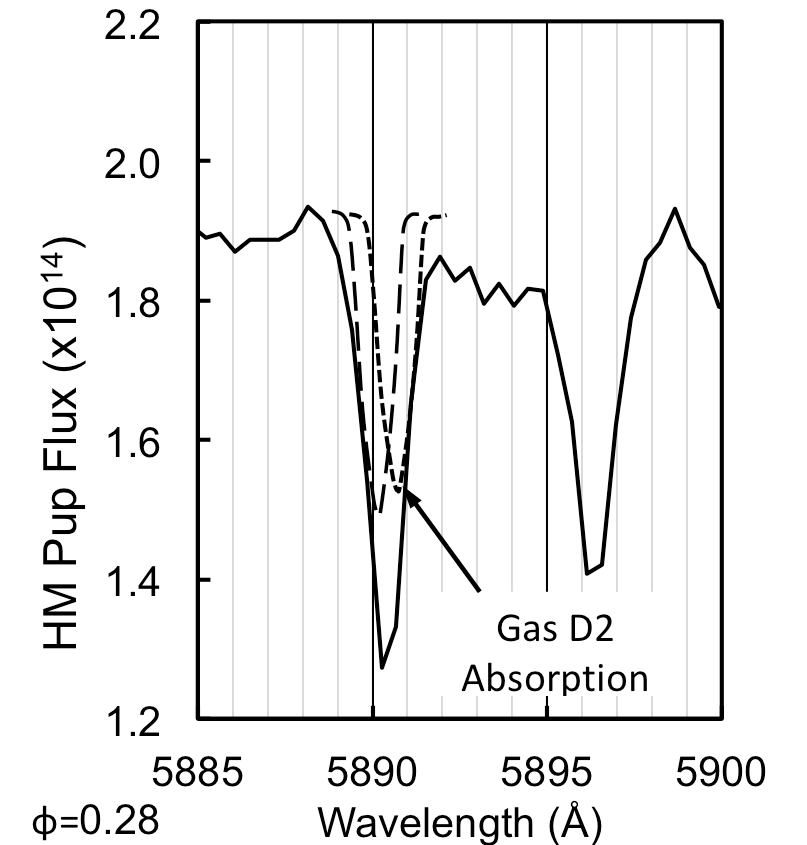}
\caption{Illustration of the method used to estimate the wavelength of the D2 line of the gas at phase 0.28. The primary star absorption line is long-dashed, the gas is short-dashed.}
\label{fig:PupNaDgas}
\centering
\includegraphics[width=8.4cm]{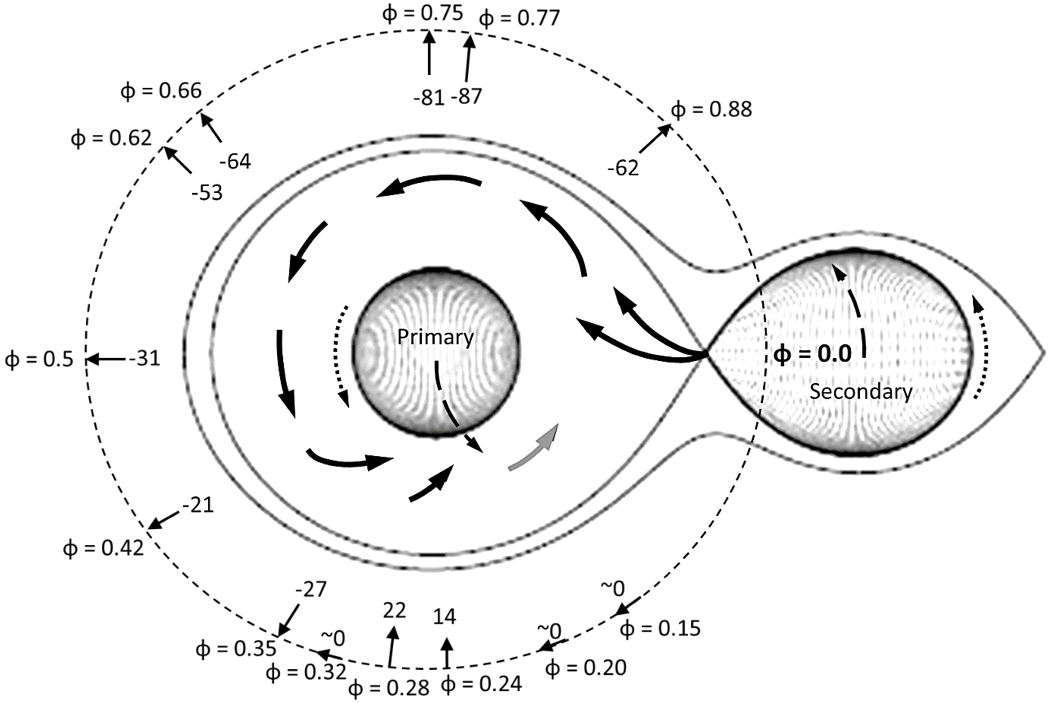}
\caption{Estimated path of the HM~Pup gas stream. The phases of the observations are marked around the periphery of the dashed circle, the corresponding observed gas velocity and direction relative to the primary star are indicated inside the circle. The orbital direction and direction of rotation of the two stars are shown by the long-dashed and short-dashed arrows respectively. The wavelengths and velocities were adjusted, for clarity, to remove the fixed systemic velocity of 58~km~s$^{-1}$.}
\label{fig:PupHaGasflow}
\end{figure}

The Na~I~D absorption lines of circumstellar gas, seen against the continuum of the primary star of HM~Pup, are distinctly separate from the Na~I~D lines of the stellar photosphere at phases between 0.6 and 0.9, as shown in Fig.~\ref{fig:PupNa}. The Full Width at Half Maximum (FWHM) of the primary component's absorption lines is 1.5~{\AA}, based on the primary star's rotational surface velocity; its velocity relative to the observer determines its central wavelength. Using these values, we were able to estimate the central position of the gas stream Na~I~D absorption lines at phases between 0.24 and 0.5 within the blended lines. This is illustrated in Fig.~\ref{fig:PupNaDgas} for phase 0.28; the centre of the stellar Na~I~D2 line was determined to be 5890.24~{\AA}. The shape of the blended absorption line indicates the line centre of the gas stream is 5890.68~{\AA}.

This method was used to estimate the observed velocities of the gas along the observer's line of sight, and thence its velocity relative the primary star at a range of phases. These velocities (listed in Table~\ref{tab:gas}) were used to determine an approximate path for the gas streaming from the secondary star and around the primary (Fig.~\ref{fig:PupHaGasflow}). The gas is shown as leaving the secondary star at the first Lagrangian point and orbiting the primary star (at least in part). The velocity and direction of the gas flow within this region are reflected in the observed gas velocities, particularly those between phases 0.24 and 0.32, where the gas appears to fall on to the primary star. A transient accretion cloud or annulus could be present, depending on the rate of mass loss from the secondary star. As discussed by \citet{RIC99}, accretion disks do not develop around EBs with short orbital periods.


\section{Light and radial velocities curves modelling and absolute parameters calculation}
\label{sec:LCRVMDL}

The light and radial velocity curves of the three systems were analysed with `PHOEBE' v.0.29d software \citep{PRS05}, which uses the 2003 version of the Wilson–Devinney code \citep{WIL71, WIL79, WIL90}. Temperatures, $T$, of the components were given values derived from the spectral classification (see Section~\ref{sec:SPECCLASS}) and the spectral type-temperature correlation of \cite{COX00}. The temperatures of the primary components were kept fixed during the analysis, while those of the secondaries were adjusted. For all systems, the albedos, $A$ and gravity darkening coefficients, $g$, of the components were given values according to their spectral types \citep{RUC69, ZEI24, LUC67}. As the maxima of the LCs are displaced towards the secondary minimum (typically between phases 0.25-0.45 and 0.55-0.75) as a result of reflection due to the heating of the secondary components’ atmospheres by the hot primary components, $A_2$ was adjusted. The (linear) limb darkening coefficients, $x$, for the $B$, $V$, and $I$ filters were taken from the tables of \citet{HAM93}, while those for the $TESS$ photometric system, $x_{TESS}$, from \citet[][Table~5 in that work]{CLA18}. The dimensionless potentials $\Omega$ and the fractional luminosity of the primary component $L_{1}$, were also adjusted during the modelling. The systemic radial velocity, $V_{\rm sys}$, the mass ratio, $q$, and the inclination of the system, $i$, were set as free parameters. As the three systems exhibit cyclic orbital period modulations (see Section~\ref{sec:OCMDL}) that could potentially be attributed to tertiary component(s), the third light parameter, $l_3$, was tested initially, but it converged to a zero value in each case, and was, therefore, excluded from further modelling.


\begin{table*}[h!]
\centering
\caption{Modeling and physical parameters results.}
\label{tab:LCRV}
\begin{tabular}{l cc cc cc}
\hline													
	&	\multicolumn{2}{c}{HM~Pup}			&	\multicolumn{2}{c}{V632~Sco}			&	\multicolumn{2}{c}{TT~Vel}			\\
\hline													
	&	\multicolumn{6}{c}{System parameters}											\\
\hline													
$T_0$~(HJD)	&	\multicolumn{2}{c}{2456306.998(1)}			&	\multicolumn{2}{c}{2458285.110(1)}			&	\multicolumn{2}{c}{2459329.957(13)}			\\
$P$~(d)	&	\multicolumn{2}{c}{2.5897239(2)}			&	\multicolumn{2}{c}{1.610163(1)}			&	\multicolumn{2}{c}{2.10846(2)}			\\
$i$~($\degr$)	&	\multicolumn{2}{c}{89.7(3)}			&	\multicolumn{2}{c}{78.4(1)}			&	\multicolumn{2}{c}{88.8(6)}			\\
$q$~($K_{1}$/$K_{2}$)	&	\multicolumn{2}{c}{0.243(9)}			&	\multicolumn{2}{c}{0.255(14)}			&	\multicolumn{2}{c}{0.232(8)}			\\
$V_{\rm sys}$~(km~s$^{-1}$)	&	\multicolumn{2}{c}{58(1)}			&	\multicolumn{2}{c}{-17(1)}			&	\multicolumn{2}{c}{24(1)}			\\
\hline													
	&	\multicolumn{6}{c}{Components parameters}											\\
\hline													
	&	$Primary$	&	$Secondary$	&	$Primary$	&	$Secondary$	&	$Primary$	&	$Secondary$	\\
\hline													
$T_{\rm eff}$~(K)	&	8000(150)$^a$	&	3915(35)	&	7000(150)$^a$	&	3734(186)	&	7600(150)$^a$	&	4515(66)	\\
$\Omega$	&	5.21(5)	&	2.370(1)	&	3.79(7)	&	2.369(1)	&	4.69(2)	&	2.332(9)	\\
$K$~(km~s$^{-1}$)	&	45(1)	&	185(2)	&	52(3)	&	205(4)	&	39(1)	&	167(2)	\\
$A$	&	1$^b$	&	1$^b$	&	1$^b$	&	0.92(1)	&	1$^b$	&	0.86(1)	\\
$g^b$	&	1	&	0.32	&	1	&	0.32	&	1	&	0.32	\\
$r_{\rm pole}$	&	0.202(1)	&	0.250(1)	&	0.282(1)	&	0.250(1)	&	0.225(1)	&	0.244(1)	\\
$r_{\rm point}$	&	0.204(1)	&	0.365(1)	&	0.290(1)	&	0.364(3)	&	0.227(1)	&	0.357(3)	\\
$r_{\rm side}$	&	0.203(1)	&	0.261(1)	&	0.287(1)	&	0.260(1)	&	0.226(1)	&	0.254(1)	\\
$r_{\rm back}$	&	0.204(1)	&	0.293(1)	&	0.289(1)	&	0.293(1)	&	0.227(1)	&	0.287(1)	\\
$x_{B}$	&	0.592	&	0.999	&	0.605	&	1.001	&	0.612	&	0.948	\\
$x_{V}$	&	0.516	&	0.866	&	0.497	&	0.812	&	0.533	&	0.797	\\
$x_{I}$	&	0.352	&	0.628	&	0.338	&	0.592	&	0.365	&	0.571	\\
$x_{TESS}$	&	0.229	&	0.565	&	0.298	&	0.468	& 0.237	&	0.496	\\
$L/(L_1+L_2)$~($B$)	&	0.976(1)	&	0.024(1)	&	0.976(1)	&	0.024(1)	&	0.930(1)	&	0.070(1)	\\
$L/(L_1+L_2)$~($V$)	&	0.947(1)	&	0.053(1)	&	0.973(1)	&	0.027(1)	&	0.884(1)	&	0.116(1)	\\
$L/(L_1+L_2)$~($I$)	&	0.866(1)	&	0.134(1)	&	0.938(3)	&	0.062(1)	&	0.803(1)	&	0.197(1)	\\
$L/(L_1+L_2)$~($TESS$)	&	0.877(1)	&	0.123(1)	&	0.919(1)	&	0.081(1)	&	0.816(1)	&	0.184(1)	\\
\hline													
	&	\multicolumn{6}{c}{Physical parameters}											\\
\hline													
$M$~($M_{\sun}$)	&	2.65(9)	&	0.65(4)	&	2.4(1)	&	0.61(9)	&	1.55(6)	&	0.36(3)	\\
$R$~($R_{\sun}$)	&	2.39(3)	&	2.99(3)	&	2.39(5)	&	2.11(5)	&	1.94(3)	&	2.12(3)	\\
$L$~($L_{\sun}$)	&	21(1)	&	2.4(3)	&	12.2(9)	&	0.78(9)	&	11.2(9)	&	1.7(2)	\\
$\log g$~(cm~s$^{-2}$)	&	4.11(2)	&	3.30(3)	&	4.06(3)	&	3.57(6)	&	4.05(2)	&	3.34(4)	\\
$a$~($R_{\sun}$)	&	2.31(5)	&	9.5(1)	&	1.70(8)	&	6.7(1)	&	1.62(5)	&	7.0(1)	\\
$M_{\rm bol}$	&	1.4(4)	&	4.1(3)	&	2.0(4)	&	5.0(9)	&	2.1(5)	&	4.2(4)	\\
\hline													
\end{tabular}
\tablefoot{$^a$taken from spectroscopy (Section~\ref{sec:SPECCLASS}), $^b$assumed}
\end{table*}

Intense magnetic dynamo activity is expected in close EBs with convective outer regions and rapid rotation rates locked to their orbital velocities \citep[see][]{BER05}. The magnetic activity causes variability or asymmetry in LCs which could affect the modelling results. Several LC cycles for each system were obtained from the $TESS$ data (see Table~\ref{tab:TESSlog}). We carefully examined them to determine the stability of the asymmetries in successive LCs, and then decide how to proceed with the model. There were 20 and 12 complete LCs of mid-cadence data for HM~Pup and for TT~Vel, respectively, which, although displaying some variability, were sufficiently stable for modelling over the time range of the observations. However, in the case of V632~Sco there were 12 complete LCs of short-cadence data which show noticeable changes every few days. Therefore, for this system we modelled three groups of four LCs folded into the $P_{\rm orb}$.


Given that the purpose of this work concerns mostly the asteroseismology of these systems, we excluded the long-cadence data of the systems from the modelling and kept only those with the best available time resolution. However, the long-cadence data were used for minima timing calculations (see Section~\ref{sec:OCMDL}). As with the $TESS$ data, the ground-based observations obtained in different time periods show asymmetries that cannot easily be fitted in a unique phased LC. In order to avoid this in our study, we again modelled the LCs of each system separately according to the asymmetries found. In particular, as V632~Sco is more active than the other two systems, in order to have complete LCs for the modelling, photometric observations in the $V$ and $I$ pass-bands over several years were combined. The examination of the LCs was made with great care, as the LC residuals are used later for frequency analysis. They must, therefore, be kept free of any proximity or stellar magnetic activity effects.


During the modelling process, all LCs from each pass-band were used initially with the RVs to derive the mean model of each system. Then, the values from the output of the mean model were used as initial inputs to model the individual phase-folded LCs (or groups of LCs) mostly by adjusting the spot parameters, namely Co-latitude, longitude, radius, and temperature factor. Thus, the parameters of the final model of each system are the average of the respective parameters of each LC model, and their uncertainties are the standard deviations. This procedure gives more realistic uncertainty values in comparison with the direct errors derived from `PHOEBE' \citep[c.f.][]{LIA17, LIA18}. Three modes of LC analyses were tried: mode 2 (detached system), mode 4 (semi-detached system with the primary component filling its Roche lobe) and mode 5 (conventional semi-detached binary); models for the three systems converged in mode 5. Therefore, all of them are semi-detached systems with their secondary components (less massive and cooler) filling their Roche lobes; thus, they are classical Algols (from evolutionary point of view). The amplitudes $K_1$ and $K_2$ were calculated using sinusoidal fittings on the respective RVs. The (linear) ephemeris of each system, i.e. the orbital period, $P_{\rm orb}$, and a reference time of a primary minimum, $T_0$, was based on the observed times of minima between 2012-2020 (see Section~\ref{sec:OCMDL}). The absolute parameters of the components were calculated based on the modelling results using the `AbsParEB' software \citep[][mode 3]{LIA15}. The upper and middle parts of Table~\ref{tab:LCRV} list the LCs and RVs modelling parameters, while its lower part the absolute parameters of the components. The fit on the LCs and RVs for all systems are given in Fig~\ref{fig:LCSRVS}. The spots parameters for each system are given in Table~\ref{tab:spots} in Appendix~\ref{sec:AppSpots}.


\begin{figure*}
\begin{tabular}{cc}
\includegraphics[width=8.5cm]{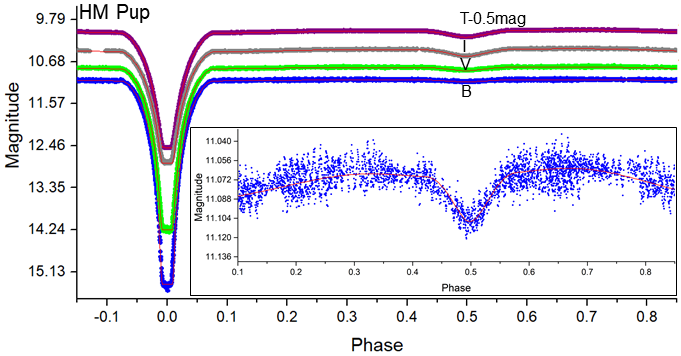}&\includegraphics[width=8.5cm]{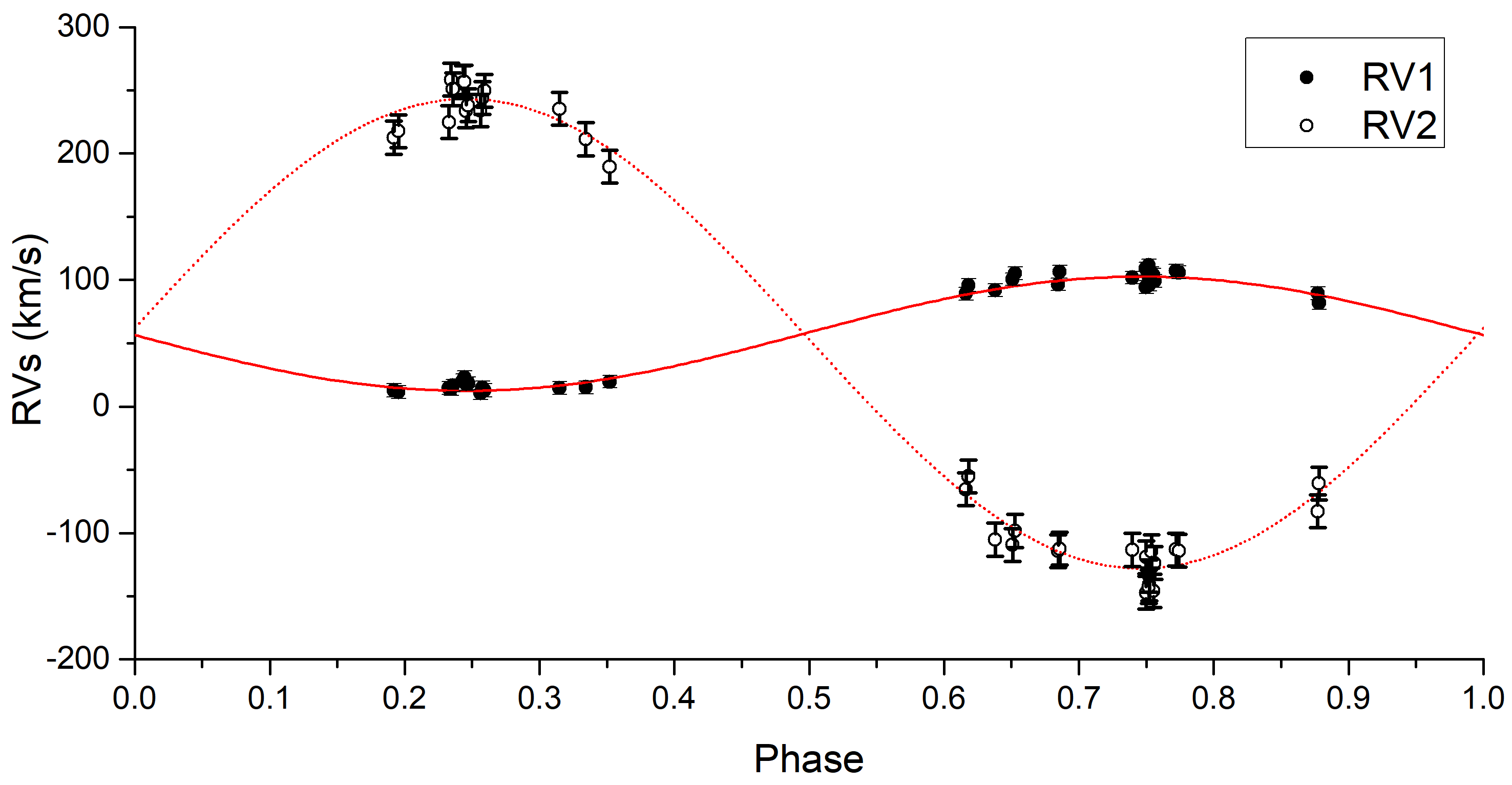}\\
\includegraphics[width=8.5cm]{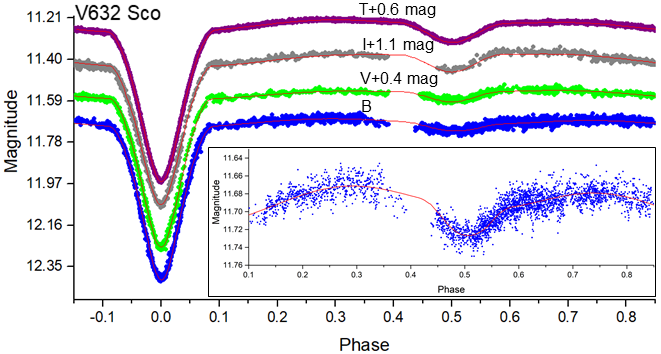}&\includegraphics[width=8.5cm]{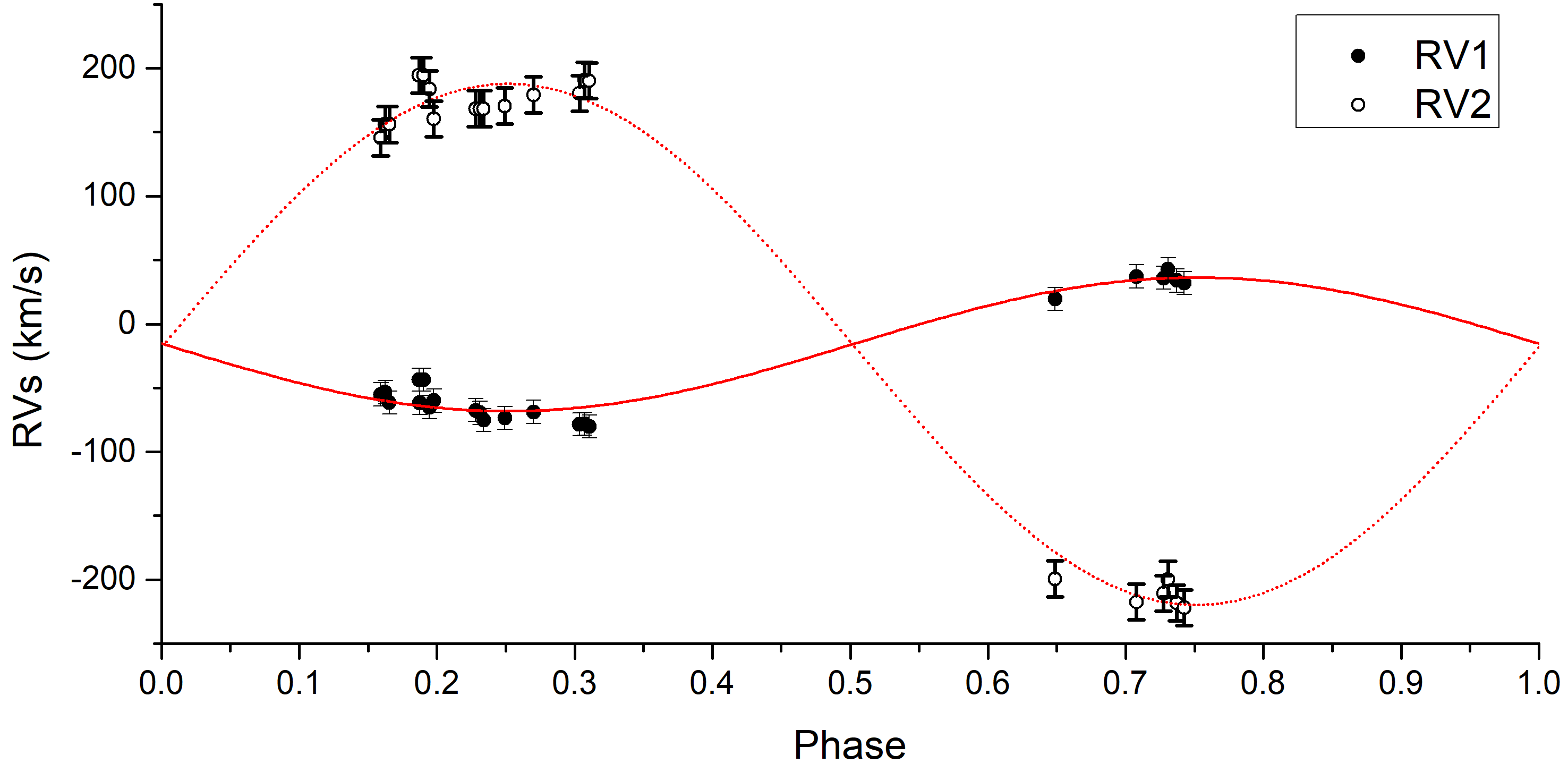}\\
\includegraphics[width=8.5cm]{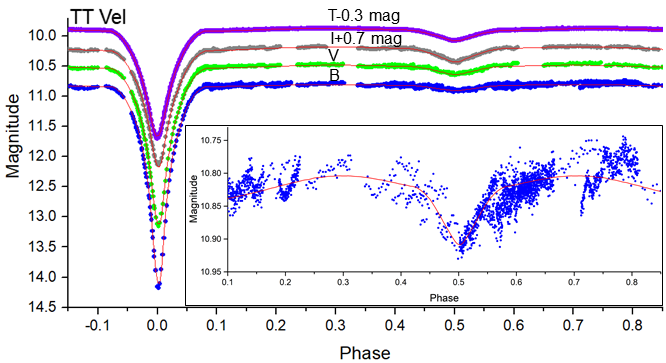}&\includegraphics[width=8.5cm]{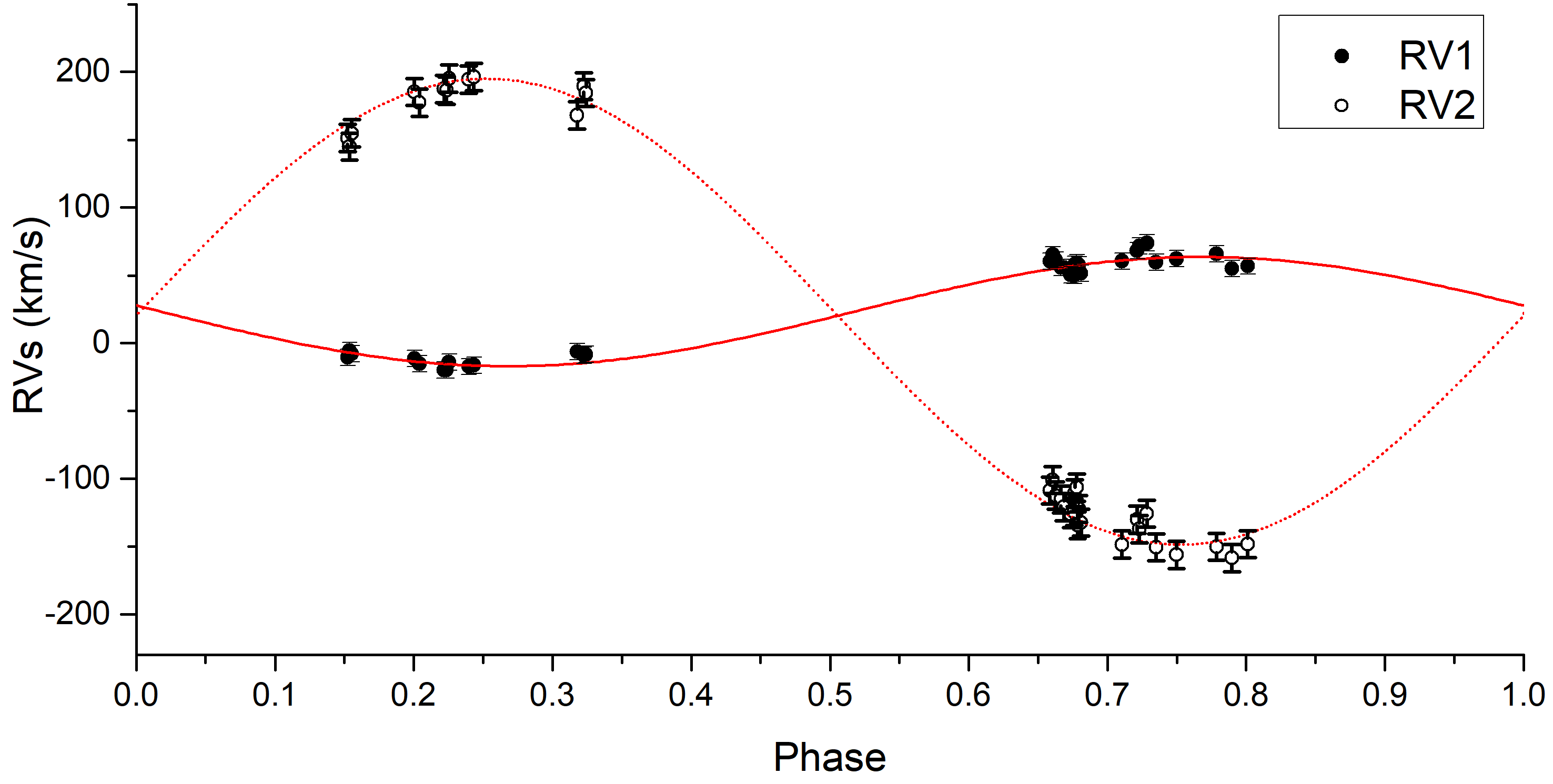}\\
\end{tabular}
\caption{Synthetic (solid lines) over observed (points) LCs (left plots) and RVs (right plots) for HM~Pup (top panels), V632~Sco (middle panels), and TT~Vel (bottom panels). Some LCs are vertically shifted for scaling reasons. The fit on the out-of-primary eclipse points of $B$ filter is re-scaled and plotted allowing for better viewing.} 
\label{fig:LCSRVS}
\end{figure*}



\section{Orbital period variation analyses}
\label{sec:OCMDL}

In order to determine whether orbital periods of the systems have changed, accurate eclipse times over a long term are necessary in order to apply an ETV analysis. For this, the first step was to gather the past times of minima from the literature and relevant minima databases and calculate new ones from our ground-based LCs as well as from the $TESS$ data. Subsequently, these minima timings were used for the construction of the ETV plots and were fitted by various curves that correspond to different orbital period modulating physical mechanisms. 

Times of minima of the systems observed at the Congarinni, El Sauce and Glen Aplin observatories were calculated in PERANSO\footnote{\url{http://www.peranso.com/}} from a 7th order polynomial fit to the light curves spanning two hours each side of the minima. Past times of minima were collected from the `O-C gateway\footnote{\url{http://var2.astro.cz/ocgate/}}', `TIDAK\footnote{\url{https://www.as.up.krakow.pl/tidak}}' \citep[][TIming DAtabase at Krakow; former O-C atlas; private communication with B.~Zakrzewski]{OGL22} and the International Variable Star Index (VSX) databases\footnote{\url{https://www.aavso.org/vsx/}} and other literature sources containing times of minima. For eclipses observed in multiple filters in the same night, the average of the timings derived from each filter was used. All available TESS data sets (see Table~\ref{tab:TESSlog}) were used for minima timing derivations. Due to the huge amount of observations in each data set, the software `B-Minima'~v.1.2 \citep{NEL05} was used; it searches and calculates minima timings automatically in big files. The calculated times of minima are given in Table~\ref{tab:MIN} in Appendix~\ref{sec:AppMin}.


The mass transfer/loss mechanisms that are expected to occur in semi-detached EBs were tested for the studied systems. Mass transfer stipulates secular changes of $P_{\rm orb}$, while the ETV points present parabolic distribution, whose curvature indicates the direction of the mass transfer (i.e. upward means mass transfer from the less to the more massive component). The parabolic coefficient $C_2$ can be determined by fitting a parabola on the data points, and is used to calculate the $P_{\rm orb}$ change rate ($\dot{P}$) using the following equation \citep[c.f.][]{KAL94}:
\begin{equation}
\dot{P}=\frac{2C_2}{P_{\rm orb}}.\\
\label{eq:PDOT}
\end{equation}
In the case of conservative mass transfer, the mass transfer rate ($\dot{M_{\rm tr}}$) is calculated with the following equation \citep[c.f.][]{KRU66, HIL01}:
\begin{equation}
\dot{M}_{\rm tr}=\frac{\dot{P}~M_1~M_2}{3P_{\rm orb}~(M_1-M_2)},\\
\label{eq:MDOT}
\end{equation}
where $M_1$, $M_2$ are the masses of the components of the EB.

Mass loss from EB systems can be due to stellar winds. This mechanism causes an increase in the orbital period, which is reflected as an upward parabolic distribution of the ETV points. The mass loss rate, $\dot{M}_{\rm loss}$, is computed according to the formula of \citet{HIL01}:
\begin{equation}
\dot{M}_{\rm loss}=-\frac{M_1+M_2}{2P_{\rm orb}}\dot{P}.\\
\label{eq:MDOT2}
\end{equation}

\begin{figure}
\centering
\includegraphics[width=8.5cm]{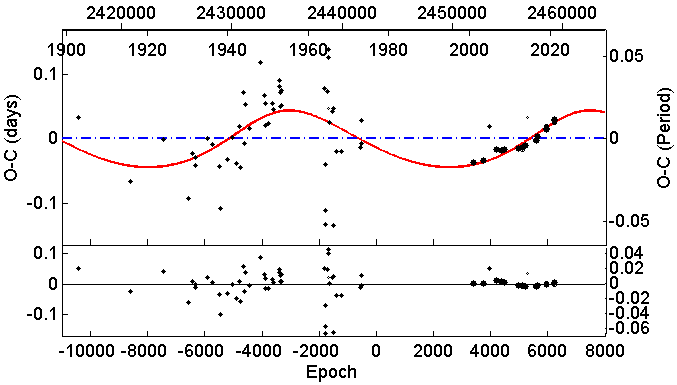}\\
\caption{Eclipse timing variations plot for HM~Pup. Top panel: The $O-C$ points of the system fitted by a LITE curve (upper part) and the residuals (lower part) of the fit. For both panels, the bigger the symbol the bigger the statistical weight of the individual points.}
\label{fig:OCPUP}
\vspace{0.5cm}
\begin{tabular}{c}
\includegraphics[width=8.5cm]{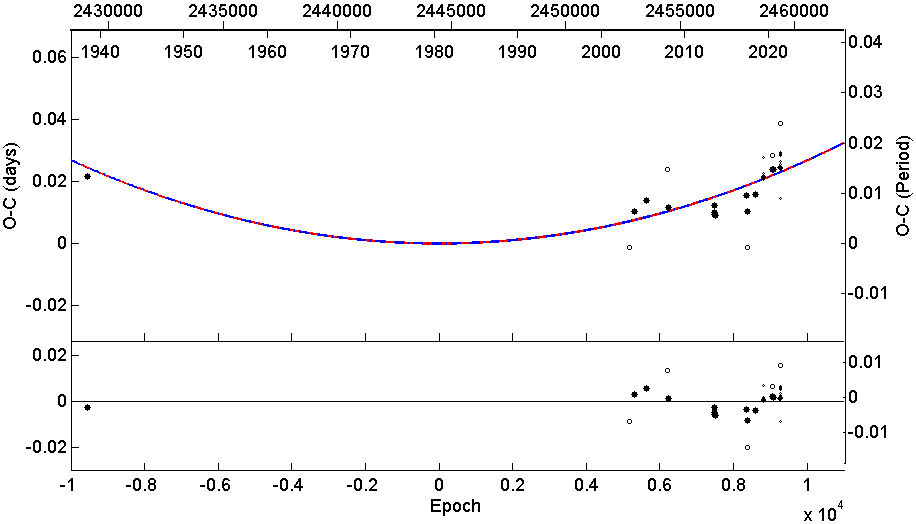}\\
\includegraphics[width=8.5cm]{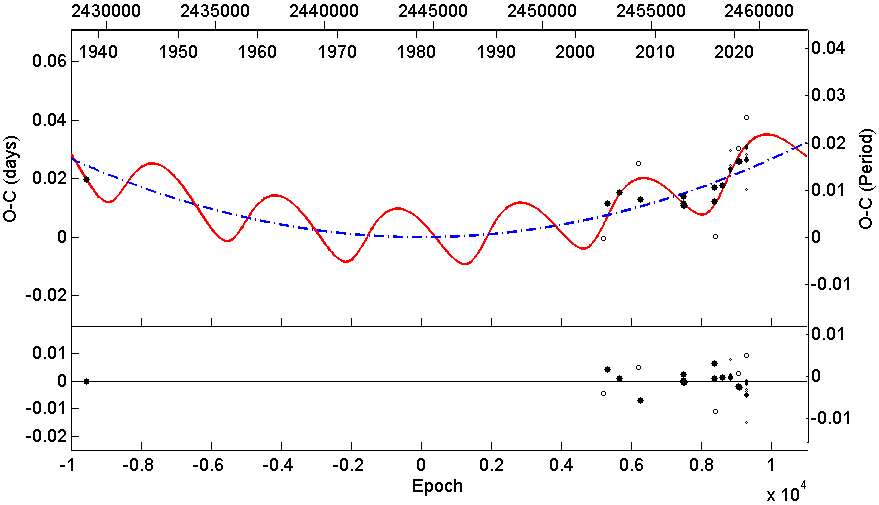}\\
\end{tabular}
\caption{Eclipse timing variations plots for V632~Sco. Top panel: The $O-C$ points of the system fitted by a parabola (upper part) and the residuals (lower part) of the fit. Bottom panel: The same as previous, but with the fit of a parabolic and a LITE curves.}
\label{fig:OCSCO}
\end{figure}


\begin{figure}
\includegraphics[width=8.5cm]{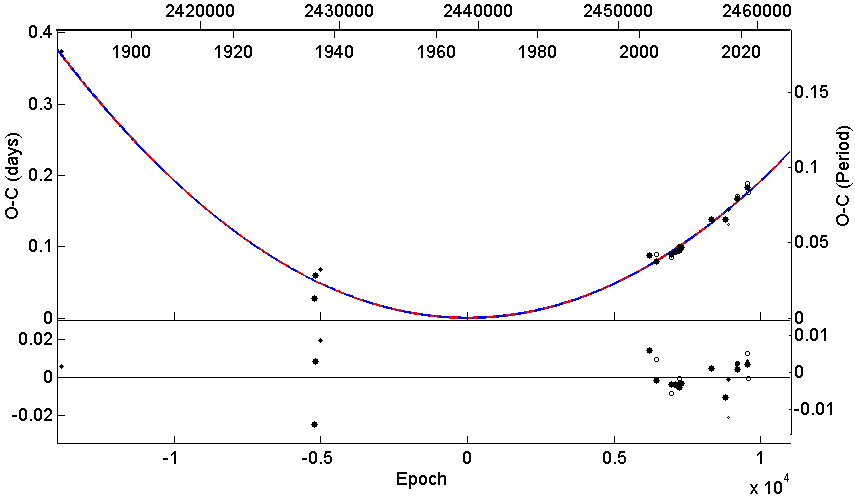}\\
\includegraphics[width=8.5cm]{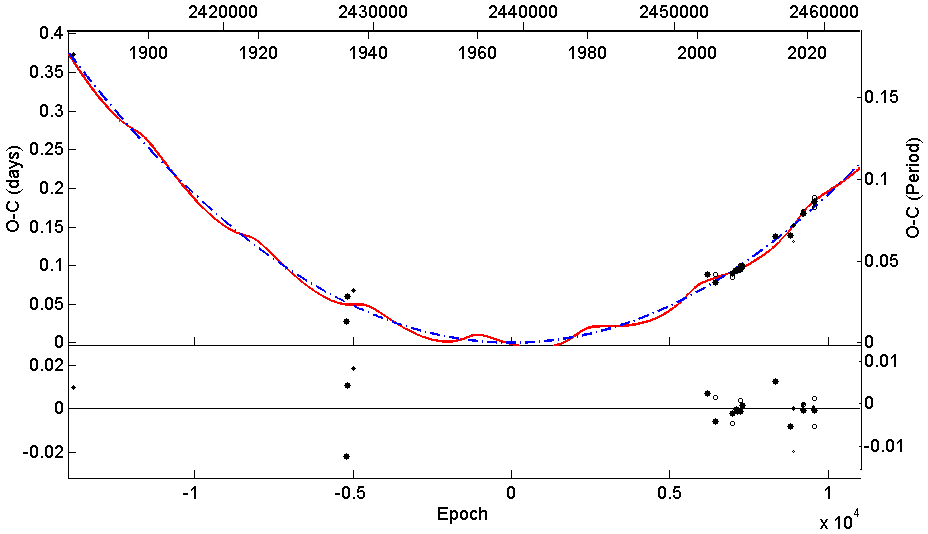}
\caption{The same as Fig.~\ref{fig:OCSCO}, but for TT~Vel.}
\label{fig:OCVEL}
\end{figure}

Cyclic changes of the orbital period can be explained by either the Light-Time effect  \citet[][LITE]{IRW59} or with variations in magnetic cycles that could also modulate changes in eclipse timing due to the effect of star spots or with changes in the magnetic quadrupole moment in the secondary star \citep[Applegate mechanism][]{APP87, APP92, LAN98, HOF06, ROV00, LAN02}. Applegate's mechanism, however, is not likely to contribute significantly to period modulation in systems like those in this study \citep{LAN06, VOL18}. An alternative mechanism involving a persistent non-axisymmetric internal magnetic field in the active component \citep{LAN20} may explain the observed ETV. However, an in deep examination of the cyclic modulations of the orbital periods of the systems are beyond the scope of this study.

Eclipse times were analysed with the $LITE$ code of \citet{ZAS09} that uses the method of statistical weights according to method used for the observations of the eclipses. Statistical weights, $w$, were applied as follows: for visually or photographically determined minima were assigned $w=1$, and the photoelectric and CCD minima $w=10$. This method works well when the minima timings are spread uniformly in time. For V632~Sco and TT~Vel, most eclipse times were obtained after the year 2000. Moreover, there are a lot of minima timings from the $TESS$ mission that cover only one or two years. In order to avoid over-weighting only one part of the ETV distribution, the $TESS$ data were given $w=1$ instead of 10. Using this method, the many data points gathered within a short time range do not drag the fitting functions on them. In addition, minima timings based on the data of ASAS \citep{POJ02}, ASAS-SN \citep{SHA14} and Catalina \citep{DRA09} surveys (included in the TIDAK database) were given $w=1$ due to their large uncertainties.

The code can fit one parabola and up to two LITE curves simultaneously. The initial ephemerides used for the ETV diagrams of the systems were taken from the `TIDAC' and `O-C gateway' databases. The ephemeris ($T_0$-reference time of minimum and $P_{\rm orb}$), the period of the modulation $P_3$, the amplitude of the ETV variation $A_{\rm ETV}$, the argument of periastron $\omega_3$, the eccentricity $e_3$, and the time of the periastron passage $T_{\rm per}$ of the additional body are the free parameters of each LITE fitting curve. Moreover, the parabolic coefficient $C_2$ is also adjusted when parabola is selected for fit \citep[for more details see][]{ZAS09}. The selection of the final model is based on the Bayes Information Criterion (BIC).


\begin{table*}
\centering
\caption{ETV analysis results.}
\label{tab:OCRES}
\begin{tabular}{l c c c}
\hline									
System	&	HM~Pup	&	V632~Sco	&	TT~Vel	\\
\hline							
$T_0$~(HJD)	&	2443091.838(4)	&	2444450.636(4)	&	2439175.396(4)	\\
$P_{\rm orb}$~(d)	&	2.5896870(8)	&	1.6101553(3)	&	2.1084195(2)	\\
\hline							
\multicolumn{4}{c}{Light-Time effect}							\\
\hline							
$P_{3}$~(yr)	&	75(2)	&	15.1(3)	&	20.2(2)	\\
$T_{\rm per}$~(HJD)	&	2679147(7450)	&	2469052(1315)	&	2776514(3786)	\\
$A_{\rm ETV}$~(d)	&	0.044(9)	&	0.010(2)	&	0.0081(8)	\\
$\omega_3$~($\degr$)	&	63(29)	&	317(92)	&	45(27)	\\
$e_3$	&	0.2(1)	&	0.3(1)	&	0.4(2)	\\
$f(m_3)$~($M_{\sun}$)	&	0.082(2)	&	0.022(1)	&	0.076(7)	\\
$M_{\rm _3, min}$~($M_{\sun}$)	&	1.18(1)	&	0.66(1)	&	0.34(1)	\\
\hline							
\multicolumn{4}{c}{Mass transfer/loss process}							\\
\hline							
$C_2$~($\times 10^{-10}$~d~cycle$^{-1}$)	&		&	2.68(2)	&	19.2(1)	\\
$\dot{P}$~($\times 10^{-7}$~d~yr$^{-1}$)	&		&	1.22(1)	&	6.7(1)	\\
$\dot{M}_{\rm tr}$~($\times 10^{-8}$~$M_{\sun}$~yr$^{-1}$)	&		&	2.1(5)	&	4.9(6)	\\
$\dot{M}_{\rm loss}$~($\times 10^{-7}$~$M_{\sun}$~yr$^{-1}$)	&		&	-1.14(8)	&	-3.0(2)	\\
\hline							
\end{tabular}
\end{table*}

Analyses of eclipse times of HM~Pup did not show evidence of mass transfer, i.e. no secular changes were detected, but one possible cyclic variation (Fig.~\ref{fig:OCPUP}). The results from a LITE curve fit are shown in Table~\ref{tab:OCRES}. The proposed mass of the additional component would require a comparable large luminosity, but there was no evidence for that in our spectroscopic observations nor in the LCs analysis. Therefore, other modulating mechanisms need to be tested for the explanation of this cyclic change. Moreover, a second LITE curve was tested because the data points after 2000 show evidence of an additional cyclic change. However, this LITE curve was proven statistically insignificant based on the BIC and was excluded.  

For V632~Sco there was only one published time of minimum before the year 2000, i.e. in 1939. A parabola was chosen to fit the data (Fig.~\ref{fig:OCSCO}-top panel), but the residuals show an additional cyclic distribution. Therefore, a LITE curve was also tested and is plotted in Fig.~\ref{fig:OCSCO} (bottom panel). Results for several possible mechanisms are given in Table~\ref{tab:OCRES} and were calculated using the `InPeVEB' software \citep{LIA15}. The upward curvature of the parabola is in good agreement with what would be expected for a conventional semi-detached configuration of this EB. It is a classical Algol system with mass transfer from the less to the more massive component and a rate of $\sim2.1\times10^{-8}~M_{\sun}$~yr$^{-1}$. The third body hypothesis suggests a minimum mass of 0.66~$M_{\sun}$ and a potential luminosity contribution of only about $\sim1.8\%$ in the system \citep[following the formalism of][]{LIA11}. This contribution would not have been detectable in the analyses of LCs and RVs. 

Most of the eclipse timings for TT~Vel were made in the past two decades; after its discovery in 1913, only a few timings were reported in the late 1930s. In the ETV analysis, a parabola fitted the data (Fig.~\ref{fig:OCVEL}-top panel). The residuals indicated a periodic trend, to which a $LITE$ curve was fitted (Fig.~\ref{fig:OCVEL}-bottom panel). The orbital period changes agree with a conservative mass transfer from the less to the more massive component of the system with a rate of $\sim4.9\times10^{-8}~M_{\sun}$~yr$^{-1}$. The cyclic changes can be explained by the $LITE$ that suggests a low mass component ($M_{3, \rm min}\sim0.34~M_{\sun}$), whose light contribution is found less than 0.5\%. Therefore, its absence from photometry and spectroscopy is plausible. 

\section{Pulsation analyses}
\label{sec:PULMDL}

\begin{figure*}[h!]
\centering
\begin{tabular}{c}
\includegraphics[width=17cm]{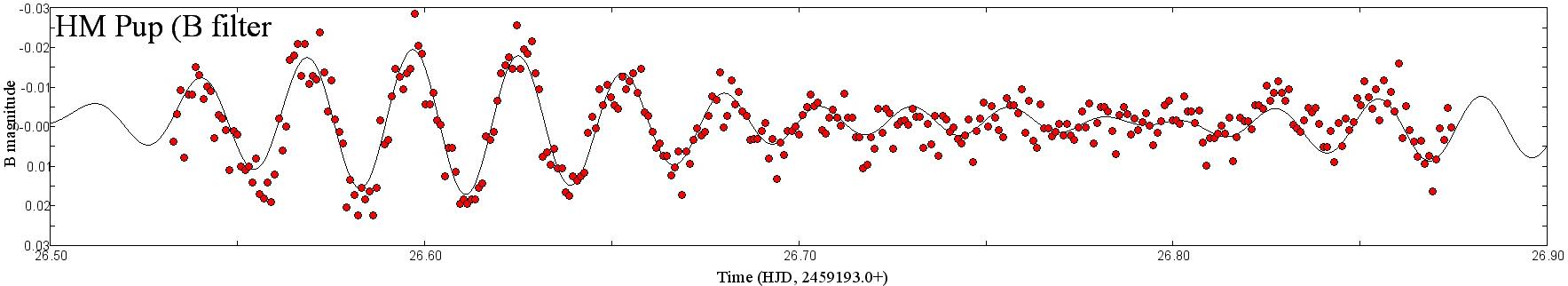}\\
\includegraphics[width=17cm]{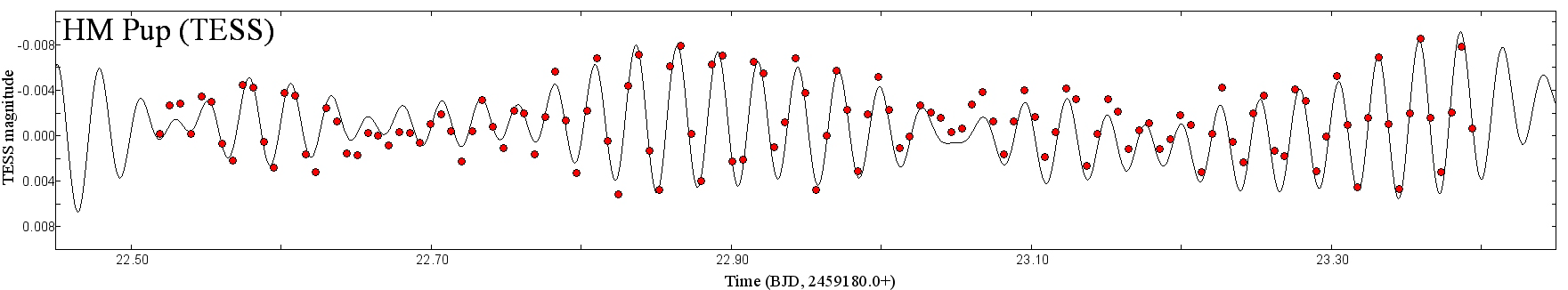}\\
\includegraphics[width=17cm]{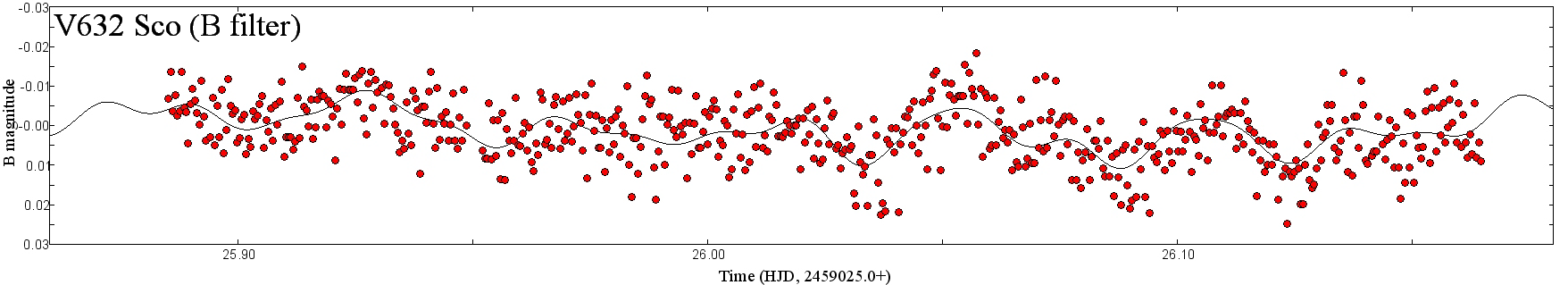}\\
\includegraphics[width=17cm]{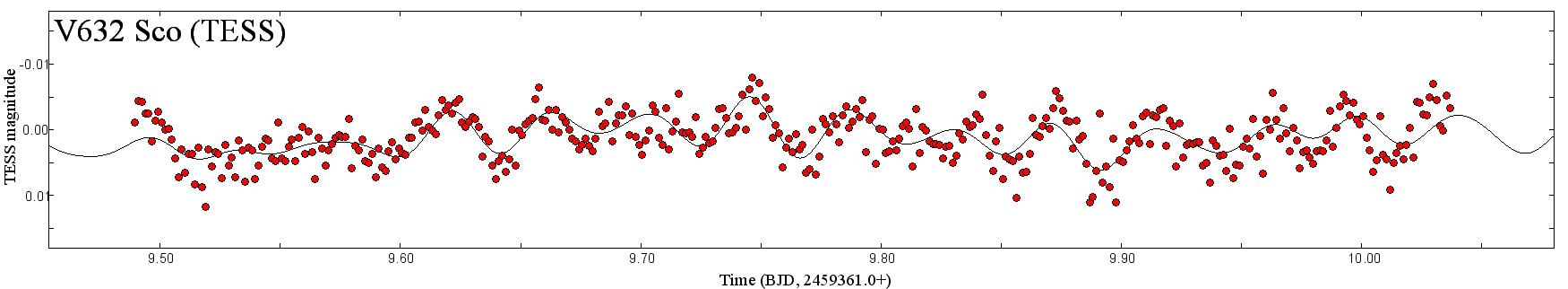}\\
\includegraphics[width=17cm]{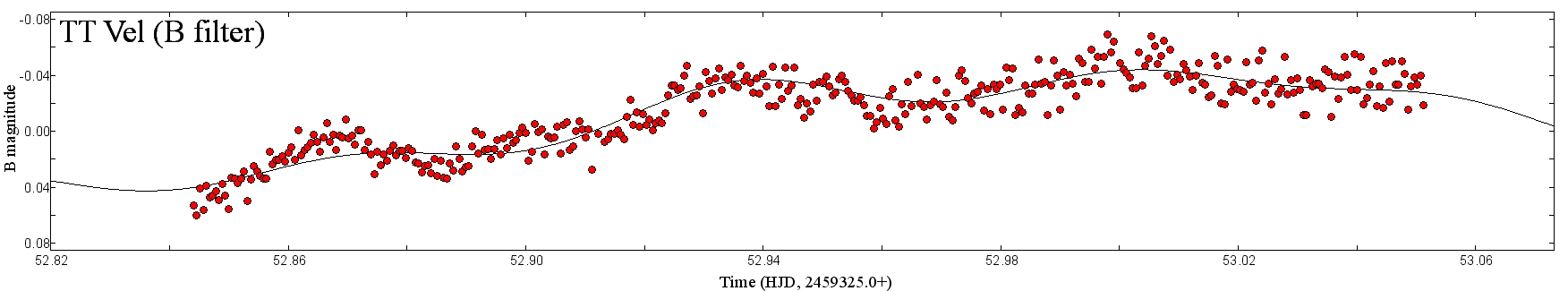}\\
\includegraphics[width=17cm]{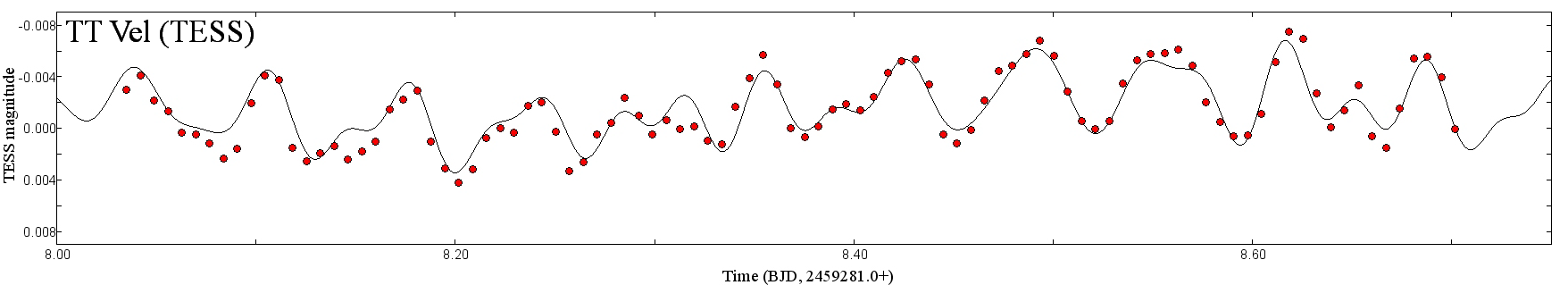}\\
\end{tabular}
\caption{Representative Fourier fit on the data of $B$~filter and $TESS$.}
\label{fig:FF}
\end{figure*}

The oscillation frequencies search was made with the software \textsc{PERIOD04} v.1.2 \citep{LEN05}, which is based on classical Fourier analysis. Given that the typical frequency range of $\delta$~Scuti stars is 4-80~cycle~d$^{-1}$ \citep{BRE00, BOW18}, but as many of them exhibit slower modes (e.g. due to $\gamma$~Doradus-$\delta$~Scuti hybrid nature and/or due to $g$-mode pulsations), the search was expanded to slower frequency regimes, that is 0-80~cycle~d$^{-1}$. For the analysis of the pulsations, only the best quality ground-based data and the $TESS$ data with the best available time resolution were used. For this analysis, the binary model was subtracted from the observed LCs (see Section~\ref{sec:LCRVMDL} for details) of each system and the Fourier method on the time-series residuals data was applied. Moreover, since the pulsation amplitudes are modified during the eclipses, and in order to keep an homogeneous data sample, only the out-of-eclipse data points were used. For the three systems, the data corresponding to phases 0.92-0.08 and 0.42-0.58 were excluded.

The S/N of the frequencies was calculated around the area of the signal \citep{BAL14} with a spacing of 2~cycle~d$^{-1}$ and a box size of 2. A S/N=4, suggested by the developers of the software \citep[see also][]{BRE93}, as the minimum value of a detected frequency to be considered as reliable was adopted. However, this threshold concerned only the ground-based data sets. For the $TESS$ data, based on the conclusions of \citet{BAR15, BAR21, BOW21}, a S/N=5 value was adopted. After the detection of a frequency, the software subtracts it from the observed points and searches for another in the new residuals data until the amplitude of the last frequency reaches the critical S/N.


\begin{table*}
\centering
\caption{Properties of the data sets used for the frequency search.}
\label{tab:PULMDLPAR}
\scalebox{0.95}{
\begin{tabular}{l ccc ccc c}
		\hline													
System	&	Filter	&	Nights	&	$\Delta T$	&	points	&	$\delta t$	&	$Nyq.$	&	$\delta f$	\\
	&		&		&	(d)	&		&	(min)	&	(cycle~d$^{-1}$)	&	(cycle~d$^{-1}$)	\\
		\hline													
\multirow{4}{*}{HM Pup}	&	$B$	&	15	&	18.3	&	3678	&	1.5	&	494.9	&	0.054	\\
	&	$V$	&	12	&	43.2	&	3168	&	1.0	&	701.4	&	0.023	\\
	&	$I$	&	10	&	12.2	&	1893	&	2.2	&	472.0	&	0.082	\\
	&	$TESS$	&	78.2	&	78.2	&	6672	&	10.0	&	72.04	&	0.013	\\
		\hline													
\multirow{3}{*}{V632 Sco}	&	$B$	&	8	&	57.2	&	2007	&	0.8	&	832.6	&	0.017	\\
	&	$V$	&	2	&	6.92	&	335	&	1.2	&	1056.4	&	0.145	\\
	&	$TESS$	&	27.8	&	27.8	&	12447	&	2.0	&	359.9	&	0.036	\\
		\hline													
\multirow{4}{*}{TT Vel}	&	$B$	&	16	&	59.0	&	1768	&	0.8	&	950.4	&	0.017	\\
	&	$V$	&	12	&	52.0	&	232	&	3.4	&	101.3	&	0.019	\\
	&	$I$	&	12	&	51.3	&	479	&	3.4	&	101.1	&	0.019	\\
	&	$TESS$	&	24.3	&	24.3	&	2060	&	10.0	&	72.04	&	0.041	\\
		\hline										
\end{tabular}}
\end{table*}

\begin{figure*}
\begin{tabular}{cc}
\centering
\includegraphics[width=8.5cm]{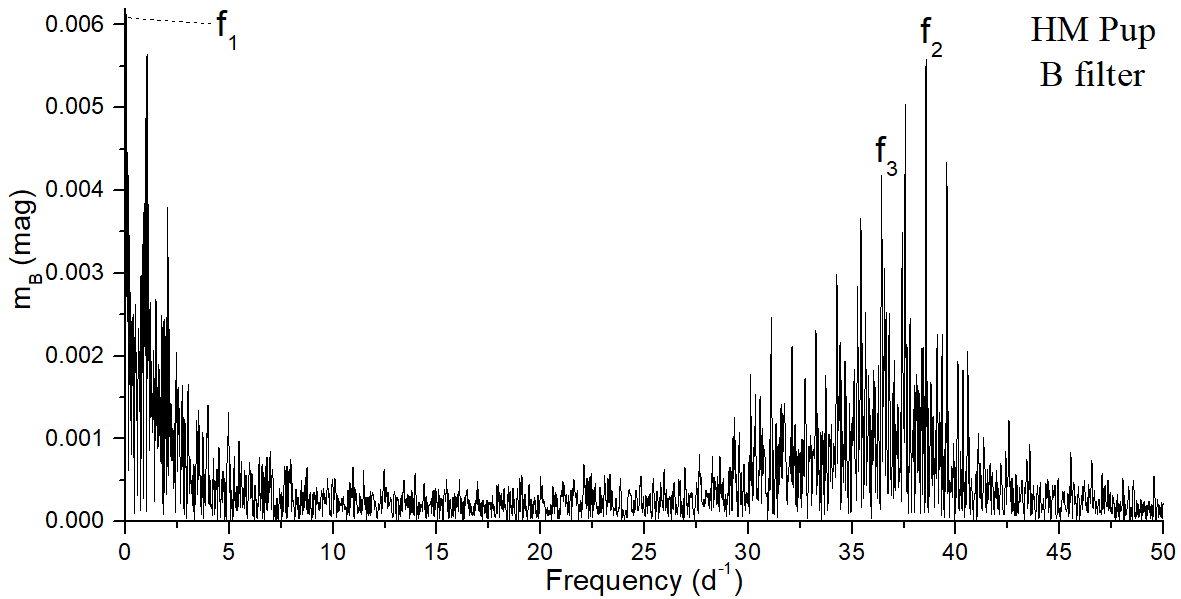}&\includegraphics[width=8.7cm]{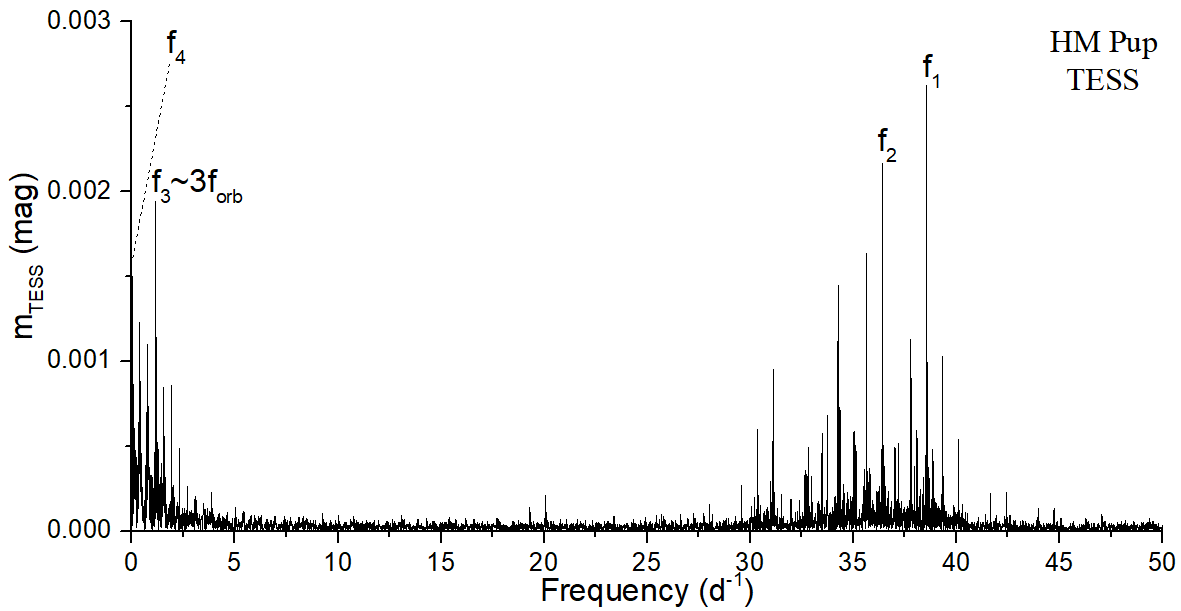}\\
\includegraphics[width=8.7cm]{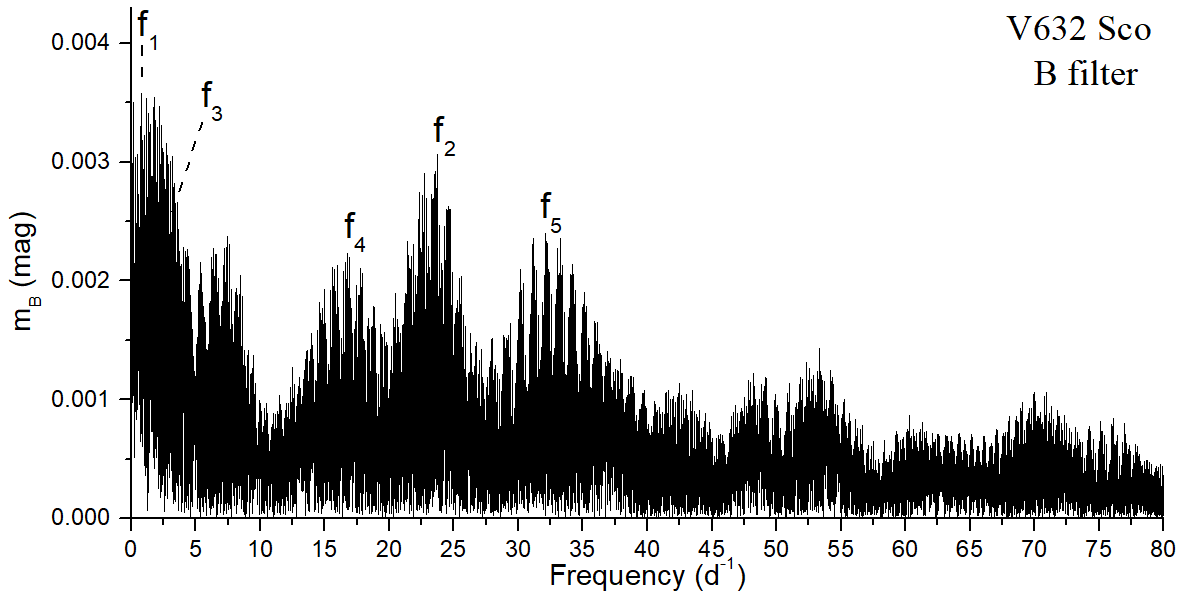}&\includegraphics[width=8.7cm]{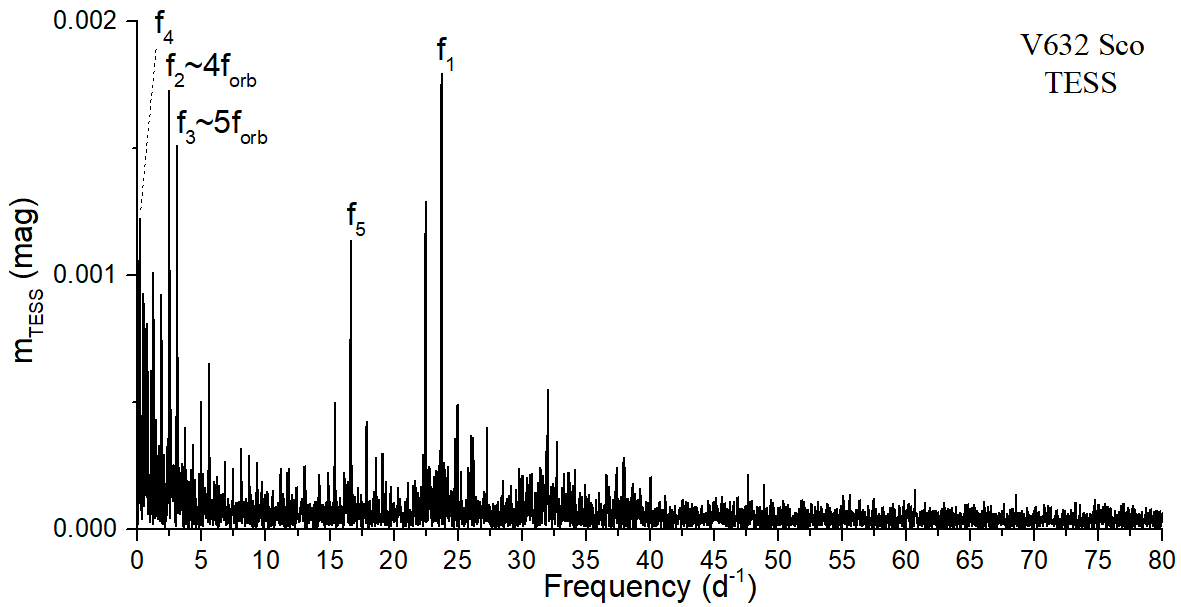}\\
\includegraphics[width=8.7cm]{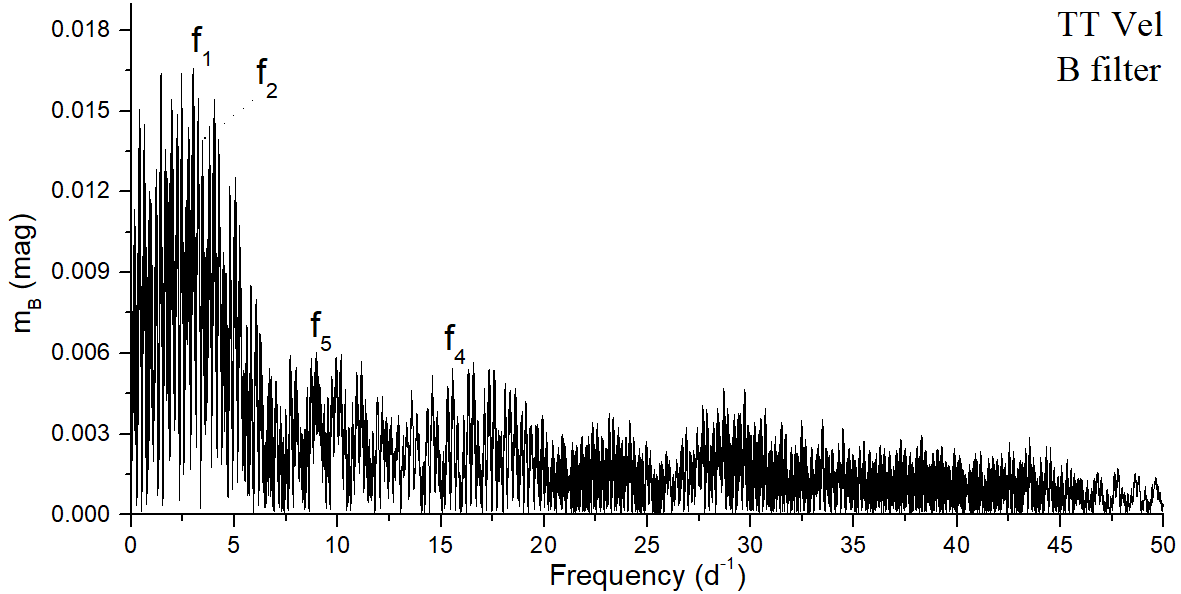}&\includegraphics[width=8.7cm]{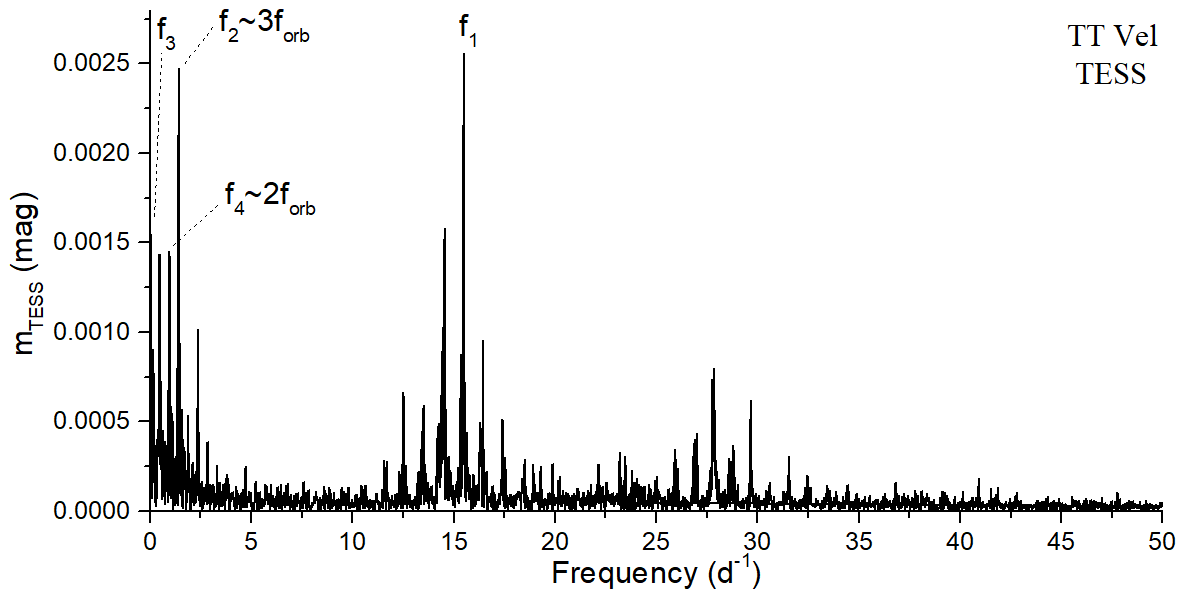}\\
\end{tabular}
\caption{Periodograms of $B$ filter (left panels) and $TESS$ data (right panels). Some of the detected strong frequencies are indicated.}
\label{fig:FS}
\end{figure*}

In Table~\ref{tab:PULMDLPAR} we list for each observed pass band: the number of nights, the range between the last and the first observed nights, the number of data points, the mean time resolution of these points ($\delta t$), the Nyquist frequency ($Nyq.$), and the frequency resolution ($\delta f$) according to the Rayleigh criterion ($1/\Delta T$). Representative Fourier fittings on the data points of $B$ filter (because the maximum amplitude of the pulsations is met in this pass band) and $TESS$ (due to their high photometric accuracy) are given for all systems in Fig.~\ref{fig:FF}. The respective periodograms are illustrated in Fig.~\ref{fig:FS}, where the strongest detected frequencies are indicated. The independent pulsation frequencies are listed in Table~\ref{tab:PULMDLind} as follows: the increasing number ($n$), the value ($f_n$), the amplitude ($A$), and the phase ($\Phi$) of each detected frequency that is characterized independent (i.e. it is not a combination of other frequencies). The complete lists of detected frequencies, including also the S/N and the most possible combination of each one, are given in Appendix~\ref{sec:AppFreqs} for each studied system.


The results from the spectral classification (Section~\ref{sec:SPECCLASS}) as well as the frequency regimes where the majority of the frequencies are detected (i.e. 0-4 and 10-40~cycle~d$^{-1}$; see Tables~\ref{tab:PULMDLind}, \ref{tab:PULMDLpup}, \ref{tab:PULMDLsco}, and \ref{tab:PULMDLvel}) suggest clearly that the primaries of the studied systems are pulsating stars of $\delta$~Scuti type. The pulsations constant, $Q$, of the independent frequencies are calculated using the formula of \citet{BRE00}:
\begin{equation}
\log Q = -\log f + 0.5 \log g + 0.1M_{\rm bol} + \log T_{\rm eff}- 6.456,\\
\label{eq:Q}
\end{equation}
where $f$ is the frequency of the pulsation mode, while $\log g$, $M_{\rm bol}$, and $T_{\rm eff}$ denote the standard quantities (see Table~\ref{tab:LCRV}-bottom part). The density of the pulsating star, $\rho_{\rm pul}$, is calculated with the pulsation constant - density relation:
\begin{equation}
\rho_{\rm pul}/\rho_{\sun}=(Q~f_{\rm dom})^2,
\label{eq:rho}
\end{equation}
where $f_{\rm dom}$ is the frequency of the dominant pulsation mode. Alternatively, the $\rho_{\rm pul}$ can be also computed using the physical parameters of the stars directly (Table~\ref{tab:LCRV}), i.e. $\rho_{\rm pul}=3M_{\rm pul}/4 \pi R_{\rm pul}^3$. For the oscillation mode identification, the theoretical MAD models for $\delta$~Scuti stars \citep{MON07} in the \textsc{FAMIAS} software v.1.01 \citep{ZIM08} were used. Given that for all systems multicolour photometry is available, FAMIAS uses the amplitude and the phase ratios in different filters for a given frequency to estimate the most likely $l$-degrees based on theoretical models grids. However, it should to be noted that, these models are based on the absolute parameters ($\log g$, $M$, and $T_{\rm eff}$) of the pulsators and they have been calculated for single $\delta$~Scuti stars. The pulsating stars of the studied systems have increased mass values due to the mass transfer \citep[i.e. their $T_{\rm eff}$ is low for their mass in comparison with main-sequence stars; see][]{COX00}, hence, the results for the $l$~degrees can be considered only as approximate.

The following subsections include comments on the analysis and results for each system, while a discussion regarding the real pulsation frequencies is also made. The latter discussion for each system is based on the comparison between the detected frequencies in the different pass bands of observations. The indices of the compared/discussed frequencies have additionally to their increasing number a filter designation in order to be more easy to be checked by the reader in Tables~\ref{tab:PULMDLind}, \ref{tab:PULMDLpup}-\ref{tab:PULMDLvel}. It is quite common in EBs with pulsating components that the imperfect fit of the LCs produces artificial slow frequencies that are connected to instrumental/observational drifts. The time gaps between the ground-based observations may also produce alias frequencies \citep{BRE00}. Furthermore, searching for pulsations in light curves with eclipses removed also creates aliases at the orbital frequency and multiples thereof. Therefore, distinguishing the real independent frequencies from the possible artefacts is done using the data obtained in bluer pass bands (i.e. $B$ and $V$) as indicators for the low-amplitude frequencies (i.e. the pulsations show larger amplitudes in these bands) and the $TESS$ data as indicator for the slow frequencies (i.e. no time gaps and higher photometric accuracy). All periodograms were checked for potential sidelobes due to the fast rotation of the pulsating stars (i.e. due to the tidal locking), but none of them showed such evidence. Finally, Table~\ref{tab:PULIND} includes the $Q$ value and the most possible oscillation mode ($l$-degrees) for the independent frequencies identified as real as well as the $\rho_{\rm pul}$ of each pulsating star.

\begin{table}
\centering
\caption{Probable independent pulsation frequencies for each observed band.}
\label{tab:PULMDLind}
\scalebox{0.99}{
\begin{tabular}{cc cc c}
		\hline							
System	&	$n$	&	  $f_{\rm n}$	&	$A$	&	  $\Phi$	\\
	&		&	     (cycle d$^{-1}$)	&	(mmag)	&	($\degr$)	\\
		\hline							
\multirow{15}{*}{HM Pup}	&	\multicolumn{4}{c}{$B$-filter}							\\
		\cline{2-5}							
	&	1	&	0.038(1)	&	5.9(1)	&	3(1)	\\
	&	2	&	38.563(1)	&	5.5(1)	&	9(1)	\\
	&	3	&	36.424(1)	&	3.8(1)	&	354(1)	\\
		\cline{2-5}							
	&	\multicolumn{4}{c}{$V$-filter}							\\
		\cline{2-5}							
	&	2	&	38.563(1)	&	4.9(1)	&	14(1)	\\
	&	3	&	36.417(1)	&	3.7(1)	&	49(2)	\\
	&	5	&	0.014(1)	&	2.8(1)	&	354(3)	\\
	&	8	&	32.806(1)	&	1.4(1)	&	88(4)	\\
		\cline{2-5}							
	&	\multicolumn{4}{c}{$I$-filter}							\\
		\cline{2-5}							
	&	1	&	38.562(2)	&	2.7(1)	&	15(3)	\\
	&	2	&	36.416(2)	&	2.4(1)	&	86(3)	\\
		\cline{2-5}							
	&	\multicolumn{4}{c}{$TESS$}							\\
		\cline{2-5}							
	&	1	&	38.5622(1)	&	2.61(3)	&	7(1)	\\
	&	2	&	36.4196(2)	&	2.10(3)	&	14(1)	\\
	&	4	&	0.0317(2)	&	1.92(3)	&	77(1)	\\
\hline		
\multirow{11}{*}{V632~Sco}	&	\multicolumn{4}{c}{$B$-filter}							\\
		\cline{2-5}							
	&	1	&	0.819(1)	&	3.6(2)	&	311(4)	\\
	&	2	&	23.539(1)	&	3.0(2)	&	141(4)	\\
	&	3	&	3.548(1)	&	2.3(2)	&	280(6)	\\
	&	4	&	16.780(1)	&	2.2(2)	&	216(6)	\\
	&	6	&	0.216(1)	&	2.2(2)	&	21(6)	\\
		\cline{2-5}							
	&	\multicolumn{4}{c}{$V$-filter}							\\
		\cline{2-5}							
	&	1	&	23.432(13)	&	2.9(5)	&	81(9)	\\
		\cline{2-5}							
&	\multicolumn{4}{c}{$TESS$}							\\
	\cline{2-5}							
&	1	&	23.7423(5)	&	1.79(4)	&	134(1)	\\
&	4	&	0.1746(7)	&	1.22(4)	&	230(2)	\\
&	5	&	16.6507(7)	&	1.14(4)	&	205(2)	\\
\hline
\multirow{18}{*}{TT~Vel}	&	\multicolumn{4}{c}{$B$-filter}							\\
		\cline{2-5}							
	&	1	&	3.0101(2)	&	17.6(4)	&	96(1)	\\
	&	4	&	15.3267(7)	&	5.0(4)	&	171(5)	\\
	&	5	&	8.9257(7)	&	5.2(4)	&	137(4)	\\
	&	7	&	16.5918(10)	&	3.6(4)	&	151(6)	\\
		\cline{2-5}							
	&	\multicolumn{4}{c}{$V$-filter}							\\
		\cline{2-5}							
	&	2	&	3.5354(5)	&	10.9(5)	&	9(3)	\\
	&	3	&	0.7564(7)	&	7.1(5)	&	177(4)	\\
	&	5	&	15.7337(14)	&	3.7(5)	&	71(7)	\\
		\cline{2-5}							
	&	\multicolumn{4}{c}{$I$-filter}							\\
		\cline{2-5}							
	&	1	&	4.0502(6)	&	8.7(5)	&	125(3)	\\
	&	2	&	0.6947(8)	&	6.6(5)	&	40(4)	\\
	&	3	&	3.1096(9)	&	5.7(5)	&	134(5)	\\
		\cline{2-5}							
	&	\multicolumn{4}{c}{$TESS$}							\\
		\cline{2-5}							
	&	1	&	15.5020(4)	&	2.64(5)	&	36(1)	\\
	&	3	&	0.0288(8)	&	1.36(5)	&	198(2)	\\
		\hline							
\end{tabular}}
\tablefoot{The temporal zeropoints for the computation of the phases for each system and per pass-band were: HM~Pup: JD2459180.0 ($B$, $V$, $I$) and JD2459193.0 ($TESS$), V632~Sco: JD2459025.0 ($B$, $V$) and JD2459361.0 ($TESS$), and TT~Vel: JD2459325.0 ($B$, $V$, $I$) and JD2459281.0 ($TESS$).}
\end{table}	

\subsection{HM~Pup}
\label{sec:PULPUP}
For this system, $B$, $V$, $I$, and $TESS$ data were used for the pulsations search. As listed in Table~\ref{tab:PULMDLPAR}, many data points in $B$, $V$, and $I$ filters were available; they have a very good time resolution ($<2$~min) and cover a time span of more than 40~days. On the other hand, the available $TESS$ data have a 10~min time resolution but they cover $\sim78$ consecutive days. Moreover, and clearly by chance, the majority of the ground-based observations in $B$ and $V$ filters and all the data of $I$ filter for this system were acquired during the $TESS$ observations.

The independent frequencies $38.56$~cycle~d$^{-1}$ and $36.42$~cycle~d$^{-1}$, as can be seen in Table~\ref{tab:PULMDLind}, are detected in all data sets, therefore, they are valid. The $f_{8, V}$ was identified as independent only in $V$ filter, while its value is very close to that of $f_{13, B}$, $f_{21, B}$ and $f_{22, T}$ (see Table~\ref{tab:PULMDLpup}), thus, probably, it is not an independent frequency. The frequency $f_{5, V}$ was not detected in any of the other filters, therefore, it is probably an artefact. Finally, $f_{1, B}$ and $f_{4, T}$ concern a slow frequency of the order of 0.03~cycle~d$^{-1}$. Although its absence in $V$ and $I$ filters is spurious, especially since it has the highest amplitude in $B$, we cannot neglect its presence in $TESS$ data. Therefore, this frequency can be characterized as suspected-independent frequency. According to the MAD models, the most probable $l$-degrees for the $f_{2, B}$ ($=f_{1, V}$=$f_{1, I}$=$f_{1, T}$) and $f_{3, B}$ ($=f_{2, V}$=$f_{2, I}$=$f_{2, T}$) are $l=2$ and $l=1$, respectively. The complete list of detected frequencies is given in Table~\ref{tab:PULMDLpup}.

\subsection{V632 Sco}
\label{sec:PULSCO}

For the pulsation analysis of V632~Sco we used the short-cadence data set of $TESS$ (see Table~\ref{tab:TESSlog}) and only the data set obtained from Congarinni observatory (see Table~\ref{tab:GBPhotlog}), because it was obtained over a much shorter time period and has higher time resolution in comparison with that obtained from Glen Aplin observatory. Eight nights of observations in $B$ filter within a time range of $\sim57$~d were available, only two in $V$ filter and none in $I$ pass band. The $TESS$ data cover $\sim28$~d of continuous monitoring, therefore, the analysis relies mostly on them.

The frequency $\sim23$~cycle~d$^{-1}$ was detected in all data sets, therefore, it is considered as the dominant independent frequency. The second frequency detected both in $B$ and $TESS$ data is the $f_{4, B}$ ($=f_{5, T}$), thus, we conclude that this is also an independent frequency. The frequency $f_{1, B}$, which is found as the strongest frequency, is not detected in the $TESS$ data, therefore its reliability is strongly questioned and it may be attributed to instrumental/observational drifts or imperfect LC fitting. The frequency $f_{3, B}$ is close to the value of 6$f_{\rm orb}$ (=3.726~cycle~d$^{-1}$) but it can also be the alias frequency of $\sim2.548$~cycle~d$^{-1}$ (i.e. $f_{3, B}$-1), which is found as $f_{2, T}$ ($=4f_{\rm orb}$). So, this frequency can be considered only as a suspected independent frequency. Similar to that found for HM~Pup, a slow frequency is detected in $B$ and $TESS$ data (i.e. $f_{6, B}$ and $f_{4, T}$), and it is of the order of 0.2~cycle~d$^{-1}$. Although $f_{6, B}$ and $f_{4, T}$ have relatively similar values, it cannot be concluded that this frequency is indeed independent. In conclusion, $f_{2, B}$ ($=f_{1, V}=f_{1, T}$) and $f_{4, B}$ ($=f_{5, T}$) are certainly the dominant and independent frequencies of the pulsating star of V632~Sco. It should be noted that the ground-based and the $TESS$ observations have a time difference of approximately 11~months. Thus, for the $l$-degrees calculation of the first independent frequency, i.e. $f_{2, B}$, the amplitude and phase ratios of the $B$ and $V$ filter sets were used. Moreover, $f_{4, B}$ was not detected in the $V$-filter data set. Therefore, for the $l$-degrees of the second independent frequency, we assumed that a) no modulation in the amplitude and phase of this frequency was made within these 11 months and b) the amplitude of a frequency in $TESS$ pass-band is similar to that of the $I$~filter. The latter assumption can be supported by the similarity of the wavelength coverage of $TESS$ \citep[i.e. 600-1000~nm, with a peak at $\sim$880~nm;][]{RIC15} and that of the $I$~filter \citep[i.e. 720-850~nm, with a peak at $\sim$790~nm;][]{RIC15} and can be checked also in the results of HM~Pup (Table~\ref{tab:PULMDLpup2}) for the $I$~filter and $TESS$ data sets. Finally, based on MAD models and the aforementioned assumptions, the $l$-degrees are estimated as 3 or 1 for the $f_{2, B}$ and 2 or 3 for $f_{4, B}$. In total, 11 frequencies were detected in $B$ filter and 12 in the $TESS$ data (Table~\ref{tab:PULMDLsco}).

\subsection{TT Vel}
\label{sec:PULVEL}

In the search for frequencies of pulsations in TT~Vel, we relied mostly on only good ground-based $B$ data set and the mid-cadence data set of $TESS$. Some observations through the $V$ and $I$ filters were obtained, but were spread over a 50 day time-period (see Table~\ref{tab:PULMDLind}). However, for reasons of clarity, a frequency search was also made on these data sets.

Four possible independent frequencies were detected in $B$ filter (Table~\ref{tab:PULIND}). The frequency $f_{4, B}$ is the same with $f_{5, V}$ and $f_{1, T}$ and can be plausibly considered as the dominant frequency in the $\delta$~Scuti regime. $f_{5, B}$ and $f_{7, B}$ are again in the frequency range of $\delta$~Scuti stars, but they are not found in all filters. The frequencies $f_{1, B}$, $f_{1, T}$, and $f_{3, T}$ are slow frequencies and cannot be verified as real ones. The remaining frequencies in $V$ filter and all frequencies in $I$ filter are not very reliable because of the poor data sets used.

Eight and 19 frequencies were detected in the $B$ and TESS data sets, respectively (Table~\ref{tab:PULMDLvel}). However, only the $f\sim15$~cycle~d$^{-1}$ is definitely an independent frequency. It should be noted that since the ground-based and the $TESS$ observations were made within a time range of approximately 100~days, we assumed that the phase and the amplitude of this frequency remained constant, and the result of the MAD models is $l=3$.


\vspace{0.5cm}	

The independent frequencies of the pulsating primary component of each system are shown in Table~\ref{tab:PULIND}, where $A_{\rm B}$ is the amplitude in the $B$ band, $Q$ is the pulsation constant and $\rho$ the density of the pulsator. Based on these $Q$ values and the models of \citet[][$M=2~M_{\sun}$ for HM~Pup and V632~Sco and $M=1.5~M_{\sun}$ for TT~Vel]{FIT81}, these frequencies are identified as non-radial pressure oscillation modes. Moreover, all of them, except the $f=38.563$~cycle~d$^{-1}$ of HM~Pup, have typical $Q$ values for $\delta$~Scuti stars \citep[0.015$<Q<$0.035;][]{BRE00}.


\begin{table}
\centering
\caption{Independent pulsation frequencies and MAD model results for the $\delta$~Scuti stars in each system.}
\label{tab:PULIND}
\scalebox{0.97}{
\begin{tabular}{cc cc c}
\hline									
$f$	&	$A_{\rm B}$	&	Q	&	$l$-deg.	&	$\rho$	\\
(d$^{-1}$)	&	(mmag)	&	(d)	&		&	($\rho_{\sun}$)	\\
\hline									
\multicolumn{5}{c}{HM~Pup}									\\
\hline									
38.563(1)	&	5.5(1)	&	0.012(1)	&	2	&	\multirow{2}{*}{0.21(3)$^a$/0.19(1)$^b$}	\\
36.424(1)	&	3.8(1)	&	0.017(1)	&	1	&		\\
\hline									
\multicolumn{5}{c}{V632~Sco}									\\
\hline									
23.539(1)	&	3.0(2)	&	0.018(2)	&	3 or 1	&	\multirow{2}{*}{0.18(3)$^a$/0.18(1)$^b$}	\\
16.780(1)	&	2.2(2)	&	0.025(3)	&	2 or 3	&		\\
\hline									
\multicolumn{5}{c}{TT~Vel}									\\
\hline									
15.327(1)	&	5.0(4)	&	0.030(3)	&	3	&	0.21(3)$^a$/0.21(1)$^b$	\\
\hline									
\end{tabular}}
\tablefoot{$^a$Based on Eq.~\ref{eq:rho}, $^b$based on the physical properties of stars (Table~\ref{tab:LCRV}).}
\end{table}


\section{Evolution and comparison with similar systems}
\label{sec:EVOL}

\begin{figure}
\includegraphics[width=8.5cm]{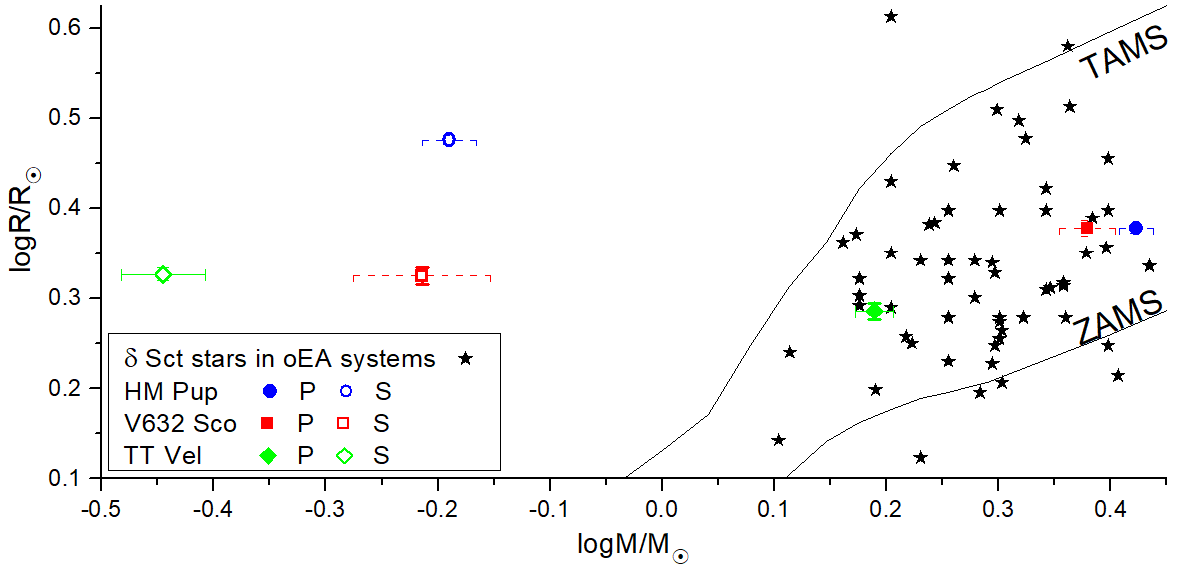}
\caption{Location of the primary (filled symbols) and secondary (empty symbols) components of HM~Pup (points), V632~Sco (squares), and TT~Vel (diamonds) within the mass-radius diagram. The stars symbols denote the  $\delta$~Scuti components of other oEA systems \citep[taken from][]{LIAN17}, while the black solid lines represent the boundaries of main-sequence.}
\label{fig:MR}
\includegraphics[width=8.5cm]{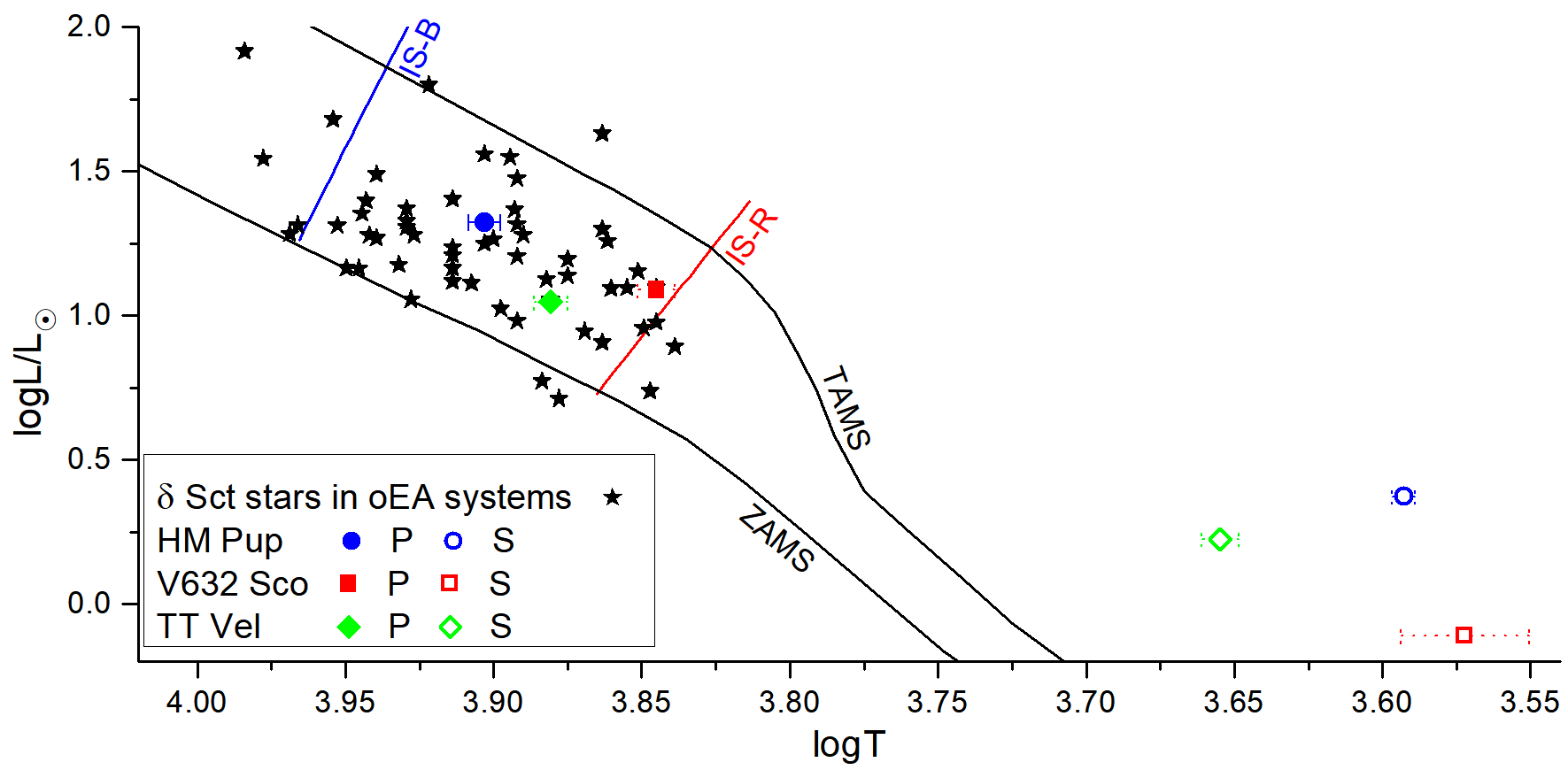}
\caption{Location of the components of the studied systems within the HR diagram. Symbols and black solid lines have the same meaning as in Fig.~\ref{fig:MR}. The coloured solid lines (B=Blue, R=Red) denote the boundaries of the instability strip \citep[IS; taken from][]{MUR19}.}
\label{fig:HR}
\end{figure}

\begin{figure}
\includegraphics[width=8.5cm]{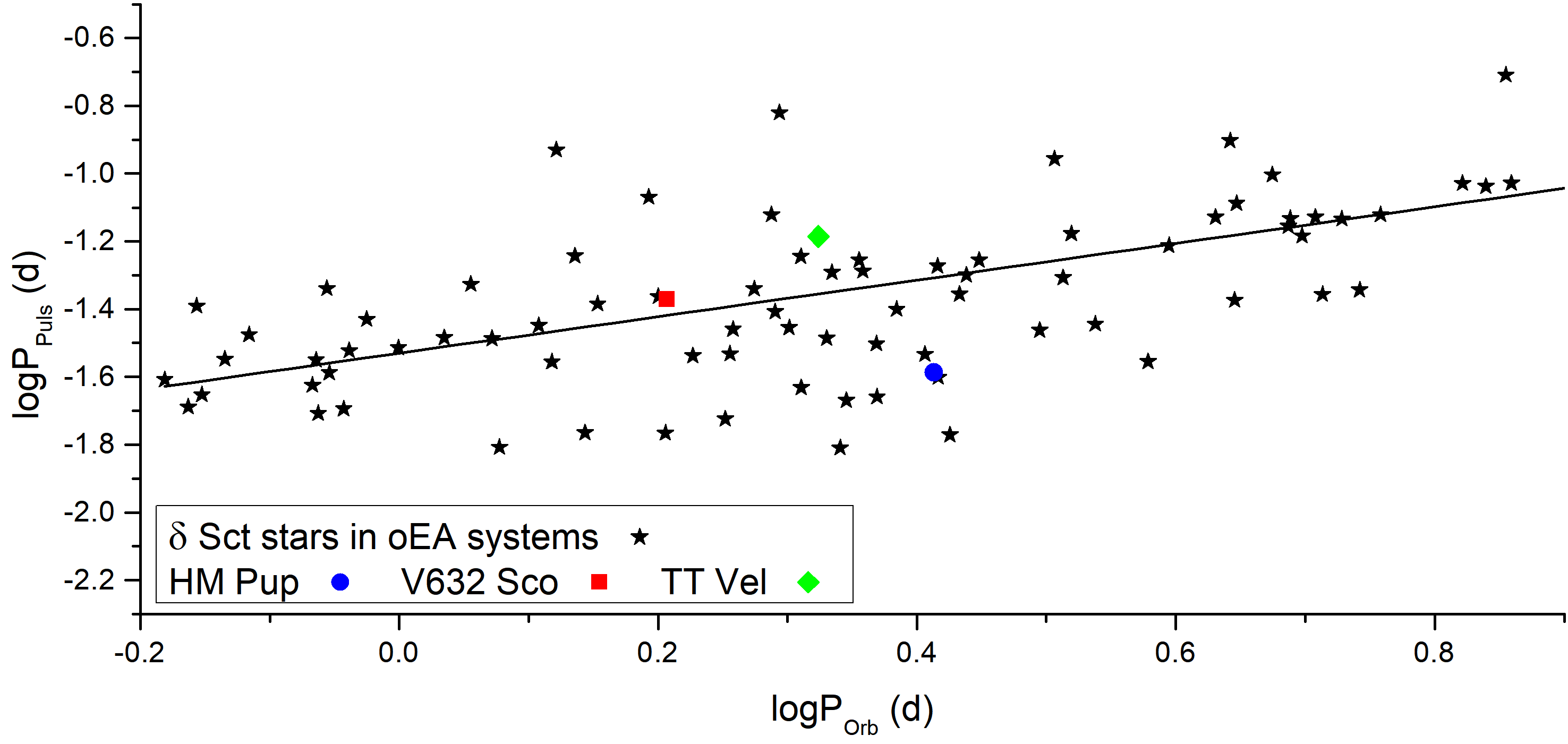}
\caption{Locations of the pulsating components of each studied system and other $\delta$~Scuti stars-members of oEA stars within the $P_{\rm orb}$-$P_{\rm pul}$ diagram. Symbols have the same meaning as in Fig.~\ref{fig:MR}, while the solid line denotes the empirical fit of \citet{LIAN17}.}
\label{fig:PP}
\includegraphics[width=8.5cm]{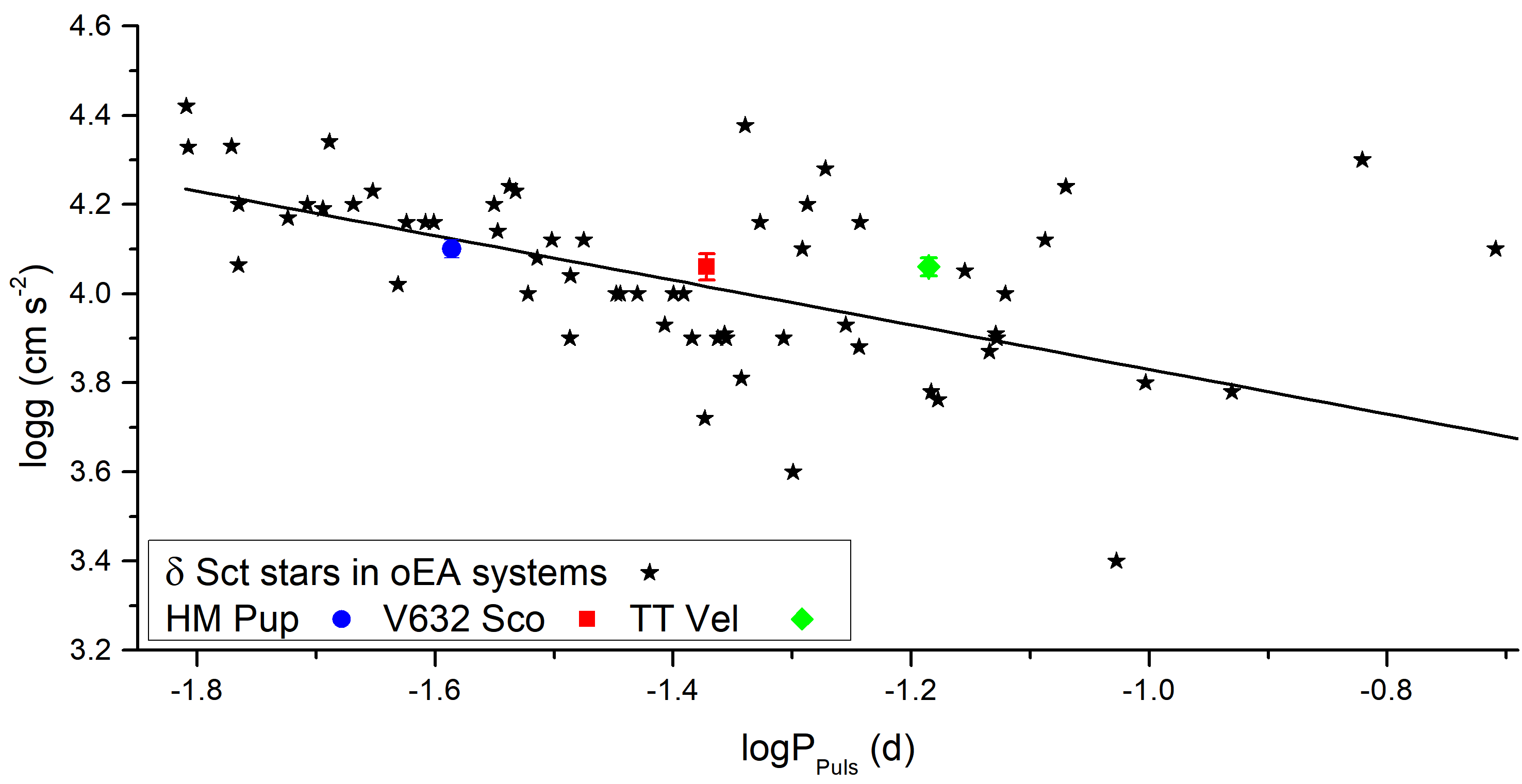}
\caption{Locations of the pulsating components of each studied system and other $\delta$~Scuti stars-members of oEA stars within the $\log g$-$P_{\rm pul}$ diagram. Symbols and line have the same meaning as in Fig.~\ref{fig:PP}.}
\label{fig:gP}
\end{figure}

The positions of the components of  HM~Pup, V632~Sco and TT~Vel on the mass-radius ($M-R$) and Hertzsprung-Russell (HR) evolutionary diagrams are shown in Figs~\ref{fig:MR} and \ref{fig:HR}, respectively. The locations of their primaries relative to pulsating components of other oEA stars in the relationship between pulsation and orbital periods ($P_{\rm orb}$-$P_{\rm pul}$) are shown in Fig.~\ref{fig:PP}. The locations of the same components in the relationship between the pulsation periods and stellar gravity $\log g$-$P_{\rm pul}$) are shown in Fig.~\ref{fig:gP} \citep[c.f.][]{LIAN17, LIA17}. The data points for these diagrams were collected from \citet[][Table~1 therein]{LIAN17} and the online catalogue\footnote{\url{http://alexiosliakos.weebly.com/catalogue.html}} of these systems (detailed lists will be published in a future study).

The primary components of each system lie in the centre of the main-sequence within the instability strip (Fig.~\ref{fig:HR}). The evolution of Algol binary systems, in which mass ratio reversal occurs, is complex and is not an aim for discussion in this paper. The primary component of V632~Sco is close to the red edge of the instability strip and the border with the $\gamma$~Doradus stars \citep{UYT11}. This evolutionary position, together with the frequency analysis indicating a probable independent frequency in the 0-4~cycle~d$^{-1}$ range (Section~\ref{sec:PULSCO}), suggests that this star could have a $\delta$~Scuti-$\gamma$~Doradus hybrid nature.

There is a relationship between the frequency of pulsations and orbital period in close binary systems \citep[c.f.][]{LIAN17}. The primary components of V632~Sco and TT~Vel fit the trend calculated by \citet{LIAN17} quite well, whereas that of HM~Pup deviates a little from it (Fig.~\ref{fig:PP}). Furthermore, a relationship has been found between the pulsation period and the evolutionary stage ($\log g$) of the $\delta$~Scuti-type components of close binary stars \citep{LIAN17}, although it is based on only a small sample of systems where radial velocities have been determined. The primaries of HM~Pup and V632~Sco are very close to the empirical fit, while that of TT~Vel shows larger deviation.



The secondary components of each system are well above terminal age on the Main Sequence (Figs.~\ref{fig:MR} and \ref{fig:HR}). Their derived $\log g$ values are 3.3-3.57~cm~s$^{-2}$ (Table~\ref{tab:LCRV}), which indicate all of them are in the subgiant/giant evolutionary phase of evolution. As the primary eclipse of HM~Pup is total, we were able to observe and classify the spectrum of its secondary component and confirm that it is a subgiant or giant, although smaller in size than typical single giant K8 stars.


\section{Summary and conclusions}
\label{sec:DIS}


In this study we present a holistic and detailed analysis of three southern EBs with a pulsating component. Multi-filter ground-based photometric observations and $TESS$ data were used. Moreover, spectroscopic observations were obtained allowing for the determination of the spectral types of the primaries of all systems and the secondary of HM~Pup as well as the RVs of both components. The combination of these data provided the means for accurate photometric modelling followed by the determination of the physical parameters and the evolutionary stages of the components of the systems. Analysis of eclipse timings collected from the literature, and our observations, offered the opportunity to determine possible orbital period modulation mechanisms and quantify the mass transfer/loss rates, which were found in the spectra. The residuals of the LCs were analysed to detect pulsation frequencies. The frequencies detected in different wavelength bands were compared with each other in order to determine the oscillation modes of the pulsators. The primary components of the systems were identified as $\delta$~Scuti type stars based on their physical (i.e. $T_{\rm eff}$, $M$, and $R$) and pulsation properties (frequency ranges). All systems were found to be in a semi-detached configuration with the primaries being main-sequence stars and with the secondaries (cooler and less massive components) filling their Roche lobes and being in evolved evolutionary stages. Therefore, these systems are classical Algols (in terms of evolution) and are oEA stars according to the definition of \citet{MKR02}. The current status of these systems confirm the conclusion of \citet{LIAN17} regarding the earlier initiation of the pulsations due to the presence of a close companion.


The primary component of HM~Pup pulsates in two independent frequencies at 36.4 and 38.6~cycle~d$^{-1}$, which are identified as $p$-mode oscillations. In total, 22 and 26 frequencies were detected in $B$ filter and $TESS$ data, respectively, but all except the two above are combinations of other frequencies.The HM~Pup orbital period has one cyclic variation which cannot be fully explained by the possible presence of a tertiary companion. No third light trace was detected in the analyses of light curves, Na~I~D spectra and broadening function values. On the other hand, our results show there is evidence for gas streaming from the secondary component and impacting on the primary component. The secondary component was found to be magnetically active and at a subgiant-giant phase of evolution.

The pulsating component of V632~Sco oscillates in two independent frequencies at 23.54 and 16.78~cycle~d$^{-1}$, which are independent non-radial pressure mode oscillations. More frequencies were detected (11 and 12 in the $B$ filter and $TESS$ data, respectively), but they are combinations of others. Another one was found in the slow-frequency regime, but its value was not exactly the same in all data sets, therefore, it might be an artefact. On the other hand, the location of the primary-pulsating component on the borders of $\delta$~Scuti-$\gamma$~Doradus stars leaves open the possibility of an hybrid nature. The secular changes in the orbital period of the system indicate a mass transfer from the secondary component at the rate of $2.1\times 10^{-8}~M_{\sun}$~yr$^{-1}$. As is the case for all three systems, mass flow from the secondary component is seen in the sodium doublet spectra. Furthermore, there is no evidence of a tertiary component in the sodium spectra, nor in the broadening function analyses. The secondary component was found to have an active spot migrating with a time period of a few days.




For the $\delta$~Scuti star of TT~Vel at least seven frequencies were detected in $B$, $V$, and $I$ filters and more than 15 in the $TESS$ data. However, only one ($\sim15.3$~cycle~d$^{-1}$) was validated as a real independent frequency and is identified as a $p$-mode oscillation. The other potential independent frequencies were either not detected in all data sets or they are combinations of other frequencies. The secular changes of the orbital period can be explained with a mass transfer rate of 4.9$\times 10^{-8}$~$M_{\sun}$~yr$^{-1}$. However, the current evolutionary status of the components gives rise to the possibility of mass loss from the system. The mass loss was calculated as $-3\times 10^{-7}~M_{\sun}$~yr$^{-1}$, and probably both mechanisms occur with slower rates. The periodic changes of the orbital period might be due to the presence of a faint tertiary component. 

More studies of similar systems will enrich the sample of oEA stars with accurately determined absolute properties, which in turn, will improve knowledge of how interactions of components in close binary systems influence pulsations.


\begin{acknowledgements}
AL acknowledges financial support by the European Space Agency (ESA) under the Consolidating Activities Regarding Moon, Earth and NEOs (CARMEN) project, NOA/SRFA no. 1084 and wishes to thank Dr~N.~Nanouris for his valuable comments on the ETV analysis. DJWM, MGB and JFW acknowledge grants for time on the ANU 2.3~m telescope from the Edward Corbould Research Fund of the Astronomical Association of Queensland. DJWM is grateful to Prof. Michael Drinkwater for his extensive assistance with the planning  and data analysis of the ANU 2.3~m telescope observations. DJWM thanks A. Mohit, Y. Rist and L. Coetzee for their assistance with the spectroscopic observations. We thank the anonymous reviewer for the valuable comments that improved the quality of this work. This research has made use of NASA's Astrophysics Data System Bibliographic Services, the SIMBAD database, operated at CDS, Strasbourg, France, the TIDAK database (B. Zakrzewski), the `O-C' gateway database, and the Mikulski Archive for Space Telescopes (MAST).
\end{acknowledgements}

%
%


\bibliography{references} 

\begin{thebibliography}{87}
\expandafter\ifx\csname natexlab\endcsname\relax\def\natexlab#1{#1}\fi

\bibitem[{{Aerts} {et~al.}(2010){Aerts}, {Christensen-Dalsgaard}, \&
  {Kurtz}}]{AER10}
{Aerts}, C., {Christensen-Dalsgaard}, J., \& {Kurtz}, D.~W. 2010,
  {Asteroseismology} (Springer Netherlands)

\bibitem[{{Antoci} {et~al.}(2014){Antoci}, {Cunha}, {Houdek}, {Kjeldsen},
  {Trampedach}, {Handler}, {L{\"u}ftinger}, {Arentoft}, \& {Murphy}}]{ANT14}
{Antoci}, V., {Cunha}, M., {Houdek}, G., {et~al.} 2014, \apj, 796, 118

\bibitem[{{Applegate}(1992)}]{APP92}
{Applegate}, J.~H. 1992, \apj, 385, 621

\bibitem[{{Applegate} \& {Patterson}(1987)}]{APP87}
{Applegate}, J.~H. \& {Patterson}, J. 1987, \apjl, 322, L99

\bibitem[{{Balona}(2014)}]{BAL14}
{Balona}, L.~A. 2014, \mnras, 439, 3453

\bibitem[{{Balona} {et~al.}(2015){Balona}, {Daszy{\'n}ska-Daszkiewicz}, \&
  {Pamyatnykh}}]{BAL15}
{Balona}, L.~A., {Daszy{\'n}ska-Daszkiewicz}, J., \& {Pamyatnykh}, A.~A. 2015,
  \mnras, 452, 3073

\bibitem[{{Baran} \& {Koen}(2021)}]{BAR21}
{Baran}, A.~S. \& {Koen}, C. 2021, \actaa, 71, 113

\bibitem[{{Baran} {et~al.}(2015){Baran}, {Koen}, \& {Pokrzywka}}]{BAR15}
{Baran}, A.~S., {Koen}, C., \& {Pokrzywka}, B. 2015, \mnras, 448, L16

\bibitem[{{Berdyugina}(2005)}]{BER05}
{Berdyugina}, S.~V. 2005, Living Reviews in Solar Physics, 2, 8

\bibitem[{{Borucki} {et~al.}(2010){Borucki}, {Koch}, {Basri}, {Batalha},
  {Brown}, {Caldwell}, {Caldwell}, {Christensen-Dalsgaard}, {Cochran},
  {DeVore}, {Dunham}, {Dupree}, {Gautier}, {Geary}, {Gilliland}, {Gould},
  {Howell}, {Jenkins}, {Kondo}, {Latham}, {Marcy}, {Meibom}, {Kjeldsen},
  {Lissauer}, {Monet}, {Morrison}, {Sasselov}, {Tarter}, {Boss}, {Brownlee},
  {Owen}, {Buzasi}, {Charbonneau}, {Doyle}, {Fortney}, {Ford}, {Holman},
  {Seager}, {Steffen}, {Welsh}, {Rowe}, {Anderson}, {Buchhave}, {Ciardi},
  {Walkowicz}, {Sherry}, {Horch}, {Isaacson}, {Everett}, {Fischer}, {Torres},
  {Johnson}, {Endl}, {MacQueen}, {Bryson}, {Dotson}, {Haas}, {Kolodziejczak},
  {Van Cleve}, {Chandrasekaran}, {Twicken}, {Quintana}, {Clarke}, {Allen},
  {Li}, {Wu}, {Tenenbaum}, {Verner}, {Bruhweiler}, {Barnes}, \& {Prsa}}]{BOR10}
{Borucki}, W.~J., {Koch}, D., {Basri}, G., {et~al.} 2010, Science, 327, 977

\bibitem[{{Bowman} {et~al.}(2019){Bowman}, {Johnston}, {Tkachenko},
  {Mkrtichian}, {Gunsriwiwat}, \& {Aerts}}]{BOW19}
{Bowman}, D.~M., {Johnston}, C., {Tkachenko}, A., {et~al.} 2019, \apjl, 883,
  L26

\bibitem[{{Bowman} \& {Kurtz}(2018)}]{BOW18}
{Bowman}, D.~M. \& {Kurtz}, D.~W. 2018, \mnras, 476, 3169

\bibitem[{{Bowman} \& {Michielsen}(2021)}]{BOW21}
{Bowman}, D.~M. \& {Michielsen}, M. 2021, \aap, 656, A158

\bibitem[{{Breger}(2000)}]{BRE00}
{Breger}, M. 2000, in Astronomical Society of the Pacific Conference Series,
  Vol. 210, Delta Scuti and Related Stars, ed. M.~{Breger} \& M.~{Montgomery},
  3

\bibitem[{{Breger} {et~al.}(1993){Breger}, {Stich}, {Garrido}, {Martin},
  {Jiang}, {Li}, {Hube}, {Ostermann}, {Paparo}, \& {Scheck}}]{BRE93}
{Breger}, M., {Stich}, J., {Garrido}, R., {et~al.} 1993, \aap, 271, 482

\bibitem[{{Claret}(2018)}]{CLA18}
{Claret}, A. 2018, \aap, 618, A20

\bibitem[{{Coelho} {et~al.}(2005){Coelho}, {Barbuy}, {Mel{\'e}ndez},
  {Schiavon}, \& {Castilho}}]{COE05}
{Coelho}, P., {Barbuy}, B., {Mel{\'e}ndez}, J., {Schiavon}, R.~P., \&
  {Castilho}, B.~V. 2005, \aap, 443, 735

\bibitem[{{Collins} {et~al.}(2017){Collins}, {Kielkopf}, {Stassun}, \&
  {Hessman}}]{COL17}
{Collins}, K.~A., {Kielkopf}, J.~F., {Stassun}, K.~G., \& {Hessman}, F.~V.
  2017, \aj, 153, 77

\bibitem[{{Cox}(2000)}]{COX00}
{Cox}, A.~N. 2000, {Allen's astrophysical quantities} (New York:
  Springer-Verlag)

\bibitem[{{Dopita} {et~al.}(2007){Dopita}, {Hart}, {McGregor}, {Oates},
  {Bloxham}, \& {Jones}}]{DOP07}
{Dopita}, M., {Hart}, J., {McGregor}, P., {et~al.} 2007, \apss, 310, 255

\bibitem[{{Dopita} {et~al.}(2010){Dopita}, {Rhee}, {Farage}, {McGregor},
  {Bloxham}, {Green}, {Roberts}, {Neilson}, {Wilson}, {Young}, {Firth},
  {Busarello}, \& {Merluzzi}}]{DOP10}
{Dopita}, M., {Rhee}, J., {Farage}, C., {et~al.} 2010, \apss, 327, 245

\bibitem[{{Drake} {et~al.}(2009){Drake}, {Djorgovski}, {Mahabal}, {Beshore},
  {Larson}, {Graham}, {Williams}, {Christensen}, {Catelan}, {Boattini},
  {Gibbs}, {Hill}, \& {Kowalski}}]{DRA09}
{Drake}, A.~J., {Djorgovski}, S.~G., {Mahabal}, A., {et~al.} 2009, \apj, 696,
  870

\bibitem[{{Fitch}(1981)}]{FIT81}
{Fitch}, W.~S. 1981, \apj, 249, 218

\bibitem[{{Gray} \& {Corbally}(2014)}]{GRA14}
{Gray}, R.~O. \& {Corbally}, C.~J. 2014, \aj, 147, 80

\bibitem[{{Gray} \& {Corbally}(2009)}]{GRA09}
{Gray}, R.~O. \& {Corbally}, Christopher, J. 2009, {Stellar Spectral
  Classification} (Princeton University Press)

\bibitem[{{Handler}(1999)}]{HAN99}
{Handler}, G. 1999, \mnras, 309, L19

\bibitem[{{Henden} {et~al.}(2015){Henden}, {Levine}, {Terrell}, \&
  {Welch}}]{HEN15}
{Henden}, A.~A., {Levine}, S., {Terrell}, D., \& {Welch}, D.~L. 2015, in
  American Astronomical Society Meeting Abstracts, Vol. 225, American
  Astronomical Society Meeting Abstracts \#225, 336.16

\bibitem[{{Hilditch}(2001)}]{HIL01}
{Hilditch}, R.~W. 2001, {An Introduction to Close Binary Stars} (Cambridge
  University Press)

\bibitem[{{Hoffman} {et~al.}(2006){Hoffman}, {Harrison}, {McNamara},
  {Vestrand}, {Holtzman}, \& {Barker}}]{HOF06}
{Hoffman}, D.~I., {Harrison}, T.~E., {McNamara}, B.~J., {et~al.} 2006, \aj,
  132, 2260

\bibitem[{{Howell} {et~al.}(2014){Howell}, {Sobeck}, {Haas}, {Still},
  {Barclay}, {Mullally}, {Troeltzsch}, {Aigrain}, {Bryson}, {Caldwell},
  {Chaplin}, {Cochran}, {Huber}, {Marcy}, {Miglio}, {Najita}, {Smith},
  {Twicken}, \& {Fortney}}]{HOW14}
{Howell}, S.~B., {Sobeck}, C., {Haas}, M., {et~al.} 2014, \pasp, 126, 398

\bibitem[{{Irwin}(1959)}]{IRW59}
{Irwin}, J.~B. 1959, \aj, 64, 149

\bibitem[{{Kalimeris} {et~al.}(1994){Kalimeris}, {Rovithis-Livaniou}, \&
  {Rovithis}}]{KAL94}
{Kalimeris}, A., {Rovithis-Livaniou}, H., \& {Rovithis}, P. 1994, \aap, 282,
  775

\bibitem[{{Kim} {et~al.}(2021){Kim}, {Lee}, {Lee}, {Lee}, {Lee}, {Hong}, {Cha},
  {Kim}, \& {Park}}]{KIM21}
{Kim}, S.-L., {Lee}, J.~W., {Lee}, C.-U., {et~al.} 2021, \aj, 162, 212

\bibitem[{{Koch} {et~al.}(2010){Koch}, {Borucki}, {Basri}, {Batalha}, {Brown},
  {Caldwell}, {Christensen-Dalsgaard}, {Cochran}, {DeVore}, {Dunham},
  {Gautier}, {Geary}, {Gilliland}, {Gould}, {Jenkins}, {Kondo}, {Latham},
  {Lissauer}, {Marcy}, {Monet}, {Sasselov}, {Boss}, {Brownlee}, {Caldwell},
  {Dupree}, {Howell}, {Kjeldsen}, {Meibom}, {Morrison}, {Owen}, {Reitsema},
  {Tarter}, {Bryson}, {Dotson}, {Gazis}, {Haas}, {Kolodziejczak}, {Rowe}, {Van
  Cleve}, {Allen}, {Chand rasekaran}, {Clarke}, {Li}, {Quintana}, {Tenenbaum},
  {Twicken}, \& {Wu}}]{KOC10}
{Koch}, D.~G., {Borucki}, W.~J., {Basri}, G., {et~al.} 2010, \apjl, 713, L79

\bibitem[{{Kruszewski}(1966)}]{KRU66}
{Kruszewski}, A. 1966, Advances in Astronomy and Astrophysics, 4, 233

\bibitem[{{Kurtz} {et~al.}(2020){Kurtz}, {Handler}, {Rappaport}, {Saio},
  {Fuller}, {Jacobs}, {Schmitt}, {Jones}, {Vand erburg}, {LaCourse}, {Nelson},
  {Kahraman Ali{\c{c}}avu{\textcommabelow s}}, \& {Giarrusso}}]{KUR20}
{Kurtz}, D.~W., {Handler}, G., {Rappaport}, S.~A., {et~al.} 2020, \mnras
  [\eprint[arXiv]{2004.03471}]

\bibitem[{{Lanza}(2006)}]{LAN06}
{Lanza}, A.~F. 2006, \mnras, 369, 1773

\bibitem[{{Lanza}(2020)}]{LAN20}
{Lanza}, A.~F. 2020, \mnras, 491, 1820

\bibitem[{{Lanza} {et~al.}(1998){Lanza}, {Catalano}, {Cutispoto}, {Pagano}, \&
  {Rodono}}]{LAN98}
{Lanza}, A.~F., {Catalano}, S., {Cutispoto}, G., {Pagano}, I., \& {Rodono}, M.
  1998, \aap, 332, 541

\bibitem[{{Lanza} \& {Rodon{\`o}}(2002)}]{LAN02}
{Lanza}, A.~F. \& {Rodon{\`o}}, M. 2002, Astronomische Nachrichten, 323, 424

\bibitem[{{Lenz} \& {Breger}(2005)}]{LEN05}
{Lenz}, P. \& {Breger}, M. 2005, Communications in Asteroseismology, 146, 53

\bibitem[{{Liakos}(2015)}]{LIA15}
{Liakos}, A. 2015, in Astronomical Society of the Pacific Conference Series,
  Vol. 496, Living Together: Planets, Host Stars and Binaries, ed. S.~M.
  {Rucinski}, G.~{Torres}, \& M.~{Zejda}, 286

\bibitem[{{Liakos}(2017)}]{LIA17}
{Liakos}, A. 2017, \aap, 607, A85

\bibitem[{{Liakos}(2018)}]{LIA18}
{Liakos}, A. 2018, \aap, 616, A130

\bibitem[{{Liakos}(2020{\natexlab{a}})}]{LIA20a}
{Liakos}, A. 2020{\natexlab{a}}, \aap, 642, A91

\bibitem[{{Liakos}(2020{\natexlab{b}})}]{LIA20b}
{Liakos}, A. 2020{\natexlab{b}}, \actaa, 70, 265

\bibitem[{{Liakos} \& {Caga{\v s}}(2014)}]{LIA14}
{Liakos}, A. \& {Caga{\v s}}, P. 2014, \apss, 353, 559

\bibitem[{{Liakos} \& {Niarchos}(2013)}]{LIA13}
{Liakos}, A. \& {Niarchos}, P. 2013, \apss, 343, 123

\bibitem[{{Liakos} \& {Niarchos}(2017)}]{LIAN17}
{Liakos}, A. \& {Niarchos}, P. 2017, \mnras, 465, 1181

\bibitem[{{Liakos} \& {Niarchos}(2020)}]{LIAN20}
{Liakos}, A. \& {Niarchos}, P. 2020, Galaxies, 8, 75

\bibitem[{{Liakos} {et~al.}(2012){Liakos}, {Niarchos}, {Soydugan}, \&
  {Zasche}}]{LIA12}
{Liakos}, A., {Niarchos}, P., {Soydugan}, E., \& {Zasche}, P. 2012, \mnras,
  422, 1250

\bibitem[{{Liakos} {et~al.}(2011){Liakos}, {Zasche}, \& {Niarchos}}]{LIA11}
{Liakos}, A., {Zasche}, P., \& {Niarchos}, P. 2011, \na, 16, 530

\bibitem[{{Lucy}(1967)}]{LUC67}
{Lucy}, L.~B. 1967, \zap, 65, 89

\bibitem[{{Mkrtichian} {et~al.}(2018{\natexlab{a}}){Mkrtichian}, {Gunsriwiwat},
  {Reichart}, {Haislip}, {Kouprianov}, \& {Poshyachinda}}]{MKR18b}
{Mkrtichian}, D.~E., {Gunsriwiwat}, K., {Reichart}, D.~E., {et~al.}
  2018{\natexlab{a}}, Information Bulletin on Variable Stars, 6238, 1

\bibitem[{{Mkrtichian} {et~al.}(2002){Mkrtichian}, {Kusakin}, {Gamarova}, \&
  {Nazarenko}}]{MKR02}
{Mkrtichian}, D.~E., {Kusakin}, A.~V., {Gamarova}, A.~Y., \& {Nazarenko}, V.
  2002, in Astronomical Society of the Pacific Conference Series, Vol. 259, IAU
  Colloq. 185: Radial and Nonradial Pulsationsn as Probes of Stellar Physics,
  ed. C.~{Aerts}, T.~R. {Bedding}, \& J.~{Christensen-Dalsgaard}, 96

\bibitem[{{Mkrtichian} {et~al.}(2018{\natexlab{b}}){Mkrtichian}, {Lehmann},
  {Rodr{\'{\i}}guez}, {Olson}, {Kim}, {Kusakin}, {Lee}, {Youn}, {Kwon},
  {L{\'o}pez-Gonz{\'a}lez}, {Janiashvili}, {Tiwari}, {Joshi}, {Lampens}, {Van
  Cauteren}, {Glazunova}, {Gamarova}, {Grankin}, {Rovithis-Livaniou},
  {Svoboda}, {Uhlar}, {Tsymbal}, {Kokumbaeva}, {Urushadze}, {Kuratov}, {Shin},
  {Kang}, \& {Soonthornthum}}]{MKR18a}
{Mkrtichian}, D.~E., {Lehmann}, H., {Rodr{\'{\i}}guez}, E., {et~al.}
  2018{\natexlab{b}}, \mnras, 475, 4745

\bibitem[{{Montalb{\'a}n} \& {Dupret}(2007)}]{MON07}
{Montalb{\'a}n}, J. \& {Dupret}, M.~A. 2007, \aap, 470, 991

\bibitem[{{Moriarty} {et~al.}(2013){Moriarty}, {Bohlsen}, {Heathcote},
  {Richards}, \& {Streamer}}]{MOR13}
{Moriarty}, D.~J.~W., {Bohlsen}, T., {Heathcote}, B., {Richards}, T., \&
  {Streamer}, M. 2013, Journal of the American Association of Variable Star
  Observers (JAAVSO), 41, 182

\bibitem[{{Moriarty} {et~al.}(2019){Moriarty}, {Liakos}, {Drinkwater}, {Mohit},
  {Sweet}, \& {West}}]{MOR19}
{Moriarty}, D.~J.~W., {Liakos}, A., {Drinkwater}, M.~J., {et~al.} 2019, \apss,
  365, 1

\bibitem[{{Murphy} {et~al.}(2019){Murphy}, {Hey}, {Van Reeth}, \&
  {Bedding}}]{MUR19}
{Murphy}, S.~J., {Hey}, D., {Van Reeth}, T., \& {Bedding}, T.~R. 2019, \mnras,
  485, 2380

\bibitem[{{Murphy} {et~al.}(2013){Murphy}, {Pigulski}, {Kurtz}, {Su{\'a}rez},
  {Handler}, {Balona}, {Smalley}, {Uytterhoeven}, {Szab{\'o}}, {Thygesen},
  {Elkin}, {Breger}, {Grigahc{\`e}ne}, {Guzik}, {Nemec}, \&
  {Southworth}}]{MUR13}
{Murphy}, S.~J., {Pigulski}, A., {Kurtz}, D.~W., {et~al.} 2013, \mnras, 432,
  2284

\bibitem[{Nelson(2005)}]{NEL05}
Nelson, Robert, H. 2005, B-Minima

\bibitem[{{Ogloza} {et~al.}(2022){Ogloza}, {Stachowski}, {Zakrzewski}, \&
  {Zejmo}}]{OGL22}
{Ogloza}, W., {Stachowski}, G., {Zakrzewski}, B., \& {Zejmo}, M. 2022, in XL
  Polish Astronomical Society Meeting proc, Vol. 123, XL Polish Astronomical
  Society Meeting, in press

\bibitem[{{Pilecki} {et~al.}(2017){Pilecki}, {Gieren}, {Smolec},
  {Pietrzy{\'n}ski}, {Thompson}, {Anderson}, {Bono}, {Soszy{\'n}ski},
  {Kervella}, {Nardetto}, {Taormina}, {St{\c{e}}pie{\'n}}, \&
  {Wielg{\'o}rski}}]{PIL17}
{Pilecki}, B., {Gieren}, W., {Smolec}, R., {et~al.} 2017, \apj, 842, 110

\bibitem[{{Pojmanski}(2002)}]{POJ02}
{Pojmanski}, G. 2002, \actaa, 52, 397

\bibitem[{{Pr{\v s}a} \& {Zwitter}(2005)}]{PRS05}
{Pr{\v s}a}, A. \& {Zwitter}, T. 2005, \apj, 628, 426

\bibitem[{{Richards} \& {Albright}(1999)}]{RIC99}
{Richards}, M.~T. \& {Albright}, G.~E. 1999, \apjs, 123, 537

\bibitem[{{Ricker} {et~al.}(2009){Ricker}, {Latham}, {Vanderspek}, {Ennico},
  {Bakos}, {Brown}, {Burgasser}, {Charbonneau}, {Deming}, {Doty}, {Dunham},
  {Elliot}, {Holman}, {Ida}, {Jenkins}, {Jernigan}, {Kawai}, {Laughlin},
  {Lissauer}, {Martel}, {Sasselov}, {Schingler}, {Seager}, {Torres}, {Udry},
  {Villasenor}, {Winn}, \& {Worden}}]{RIC09}
{Ricker}, G.~R., {Latham}, D.~W., {Vanderspek}, R.~K., {et~al.} 2009, in
  American Astronomical Society Meeting Abstracts, Vol. 213, American
  Astronomical Society Meeting Abstracts \#213, 403.01

\bibitem[{{Ricker} {et~al.}(2015){Ricker}, {Winn}, {Vanderspek}, {Latham},
  {Bakos}, {Bean}, {Berta-Thompson}, {Brown}, {Buchhave}, {Butler}, {Butler},
  {Chaplin}, {Charbonneau}, {Christensen-Dalsgaard}, {Clampin}, {Deming},
  {Doty}, {De Lee}, {Dressing}, {Dunham}, {Endl}, {Fressin}, {Ge}, {Henning},
  {Holman}, {Howard}, {Ida}, {Jenkins}, {Jernigan}, {Johnson}, {Kaltenegger},
  {Kawai}, {Kjeldsen}, {Laughlin}, {Levine}, {Lin}, {Lissauer}, {MacQueen},
  {Marcy}, {McCullough}, {Morton}, {Narita}, {Paegert}, {Palle}, {Pepe},
  {Pepper}, {Quirrenbach}, {Rinehart}, {Sasselov}, {Sato}, {Seager},
  {Sozzetti}, {Stassun}, {Sullivan}, {Szentgyorgyi}, {Torres}, {Udry}, \&
  {Villasenor}}]{RIC15}
{Ricker}, G.~R., {Winn}, J.~N., {Vanderspek}, R., {et~al.} 2015, Journal of
  Astronomical Telescopes, Instruments, and Systems, 1, 014003

\bibitem[{{Rovithis-Livaniou} {et~al.}(2000){Rovithis-Livaniou}, {Kranidiotis},
  {Rovithis}, \& {Athanassiades}}]{ROV00}
{Rovithis-Livaniou}, H., {Kranidiotis}, A.~N., {Rovithis}, P., \&
  {Athanassiades}, G. 2000, \aap, 354, 904

\bibitem[{{Ruci{\'n}ski}(1969)}]{RUC69}
{Ruci{\'n}ski}, S.~M. 1969, \actaa, 19, 245

\bibitem[{{Ruci{\'n}ski}(1992)}]{RUC92}
{Ruci{\'n}ski}, S.~M. 1992, \aj, 104, 1968

\bibitem[{{Ruci{\'n}ski}(2002)}]{RUC02}
{Ruci{\'n}ski}, S.~M. 2002, \aj, 124, 1746

\bibitem[{{Shappee} {et~al.}(2014){Shappee}, {Prieto}, {Stanek}, {Kochanek},
  {Holoien}, {Jencson}, {Basu}, {Beacom}, {Szczygiel}, {Pojmanski},
  {Brimacombe}, {Dubberley}, {Elphick}, {Foale}, {Hawkins}, {Mullins},
  {Rosing}, {Ross}, \& {Walker}}]{SHA14}
{Shappee}, B., {Prieto}, J., {Stanek}, K.~Z., {et~al.} 2014, in American
  Astronomical Society Meeting Abstracts, Vol. 223, American Astronomical
  Society Meeting Abstracts \#223, 236.03

\bibitem[{{Soydugan} {et~al.}(2013){Soydugan}, {Soydugan}, {Kanvermez}, \&
  {Liakos}}]{SOY13}
{Soydugan}, F., {Soydugan}, E., {Kanvermez}, {\c C}., \& {Liakos}, A. 2013,
  \mnras, 432, 3278

\bibitem[{{Stassun} {et~al.}(2019){Stassun}, {Oelkers}, {Paegert}, {Torres},
  {Pepper}, {De Lee}, {Collins}, {Latham}, {Muirhead}, {Chittidi},
  {Rojas-Ayala}, {Fleming}, {Rose}, {Tenenbaum}, {Ting}, {Kane}, {Barclay},
  {Bean}, {Brassuer}, {Charbonneau}, {Ge}, {Lissauer}, {Mann}, {McLean},
  {Mullally}, {Narita}, {Plavchan}, {Ricker}, {Sasselov}, {Seager}, {Sharma},
  {Shiao}, {Sozzetti}, {Stello}, {Vanderspek}, {Wallace}, \& {Winn}}]{STA19}
{Stassun}, K.~G., {Oelkers}, R.~J., {Paegert}, M., {et~al.} 2019, \aj, 158, 138

\bibitem[{{Streamer} {et~al.}(2016){Streamer}, {Bohlsen}, \& {Ogmen}}]{STR16}
{Streamer}, M., {Bohlsen}, T., \& {Ogmen}, Y. 2016, Journal of the American
  Association of Variable Star Observers (JAAVSO), 44, 39

\bibitem[{{Ula{\c{s}}} {et~al.}(2020){Ula{\c{s}}}, {Gazeas}, {Liakos},
  {Ulusoy}, {Stateva}, {Erkan}, {Napetova}, \& {Iliev}}]{ULA20}
{Ula{\c{s}}}, B., {Gazeas}, K., {Liakos}, A., {et~al.} 2020, \actaa, 70, 219

\bibitem[{{Uytterhoeven} {et~al.}(2011){Uytterhoeven}, {Moya},
  {Grigahc{\`e}ne}, {Guzik}, {Guti{\'e}rrez-Soto}, {Smalley}, {Handler},
  {Balona}, {Niemczura}, {Fox Machado}, {Benatti}, {Chapellier}, {Tkachenko},
  {Szab{\'o}}, {Su{\'a}rez}, {Ripepi}, {Pascual}, {Mathias},
  {Mart{\'{\i}}n-Ru{\'{\i}}z}, {Lehmann}, {Jackiewicz}, {Hekker},
  {Gruberbauer}, {Garc{\'{\i}}a}, {Dumusque}, {D{\'{\i}}az-Fraile}, {Bradley},
  {Antoci}, {Roth}, {Leroy}, {Murphy}, {De Cat}, {Cuypers}, {Kjeldsen},
  {Christensen-Dalsgaard}, {Breger}, {Pigulski}, {Kiss}, {Still}, {Thompson},
  \& {van Cleve}}]{UYT11}
{Uytterhoeven}, K., {Moya}, A., {Grigahc{\`e}ne}, A., {et~al.} 2011, \aap, 534,
  A125

\bibitem[{{van Hamme}(1993)}]{HAM93}
{van Hamme}, W. 1993, \aj, 106, 2096

\bibitem[{{V{\"o}lschow} {et~al.}(2018){V{\"o}lschow}, {Schleicher},
  {Banerjee}, \& {Schmitt}}]{VOL18}
{V{\"o}lschow}, M., {Schleicher}, D.~R.~G., {Banerjee}, R., \& {Schmitt},
  J.~H.~M.~M. 2018, \aap, 620, A42

\bibitem[{{von Zeipel}(1924)}]{ZEI24}
{von Zeipel}, H. 1924, \mnras, 84, 665

\bibitem[{{Wilson}(1979)}]{WIL79}
{Wilson}, R.~E. 1979, \apj, 234, 1054

\bibitem[{{Wilson}(1990)}]{WIL90}
{Wilson}, R.~E. 1990, \apj, 356, 613

\bibitem[{{Wilson} \& {Devinney}(1971)}]{WIL71}
{Wilson}, R.~E. \& {Devinney}, E.~J. 1971, \apj, 166, 605

\bibitem[{{Zasche} {et~al.}(2009){Zasche}, {Liakos}, {Niarchos}, {Wolf},
  {Manimanis}, \& {Gazeas}}]{ZAS09}
{Zasche}, P., {Liakos}, A., {Niarchos}, P., {et~al.} 2009, \na, 14, 121

\bibitem[{{Zima}(2008)}]{ZIM08}
{Zima}, W. 2008, Communications in Asteroseismology, 157, 387

\end{thebibliography}

\begin{appendix}

\onecolumn
\section{Radial velocities}
\label{sec:AppRVs}

This appendix includes the heliocentric radial velocities measurements of the studied systems. In Table~\ref{tab:RVs} we list: the heliocentric Julian date of the observations and the radial velocities of the components of the systems (1=primary, 2=secondary). Details regarding the calculation method of the RVs are given in Section~\ref{sec:LCRVMDL}.

\begin{table}[h]
\centering
\caption{Radial velocities measurements.}
\label{tab:RVs}
\begin{tabular}{ccc ccc ccc}
\hline																	
HJD-2450000	&	$RV_1$	&	$RV_2$	&	HJD-2450000	&	$RV_1$	&	$RV_2$	&	HJD-2450000	&	$RV_1$	&	$RV_2$	\\
	&	(km~s$^{-1}$)	&	(km~s$^{-1}$)	&		&	(km~s$^{-1}$)	&	(km~s$^{-1}$)	&		&	(km~s$^{-1}$)	&	(km~s$^{-1}$)	\\
\hline																	
\multicolumn{3}{c}{HM~Pup}					&	9360.8937	&	17(5)	&	251(13)	&	9301.8991	&	61(6)	&	-148(10)	\\
\cline{1-3}																	
7798.0305	&	94(5)	&	-147(13)	&	9361.8788	&	89(5)	&	-66(13)	&	9301.9506	&	60(6)	&	-151(10)	\\
7798.0371	&	96(5)	&	-141(13)	&	9361.8833	&	96(5)	&	-55(13)	&	9301.9822	&	62(6)	&	-156(10)	\\
7798.0445	&	104(5)	&	-146(13)	&	9361.8878	&	97(5)	&	-57(13)	&	9302.0427	&	66(6)	&	-150(10)	\\
\cline{4-6}																	
7798.0469	&	99(5)	&	-124(13)	&	\multicolumn{3}{c}{V632~Sco}					&	9302.0902	&	57(6)	&	-148(10)	\\
\cline{4-6}																	
7889.9488	&	21(5)	&	256(13)	&	8241.1658	&	37(9)	&	-217(14)	&	9305.0406	&	-11(6)	&	185(10)	\\
7889.9519	&	23(5)	&	257(13)	&	8241.1973	&	36(9)	&	-211(14)	&	9305.0483	&	-15(6)	&	177(10)	\\
7889.9550	&	18(5)	&	234(13)	&	8241.2025	&	43(9)	&	-200(14)	&	9305.0856	&	-20(6)	&	188(10)	\\
7889.9590	&	18(5)	&	238(13)	&	8241.2129	&	34(9)	&	-218(14)	&	9305.0894	&	-20(6)	&	186(10)	\\
7916.8682	&	92(5)	&	-105(13)	&	8270.0536	&	20(9)	&	-199(14)	&	9305.0932	&	-14(6)	&	195(10)	\\
8240.8736	&	109(5)	&	-119(13)	&	8270.2046	&	32(9)	&	-222(14)	&	9305.1232	&	-17(6)	&	194(10)	\\
8240.8788	&	111(5)	&	-143(13)	&	8270.9209	&	-62(9)	&	194(14)	&	9305.1308	&	-16(6)	&	196(10)	\\
8240.8840	&	105(5)	&	-115(13)	&	8270.9864	&	-67(9)	&	168(14)	&	9360.8630	&	58(6)	&	-111(10)	\\
8270.8744	&	15(5)	&	211(13)	&	8270.9911	&	-70(9)	&	168(14)	&	9360.8661	&	59(6)	&	-107(10)	\\
8924.8909	&	89(5)	&	-83(13)	&	8270.9956	&	-75(9)	&	168(14)	&	9360.8692	&	58(6)	&	-122(10)	\\
8924.8934	&	82(5)	&	-61(13)	&	8271.0203	&	-74(9)	&	170(14)	&	9360.9579	&	68(6)	&	-130(10)	\\
9155.1041	&	107(5)	&	-113(13)	&	8271.1079	&	-79(9)	&	180(14)	&	9360.9611	&	72(6)	&	-137(10)	\\
9155.1093	&	106(5)	&	-114(13)	&	8271.1138	&	-78(9)	&	191(14)	&	9360.9725	&	74(6)	&	-126(10)	\\
9156.1914	&	13(5)	&	213(13)	&	8271.1190	&	-80(9)	&	190(14)	&	9361.8664	&	-11(6)	&	151(10)	\\
9156.2005	&	11(5)	&	218(13)	&	9034.1377	&	-44(9)	&	194(14)	&	9361.8695	&	-6(6)	&	145(10)	\\
9184.9968	&	15(5)	&	235(13)	&	9034.1429	&	-44(9)	&	194(14)	&	9361.8726	&	-8(6)	&	155(10)	\\
9185.0934	&	20(5)	&	190(13)	&	9038.9233	&	-55(9)	&	146(14)	&	9362.9335	&	60(6)	&	-109(10)	\\
9186.0966	&	102(5)	&	-113(13)	&	9038.9285	&	-53(9)	&	156(14)	&	9362.9380	&	65(6)	&	-101(10)	\\
9186.1297	&	100(5)	&	-134(13)	&	9038.9337	&	-62(9)	&	156(14)	&	9362.9426	&	61(6)	&	-112(10)	\\
9329.8694	&	11(5)	&	234(13)	&	9038.9803	&	-65(9)	&	184(14)	&	9362.9504	&	56(6)	&	-115(10)	\\
9329.8732	&	15(5)	&	244(13)	&	9038.9855	&	-60(9)	&	160(14)	&	9362.9549	&	56(6)	&	-121(10)	\\
9329.8770	&	13(5)	&	250(13)	&	9039.1021	&	-69(9)	&	179(14)	&	9362.9653	&	50(6)	&	-126(10)	\\
9330.8914	&	101(5)	&	-109(13)	&	9039.1073	&	-63(9)	&	193(14)	&	9362.9705	&	50(6)	&	-125(10)	\\
\cline{4-6}																	
9330.8959	&	105(5)	&	-98(13)	&	\multicolumn{3}{c}{TT~Vel}					&	9362.9757	&	53(6)	&	-134(10)	\\
\cline{4-6}																	
9330.9790	&	97(5)	&	-114(13)	&	9185.1077	&	-6(6)	&	168(10)	&	9362.9811	&	52(6)	&	-132(10)	\\
9330.9822	&	106(5)	&	-113(13)	&	9185.1182	&	-9(6)	&	190(10)	&	9362.9863	&	51(6)	&	-140(10)	\\
9360.8861	&	15(5)	&	225(13)	&	9185.1213	&	-8(6)	&	185(10)	&		&		&		\\
9360.8899	&	14(5)	&	258(13)	&	9186.1031	&	55(6)	&	-159(10)	&		&		&		\\
\hline																	
\end{tabular}
\end{table}

\section{Gas velocities of HM~Pup}
\label{sec:AppGas}

This appendix (Table~\ref{tab:gas}) includes for various orbital phases of HM~Pup the velocities of the primary component ($V_1$), based on the centered wavelength ($\lambda_1$) of its Na~I~D line, the centered wavelength of the line of the gas ($\lambda_{\rm gas}$), and its corresponding velocity ($V_{\rm gas}$) as well as the relative velocity of the gas to the primary component ($V_{\rm 1,~gas}$). The wavelengths and velocities were adjusted, for clarity, to remove the fixed systemic velocity of 58~km~s$^{-1}$.

\begin{table}[h]
\centering
\caption{Velocities of the HM~Pup primary star and wavelengths and velocities of its Na~I~D lines and the gas stream from the secondary star at a range of orbital phases (with systemic velocity subtracted).}
\label{tab:gas}
\scalebox{0.99}{
\begin{tabular}{ccc ccc ccc ccc}
\hline																							
Phase 	&	$V_1$	&	$\lambda_1$	&	$\lambda_{\rm gas}$	&	$V_{\rm gas}$	&	$V_{\rm 1,~gas}$	&	Phase 	&	$V_1$	&	$\lambda_1$	&	$\lambda_{\rm gas}$	&	$V_{\rm gas}$	&	$V_{\rm 1,~gas}$	\\
	&	(km~s$^{-1}$)	&	({\AA})	&	({\AA})	&	(km~s$^{-1}$)	&	(km~s$^{-1}$)	&		&	(km~s$^{-1}$)	&	({\AA})	&	({\AA})	&	(km~s$^{-1}$)	&	(km~s$^{-1}$)	\\
\hline																							
0.24	&	-44.91	&	5890.23	&	5890.50	&	-31.02	&	13.89	&	0.62	&	30.80	&	5891.71	&	5890.68	&	-21.86	&	-52.67	\\
0.28	&	-44.20	&	5890.24	&	5890.68	&	-21.86	&	22.34	&	0.66	&	37.99	&	5891.86	&	5890.60	&	-25.93	&	-63.93	\\
0.32	&	-40.72	&	5890.31	&	5890.30	&	-41.20	&	-0.49	&	0.75	&	45.00	&	5891.99	&	5890.40	&	-36.11	&	-81.11	\\
0.35	&	-36.41	&	5890.39	&	5889.86	&	-63.60	&	-27.19	&	0.77	&	44.65	&	5891.99	&	5890.28	&	-42.22	&	-86.87	\\
0.42	&	-21.68	&	5890.68	&	5890.28	&	-42.22	&	-20.54	&	0.88	&	30.80	&	5891.71	&	5890.50	&	-31.02	&	-61.83	\\
0.50	&	0.00	&	5891.11	&	5890.50	&	-31.02	&	-31.02	&		&		&		&		&		&		\\
\hline																		
\end{tabular}}
\end{table}

\newpage
\section{Spots}
\label{sec:AppSpots}

The location of the spots on the secondary components of the systems, assumed in the LCs and RVs modelling (Section~\ref{sec:LCRVMDL}), are given in Table~\ref{tab:spots}. This table lists: The pass bands used for modelling LCs and RVs, the number of spots for each model, and the parameters of the spots (Co-latitude, longitude, radius, and temperature factor). The letter `G' given for the models of V632~Sco using the $TESS$ data stands for group of LCs (see Section~\ref{sec:LCRVMDL} for details).


\begin{table}[h]
\centering
\caption{Spot location on the surface of the secondary components.}
\label{tab:spots}
\begin{tabular}{l ccccccccc}
\hline																			
	&	\multicolumn{3}{c}{HM~Pup}					&	TT~Vel	&	\multicolumn{5}{c}{V632~Sco}									\\
\hline																			
Filters	&	$BVI$	&	\multicolumn{2}{c}{$TESS$}			&	$TESS$	&	$B$	&	$VI$	&	$TESS-G1$	&	$TESS-G2$	&	$TESS-G3$	\\
Spot	&	1	&	1	&	2	&	1	&	1	&	1	&	1	&	1	&	1	\\
$Co-latitude$~($\degr$)	&	59(4)	&	57(4)	&	62(4)	&	111(15)	&	84(9)	&	81(10)	&	73(5)	&	68(4)	&	67(5)	\\
$Longitude$~($\degr$)	&	59(4)	&	52(4)	&	225(4)	&	262(8)	&	329(2)	&	90(13)	&	259(2)	&	261(2)	&	260(3)	\\
$Radius$~($\degr$)	&	26(3)	&	30(3)	&	16(4)	&	11(2)	&	20(1)	&	17(2)	&	24(1)	&	26(1)	&	27(1)	\\
$T_{\rm f}$~($_{\rm spot}/T_{\rm sur}$)	&	0.8(1)	&	0.9(1)	&	1.1(1)	&	0.7(1)	&	0.7(1)	&	0.8(1)	&	0.7(1)	&	0.8(1)	&	0.8(1)	\\
\hline
\end{tabular}
\tablefoot{G=Group of LCs, $T_{\rm spot}$=temperature of the spot, $T_{\rm sur}$=temperature of the surface}
\end{table}

\section{Times of minima}
\label{sec:AppMin}

Table~\ref{tab:MIN} hosts the times of minima calculated from personal ground-based observations and the available data from the $TESS$ mission. We list: the heliocentric Julian date and the type (I=primary eclipse, II=secondary eclipse) of each minimum and the filters used for the individual observations. In cases where two or more filters are listed for a given minimum time, the latter is the average of the times of minima calculated for the data of each filter. More details about the calculation method are given in Section~\ref{sec:OCMDL}.

\begin{table}[h]
\centering
\caption{Times of minima derived from the ground-based and $TESS$ data.}
\label{tab:MIN}
\begin{tabular}{ccc ccc ccc}
\hline																	
HJD	&	Type	&	Filter	&	HJD	&	Type	&	Filter	&	HJD	&	Type	&	Filter	\\
\hline																	
\multicolumn{3}{c}{HM~Pup}					&	2459203.6011(5)	&	II	&	$TESS$	&	2459250.2161(4)	&	II	&	$TESS$	\\
\cline{1-3}																	
2458492.7218(17)	&	 I	&	$TESS$	&	2459204.8999(1)	&	 I	&	$TESS$	&	2459251.5152(1)	&	 I	&	$TESS$	\\
2458494.0194(11)	&	II	&	$TESS$	&	2459206.1902(11)	&	II	&	$TESS$	&	2459252.8054(7)	&	II	&	$TESS$	\\
2458495.3100(15)	&	 I	&	$TESS$	&	2459207.4897(1)	&	 I	&	$TESS$	&	2459256.6947(1)	&	 I	&	$TESS$	\\
2458496.6085(4)	&	II	&	$TESS$	&	2459208.7809(7)	&	II	&	$TESS$	&	2459257.9845(6)	&	II	&	$TESS$	\\
2458497.9042(65)	&	 I	&	$TESS$	&	2459210.0794(1)	&	 I	&	$TESS$	&	2459259.2844(1)	&	 I	&	$TESS$	\\
2458499.1989(3)	&	II	&	$TESS$	&	2459211.3700(5)	&	II	&	$TESS$	&	2459260.5742(10)	&	II	&	$TESS$	\\
2458500.4906(14)	&	 I	&	$TESS$	&	2459212.6691(1)	&	 I	&	$TESS$	&	2459261.8741(1)	&	 I	&	$TESS$	\\
2458501.7885(11)	&	II	&	$TESS$	&	2459216.5486(6)	&	II	&	$TESS$	&	2459263.1644(13)	&	II	&	$TESS$	\\
2458505.6704(37)	&	 I	&	$TESS$	&	2459217.8487(1)	&	 I	&	$TESS$	&	2459264.4639(1)	&	 I	&	$TESS$	\\
2458506.9692(8)	&	II	&	$TESS$	&	2459220.4383(1)	&	 I	&	$TESS$	&	2459265.7524(9)	&	II	&	$TESS$	\\
2458508.2595(18)	&	 I	&	$TESS$	&	2459221.7283(6)	&	II	&	$TESS$	&	2459273.5229(9)	&	II	&	$TESS$	\\
2458509.5573(7)	&	II	&	$TESS$	&	2459221.7288(2)	&	II	&	$V,~I$	&	2459274.8228(1)	&	 I	&	$TESS$	\\
2458510.8518(71)	&	 I	&	$TESS$	&	2459223.0281(1)	&	 I	&	$TESS$	&	2459276.1122(8)	&	II	&	$TESS$	\\
2458512.1472(5)	&	II	&	$TESS$	&	2459224.3191(12)	&	II	&	$TESS$	&	2459277.4126(1)	&	 I	&	$TESS$	\\
2458513.4389(23)	&	 I	&	$TESS$	&	2459225.6171(1)	&	I	&	$I$	&	2459278.7033(9)	&	II	&	$TESS$	\\
\cline{7-9}																	
2458514.7366(4)	&	II	&	$TESS$	&	2459225.6178(1)	&	 I	&	$TESS$	&	\multicolumn{3}{c}{V632~Sco}					\\
\cline{7-9}																	
2458518.6191(2)	&	 I	&	$TESS$	&	2459226.9085(7)	&	II	&	$TESS$	&	2458285.1077(30)	&	I	&	$B,~V,~I$	\\
2458519.9167(16)	&	II	&	$TESS$	&	2459229.4972(6)	&	II	&	$TESS$	&	2458625.6675(85)	&	II	&	$TESS$	\\
2458521.2091(8)	&	 I	&	$TESS$	&	2459230.7967(21)	&	I	&	$I$	&	2458626.4660(6)	&	 I	&	$TESS$	\\
2458522.5059(3)	&	II	&	$TESS$	&	2459230.7973(1)	&	 I	&	$TESS$	&	2458628.0767(4)	&	 I	&	$TESS$	\\
2458523.7989(19)	&	 I	&	$TESS$	&	2459232.0874(4)	&	II	&	$TESS$	&	2458628.8809(29)	&	II	&	$TESS$	\\
2458526.3884(1)	&	 I	&	$TESS$	&	2459233.3870(1)	&	 I	&	$TESS$	&	2458629.6863(1)	&	 I	&	$TESS$	\\
2458527.6862(10)	&	II	&	$TESS$	&	2459234.6758(6)	&	II	&	$TESS$	&	2458631.2966(2)	&	 I	&	$TESS$	\\
2458531.5679(13)	&	 I	&	$TESS$	&	2459235.9768(1)	&	 I	&	$TESS$	&	2458632.1010(12)	&	II	&	$TESS$	\\
2458535.4569(12)	&	II	&	$TESS$	&	2459237.2664(6)	&	II	&	$TESS$	&	2458634.5172(1)	&	 I	&	$TESS$	\\
2458536.7471(5)	&	 I	&	$TESS$	&	2459238.5666(1)	&	 I	&	$TESS$	&	2458636.9324(4)	&	II	&	$TESS$	\\
2458538.0450(4)	&	II	&	$TESS$	&	2459239.8576(9)	&	II	&	$TESS$	&	2458637.7373(7)	&	 I	&	$TESS$	\\
2458539.3370(6)	&	 I	&	$TESS$	&	2459242.4475(6)	&	II	&	$TESS$	&	2458640.9578(2)	&	 I	&	$TESS$	\\
2458540.6353(2)	&	II	&	$TESS$	&	2459243.7460(1)	&	 I	&	$TESS$	&	2458642.5679(2)	&	 I	&	$TESS$	\\
2459186.7712(20)	&	I	&	$V$	&	2459245.0356(10)	&	II	&	$TESS$	&	2458643.3735(6)	&	II	&	$TESS$	\\
2459190.6528(2)	&	II	&	$I$	&	2459246.3357(1)	&	 I	&	$TESS$	&	2458645.7883(1)	&	 I	&	$TESS$	\\
2459199.7204(15)	&	I	&	$B$	&	2459247.6254(8)	&	II	&	$TESS$	&	2458648.2045(2)	&	II	&	$TESS$	\\
2459202.3102(1)	&	 I	&	$TESS$	&	2459248.9255(1)	&	 I	&	$TESS$	&	2458649.0087(8)	&	 I	&	$TESS$	\\
\hline																	
\end{tabular}
\end{table}

\begin{table}
\centering
\caption{Table~\ref{tab:MIN} cont.}
\label{tab:MIN2}
\begin{tabular}{ccc |ccc ccc}
\hline																	
HJD	&	Type	&	Filter	&	HJD	&	Type	&	Filter	&	HJD	&	Type	&	Filter	\\
\hline																	
\multicolumn{3}{c}{V632~Sco}					&	\multicolumn{6}{c}{TT~Vel}											\\
\hline
2458650.6189(1)	&	 I	&	$TESS$	&	2458544.5609(5)	&	II	&	$TESS$	&	2458586.7312(10)	&	II	&	$TESS$	\\
2458651.4239(9)	&	II	&	$TESS$	&	2458545.6132(15)	&	 I	&	$TESS$	&	2458587.7824(1)	&	 I	&	$TESS$	\\
2458652.2289(1)	&	 I	&	$TESS$	&	2458546.6696(14)	&	II	&	$TESS$	&	2458588.8397(9)	&	II	&	$TESS$	\\
2459024.9868(2)	&	II	&	$V$	&	2458547.7215(1)	&	 I	&	$TESS$	&	2458589.8907(1)	&	 I	&	$TESS$	\\
2459032.2280(2)	&	I	&	$B,~V,~I$	&	2458548.7775(7)	&	II	&	$TESS$	&	2458590.9483(8)	&	II	&	$TESS$	\\
2459066.0411(20)	&	I	&	$B$	&	2458549.8299(1)	&	 I	&	$TESS$	&	2458591.9990(15)	&	 I	&	$TESS$	\\
2459362.3102(1)	&	 I	&	$TESS$	&	2458550.8866(5)	&	II	&	$TESS$	&	2458593.0564(8)	&	II	&	$TESS$	\\
2459363.9205(1)	&	 I	&	$TESS$	&	2458551.9385(9)	&	 I	&	$TESS$	&	2458594.1076(1)	&	 I	&	$TESS$	\\
2459364.7265(6)	&	II	&	$TESS$	&	2458552.9950(7)	&	II	&	$TESS$	&	2458595.1649(5)	&	II	&	$TESS$	\\
2459365.5306(1)	&	 I	&	$TESS$	&	2458554.0469(4)	&	 I	&	$TESS$	&	2459281.4668(1)	&	 I	&	$TESS$	\\
2459366.3397(3)	&	II	&	$TESS$	&	2458555.1034(16)	&	II	&	$TESS$	&	2459282.5225(5)	&	II	&	$TESS$	\\
2459367.1407(1)	&	 I	&	$TESS$	&	2458557.2124(11)	&	II	&	$TESS$	&	2459283.5756(1)	&	 I	&	$TESS$	\\
2459367.9505(4)	&	II	&	$TESS$	&	2458558.2638(1)	&	 I	&	$TESS$	&	2459284.6310(2)	&	II	&	$TESS$	\\
2459368.7508(1)	&	 I	&	$TESS$	&	2458559.3206(11)	&	II	&	$TESS$	&	2459285.6841(1)	&	 I	&	$TESS$	\\
2459369.5606(3)	&	II	&	$TESS$	&	2458560.3722(2)	&	 I	&	$TESS$	&	2459286.7392(4)	&	II	&	$TESS$	\\
2459370.3611(1)	&	 I	&	$TESS$	&	2458561.4293(6)	&	II	&	$TESS$	&	2459287.7925(1)	&	 I	&	$TESS$	\\
2459371.1702(3)	&	II	&	$TESS$	&	2458562.4808(8)	&	 I	&	$TESS$	&	2459288.8481(4)	&	II	&	$TESS$	\\
2459371.9711(1)	&	 I	&	$TESS$	&	2458563.5377(4)	&	II	&	$TESS$	&	2459289.9010(1)	&	 I	&	$TESS$	\\
2459372.7813(5)	&	II	&	$TESS$	&	2458564.5891(6)	&	 I	&	$TESS$	&	2459290.9564(5)	&	II	&	$TESS$	\\
2459373.5813(1)	&	 I	&	$TESS$	&	2458565.6464(11)	&	II	&	$TESS$	&	2459292.0094(1)	&	 I	&	$TESS$	\\
2459374.3905(3)	&	II	&	$TESS$	&	2458566.6978(1)	&	 I	&	$TESS$	&	2459294.1182(1)	&	 I	&	$TESS$	\\
2459376.0004(3)	&	II	&	$TESS$	&	2458567.7548(13)	&	II	&	$TESS$	&	2459295.1729(3)	&	II	&	$TESS$	\\
2459376.8016(1)	&	 I	&	$TESS$	&	2458570.9145(2)	&	 I	&	$TESS$	&	2459296.2264(1)	&	 I	&	$TESS$	\\
2459378.4117(1)	&	 I	&	$TESS$	&	2458571.9716(9)	&	II	&	$TESS$	&	2459297.2819(3)	&	II	&	$TESS$	\\
2459379.2214(2)	&	II	&	$TESS$	&	2458573.0231(6)	&	 I	&	$TESS$	&	2459298.3348(1)	&	 I	&	$TESS$	\\
2459380.0220(1)	&	 I	&	$TESS$	&	2458574.0800(6)	&	II	&	$TESS$	&	2459299.3903(5)	&	II	&	$TESS$	\\
2459380.8170(13)	&	II	&	$TESS$	&	2458575.1314(8)	&	 I	&	$TESS$	&	2459300.4433(1)	&	 I	&	$TESS$	\\
2459381.6320(1)	&	 I	&	$TESS$	&	2458576.1887(13)	&	II	&	$TESS$	&	2459301.4984(4)	&	II	&	$TESS$	\\
2459382.4392(2)	&	II	&	$TESS$	&	2458577.2401(2)	&	 I	&	$TESS$	&	2459302.5517(0)	&	 I	&	$TESS$	\\
2459383.2421(1)	&	 I	&	$TESS$	&	2458578.2971(9)	&	II	&	$TESS$	&	2459303.6070(8)	&	II	&	$TESS$	\\
2459384.8523(1)	&	 I	&	$TESS$	&	2458579.3484(1)	&	 I	&	$TESS$	&	2459304.6602(0)	&	 I	&	$TESS$	\\
2459385.6617(3)	&	II	&	$TESS$	&	2458580.4058(9)	&	II	&	$TESS$	&	2459305.7155(2)	&	II	&	$TESS$	\\
2459386.4625(1)	&	 I	&	$TESS$	&	2458581.4567(2)	&	 I	&	$TESS$	&	2459329.9612(15)	&	I	&	$B,~V,~I$	\\
2459387.2821(9)	&	II	&	$TESS$	&	2458584.6228(6)	&	II	&	$TESS$	&	2459331.0212(17)	&	II	&	$B,~V,~I$	\\
	&		&		&	2458585.6737(10)	&	 I	&	$TESS$	&	2459352.0925(17)	&	II	&	$B,~V,~I$	\\
\hline																	
																																																		
\end{tabular}
\end{table}

\newpage
\section{Complete pulsation models}
\label{sec:AppFreqs}

This appendix contains three tables which list the detected frequencies in all available data sets for each system. In each table are listed: the increasing number ($n$), the value ($f_n$), the amplitude ($A$), the phase ($\Phi$), the $S/N$, and the most possible combination of each detected frequency. Details about the frequency analysis can be found in Section~\ref{sec:PULMDL}.

\begin{table}[h]
\centering
\caption{Detected pulsation frequencies of the $\delta$~Scuti component of HM~Pup in all data sets.}
\label{tab:PULMDLpup}
\scalebox{0.93}{
\begin{tabular}{l ccccc l ccccc}
\hline
$n$	&	  $f_{\rm n}$	&	$A$	&	  $\Phi$	&	S/N	&	Combination	&	$n$	&	  $f_{\rm n}$	&	$A$	&	  $\Phi$	&	S/N	&	Combination	\\
	&	     (cycle~d$^{-1}$)	&	(mmag)	&	($\degr$)	&		&		&		&	     (cycle~d$^{-1}$)	&	(mmag)	&	($\degr$)	&		&		\\
\hline																							
\multicolumn{12}{c}{$B$-filter}																							\\
\hline																							
1	&	0.038(1)	&	5.9(1)	&	3(1)	&	33.4	&		&	12	&	34.741(3)	&	1.2(1)	&	237(5)	&	7.1	&	$f_4$+$f_8$	\\
2	&	38.563(1)	&	5.5(1)	&	9(1)	&	27.6	&		&	13	&	32.982(3)	&	1.0(1)	&	143(6)	&	6.0	&	$f_6$+$f_7$-$f_{5}$	\\
3	&	36.424(1)	&	3.8(1)	&	354(1)	&	23.3	&		&	14	&	35.641(3)	&	1.2(1)	&	311(6)	&	5.7	&	$f_3$-$f_{11}$-$f_{8}$	\\
4	&	34.265(1)	&	2.7(1)	&	174(2)	&	17.0	&	2$f_3$-$f_2$	&	15	&	0.454(3)	&	1.6(1)	&	208(7)	&	5.5	&	$\sim f_{\rm orb}$	\\
5	&	0.098(1)	&	3.1(1)	&	91(2)	&	14.5	&	$\sim f_1$	&	16	&	31.879(4)	&	1.2(1)	&	47(7)	&	5.3	&	$f_3$+$f_{16}$-$f_{14}$	\\
6	&	31.113(1)	&	2.5(1)	&	72(3)	&	12.4	&	$f_2$-$f_3$-$f_5$ or $\sim 5f_{\rm orb}$	&	17	&	31.756(4)	&	0.9(1)	&	47(8)	&	4.6	&	$f_{16}$-$f_{5}$	\\
7	&	1.997(2)	&	1.5(1)	&	206(3)	&	9.2	&	$\sim 5f_{\rm orb}$	&	18	&	2.692(4)	&	0.8(1)	&	167(8)	&	4.0	&	$f_{14}$-$f_{13}$ or $\sim 7f_{\rm orb}$	\\
8	&	0.506(2)	&	1.6(1)	&	304(4)	&	9.5	&	$f_2$-$f_3$-4$f_{\rm orb}$	&	19	&	4.542(4)	&	0.8(1)	&	97(8)	&	4.2	&	$f_{3}$-$f_{16}$ or $\sim 12f_{\rm orb}$	\\
9	&	0.156(2)	&	1.7(1)	&	63(4)	&	8.0	&	$\sim f_5$	&	20	&	38.084(4)	&	0.8(1)	&	24(8)	&	4.0	&	$f_{2}$-$f_{8}$	\\
10	&	34.355(2)	&	1.3(1)	&	37(5)	&	7.8	&	$f_4$+$f_1$	&	21	&	32.815(5)	&	0.8(1)	&	323(9)	&	4.2	&	$f_{13}$-$f_{5}$	\\
11	&	0.205(2)	&	1.4(1)	&	126(5)	&	7.1	&	$\sim f_9$	&	22	&	37.154(5)	&	0.8(1)	&	13(9)	&	4.1	&	$2f_{3}$-$f_{14}$	\\
\hline																					\end{tabular}}
\tablefoot{The temporal zeropoints for the computation of the phases for each pass-band were: JD2459180.0 ($B$, $V$, $I$) and JD2459193.0 ($TESS$).}
\end{table}

\begin{table}[h]
\centering
\caption{Table~\ref{tab:PULMDLpup} cont.}
\label{tab:PULMDLpup2}
\scalebox{0.95}{
\begin{tabular}{l ccccc l ccccc}
\hline																							
$n$	&	  $f_{\rm n}$	&	$A$	&	  $\Phi$	&	S/N	&	Combination	&	$n$	&	  $f_{\rm n}$	&	$A$	&	  $\Phi$	&	S/N	&	Combination	\\
	&	     (cycle~d$^{-1}$)	&	(mmag)	&	($\degr$)	&		&		&		&	     (cycle~d$^{-1}$)	&	(mmag)	&	($\degr$)	&		&		\\
\hline																							
\multicolumn{6}{c}{$V$-filter}											&	\multicolumn{6}{c}{$TESS$}											\\
\hline																							
1	&	0.421(1)	&	5.5(1)	&	328(1)	&	31.5	&	$\sim f_{\rm orb}$	&	1	&	38.5622(1)	&	2.61(3)	&	7(1)	&	81.2	&		\\
2	&	38.563(1)	&	4.9(1)	&	14(1)	&	19.3	&		&	2	&	36.4196(2)	&	2.10(3)	&	14(1)	&	62.3	&		\\
3	&	36.417(1)	&	3.7(1)	&	49(2)	&	20.1	&		&	3	&	1.1584(2)	&	2.02(3)	&	156(1)	&	13.7	&	3$f_{\rm orb}$	\\
4	&	0.568(1)	&	3.7(1)	&	319(2)	&	15.7	&	$f_2$-$f_3$-4$f_{\rm orb}$	&	4	&	0.0317(2)	&	1.92(3)	&	77(1)	&	8.9	&		\\
5	&	0.014(1)	&	2.8(1)	&	354(3)	&	17.5	&		&	5	&	34.2608(2)	&	1.41(3)	&	200(1)	&	42.2	&	2$f_2$-$f_1$	\\
6	&	34.260(1)	&	2.0(1)	&	248(3)	&	8.8	&	2$f_3$-$f_2$	&	6	&	0.7733(3)	&	1.06(3)	&	286(2)	&	5.6	&	2$f_{\rm orb}$	\\
7	&	0.132(1)	&	1.8(1)	&	0(4)	&	10.9	&	$f_4$-$f_1$	&	7	&	31.1131(3)	&	0.96(3)	&	78(2)	&	37.7	&	$f_3$+2$f_5$-$f_1$	\\
8	&	32.806(1)	&	1.4(1)	&	88(4)	&	5.3	&		&	8	&	35.6479(4)	&	0.91(3)	&	246(2)	&	27.0	&	$f_2$-$f_6$	\\
9	&	31.110(1)	&	1.3(1)	&	92(5)	&	5.6	&	$f_8$-3$f_{4}$	&	9	&	34.3527(5)	&	0.70(3)	&	52(3)	&	21.0	&	$f_5$+2$f_4$	\\
10	&	38.082(1)	&	1.2(1)	&	69(5)	&	4.5	&	$f_8$+$f_{3}$-$f_9$	&	10	&	33.7391(5)	&	0.63(3)	&	100(3)	&	21.0	&	$f_8$-$f_3$-$f_6$	\\
11	&	2.186(1)	&	1.2(1)	&	199(6)	&	5.0	&	$f_3$-$f_{6}$	&	11	&	2.3154(5)	&	0.62(3)	&	194(3)	&	8.1	&	2$f_3$	\\
12	&	34.353(2)	&	1.1(1)	&	51(7)	&	4.9	&	$f_6$+$f_{7}$-$f_5$	&	12	&	38.0850(7)	&	0.49(3)	&	16(4)	&	15.6	&	2$f_8$+$f_3$-$f_9$	\\
13	&	33.693(1)	&	0.9(1)	&	49(6)	&	4.5	&	$f_6$-$f_4$	&	13	&	33.5800(8)	&	0.43(3)	&	140(4)	&	14.3	&	$f_9$-$f_6$	\\
\cline{1-6}																							
\multicolumn{6}{c}{$I$~filter}											&	14	&	31.8850(8)	&	0.40(3)	&	9(5)	&	16.0	&	$f_7$+$f_6$	\\
\cline{1-6}																							
1	&	38.562(2)	&	2.7(1)	&	15(3)	&	12.6	&		&	15	&	3.0853(10)	&	0.34(3)	&	137(5)	&	6.4	&	8$f_{\rm orb}$	\\
2	&	36.416(2)	&	2.4(1)	&	86(3)	&	11.3	&		&	16	&	32.7497(11)	&	0.30(3)	&	202(6)	&	10.7	&	2$f_8$-$f_1$	\\
3	&	2.171(3)	&	2.3(1)	&	7(4)	&	9.0	&	$f_1$-$f_2$	&	17	&	36.4216(11)	&	0.29(3)	&	186(6)	&	8.7	&	$\sim f_2$	\\
4	&	1.786(3)	&	1.3(1)	&	122(4)	&	5.2	&	$f_3$+$f_{\rm orb}$	&	18	&	35.5271(11)	&	0.29(3)	&	243(6)	&	8.3	&	$f_9$+$f_3$	\\
5	&	0.376(4)	&	2.7(1)	&	246(5)	&	13.8	&	$f_3$-$f_4$ or $\sim f_{\rm orb}$	&	19	&	30.9800(15)	&	0.22(3)	&	56(8)	&	8.5	&	$f_{13}$+$f_7$-$f_{10}$	\\
6	&	31.111(5)	&	1.2(1)	&	129(7)	&	5.2	&	$f_2$-3$f_4$	&	20	&	31.5101(16)	&	0.21(3)	&	158(9)	&	8.2	&	$f_{3}$+$f_7$-$f_{6}$	\\
7	&	34.287(6)	&	1.0(1)	&	88(8)	&	5.3	&	$f_2$-$f_3$	&	21	&	20.0784(16)	&	0.20(3)	&	119(9)	&	9.9	&	3$f_{19}$-$f_{2}$	\\
8	&	0.871(6)	&	1.4(1)	&	283(8)	&	6.9	&	2$f_5$	&	22	&	32.8101(18)	&	0.19(3)	&	6(10)	&	6.5	&	$f_{16}$+$f_{4}$	\\
9	&	1.059(7)	&	1.5(1)	&	350(9)	&	6.9	&	3$f_5$ or $\sim 3f_{\rm orb}$	&	23	&	37.0601(22)	&	0.16(3)	&	310(12)	&	4.8	&	2$f_{8}$-$f_{5}$	\\
10	&	37.324(7)	&	1.1(1)	&	66(9)	&	4.7	&	$f_2$+$f_8$	&	24	&	33.7526(22)	&	0.15(3)	&	142(12)	&	5.1	&	$\sim f_{10}$	\\
11	&	34.753(7)	&	0.9(1)	&	299(10)	&	4.6	&	$f_5$+$f_7$	&	25	&	39.8768(22)	&	0.15(3)	&	137(12)	&	5.4	&	$f_{3}$+$f_2$+$f_{11}$	\\
	&		&		&		&		&		&	26	&	31.1142(22)	&	0.15(3)	&	75(12)	&	5.9	&	$\sim f_{7}$	\\
\hline
\end{tabular}}
\tablefoot{The temporal zeropoints for the computation of the phases for each pass-band were: JD2459180.0 ($B$, $V$, $I$) and JD2459193.0 ($TESS$).}
\end{table}

\begin{table}
\centering
\caption{The same as Table~\ref{tab:PULMDLpup}, but for V632~Sco.}
\label{tab:PULMDLsco}
\begin{tabular}{l ccccc l ccccc}
\hline																							
$n$	&	$f_{\rm n}$	&	$A$	&	$\Phi$	&	S/N	&	Combination	&	$n$	&	$f_{\rm n}$	&	$A$	&	$\Phi$	&	S/N	&	Combination	\\
	&	(cycle~d$^{-1}$)	&	(mmag)	&	($\degr$)	&		&		&		&	(cycle~d$^{-1}$)	&	(mmag)	&	($\degr$)	&		&		\\
\hline																				
	\multicolumn{6}{c}{$B$-filter}										&		\multicolumn{6}{c}{$TESS$}										\\
\hline																							
1	&	0.819(1)	&	3.6(2)	&	311(4)	&	8.9	&		&	1	&	23.7423(5)	&	1.79(4)	&	134(1)	&	23.5	&		\\
2	&	23.539(1)	&	3.0(2)	&	141(4)	&	7.8	&		&	2	&	2.4846(5)	&	1.72(4)	&	181(1)	&	13.2	&	4$f_{\rm orb}$	\\
3	&	3.548(1)	&	2.3(2)	&	280(6)	&	5.0	&		&	3	&	3.1069(6)	&	1.50(4)	&	273(2)	&	14.0	&	5$f_{\rm orb}$	\\
4	&	16.780(1)	&	2.2(2)	&	216(6)	&	6.1	&		&	4	&	0.1746(7)	&	1.22(4)	&	230(2)	&	5.6	&		\\
5	&	33.064(1)	&	2.3(2)	&	144(6)	&	7.2	&	2$f_4$-$f_{\rm orb}$	&	5	&	16.6507(7)	&	1.14(4)	&	205(2)	&	17.3	&		\\
6	&	0.216(1)	&	2.2(2)	&	21(6)	&	5.5	&		&	6	&	4.3486(13)	&	0.63(4)	&	94(4)	&	7.8	&	3$f_2$-$f_3$	\\
7	&	7.115(1)	&	1.9(2)	&	149(7)	&	5.0	&	2$f_3$	&	7	&	22.5038(14)	&	0.58(4)	&	164(4)	&	7.5	&	$f_1$+$f_3$-$f_6$	\\
8	&	31.322(1)	&	1.5(2)	&	35(9)	&	4.6	&	$f_5$-$f_{\rm orb}$	&	8	&	32.0532(15)	&	0.56(4)	&	70(4)	&	5.9	&	$f_7$+2$f_5$-$f_1$	\\
9	&	53.324(2)	&	1.3(2)	&	60(10)	&	4.0	&	2$f_7$+2$f_8$-$f_2$	&	9	&	5.5894(19)	&	0.43(4)	&	283(6)	&	5.8	&	9$f_{\rm orb}$	\\
10	&	37.904(2)	&	1.3(2)	&	127(10)	&	4.2	&	$f_1$+$f_3$+2$f_4$	&	10	&	24.8052(21)	&	0.40(4)	&	335(6)	&	5.1	&	$f_7$+$f_2$-$f_4$	\\
11	&	20.196(2)	&	1.3(2)	&	128(10)	&	5.0	&	$f_2$+$f_6$-$f_3$	&	11	&	26.0437(16)	&	0.54(4)	&	125(4)	&	7.9	&	$f_1$+$f_2$-$f_4$	\\
\cline{1-6}																								\multicolumn{6}{c}{$V$-filter}										&	12	&	32.0127(21)	&	0.40(4)	&	326(6)	&	5.3	&	$\sim f_8$	\\
\cline{1-6}																							
1	&	23.432(13)	&	2.9(5)	&	81(9)	&	7.1	&		&		&		&		&		&		&		\\
\hline																					
\end{tabular}
\tablefoot{The temporal zeropoints for the computation of the phases for each pass-band were: JD2459025.0 ($B$, $V$) and JD2459361.0 ($TESS$).}
\end{table}

\begin{table}
\centering
\caption{The same as Table~\ref{tab:PULMDLpup}, but for TT~Vel.}
\label{tab:PULMDLvel}
\scalebox{0.98}{
\begin{tabular}{l ccccc l ccccc}
\hline																							
$n$	&	$f_{\rm n}$	&	$A$	&	$\Phi$	&	S/N	&	Combination	&	$n$	&	$f_{\rm n}$	&	$A$	&	$\Phi$	&	S/N	&	Combination	\\
	&	(cycle~d$^{-1}$)	&	(mmag)	&	($\degr$)	&		&		&		&	(cycle~d$^{-1}$)	&	(mmag)	&	($\degr$)	&		&		\\
\hline																							
	\multicolumn{6}{c}{$B$-filter}										&	4	&	11.6748(11)	&	4.8(5)	&	2(6)	&	4.8	&	$f_1$+$f_2$+$f_3$	\\
\cline{1-6}																							
1	&	3.0101(2)	&	17.6(4)	&	96(1)	&	25.8	&		&	5	&	6.0325(13)	&	4.0(5)	&	242(7)	&	4.7	&	2$f_1$-3$f_2$	\\
2	&	3.2725(3)	&	10.8(4)	&	122(2)	&	15.1	&	3$f_{\rm orb}$	&	6	&	12.3723(15)	&	3.5(5)	&	117(8)	&	4.5	&	$f_2$+$f_4$	\\
3	&	1.9443(4)	&	8.3(4)	&	222(3)	&	12.5	&		&	7	&	0.9916(15)	&	3.7(5)	&	265(8)	&	4.2	&	$f_2$+$f_6$-2$f_5$	\\
\cline{7-12}																							
4	&	15.3267(7)	&	5.0(4)	&	171(5)	&	6.6	&	2$f_{\rm orb}$	&		\multicolumn{6}{c}{$TESS$}										\\
\cline{7-12}																							
5	&	8.9257(7)	&	5.2(4)	&	137(4)	&	5.3	&		&	1	&	15.5020(4)	&	2.64(5)	&	36(1)	&	37.8	&		\\
6	&	12.0160(8)	&	4.5(4)	&	212(5)	&	6.7	&	4$f_1$	&	2	&	1.4248(4)	&	2.34(5)	&	214(1)	&	14.4	&	3$f_{\rm orb}$	\\
7	&	16.5918(10)	&	3.6(4)	&	151(6)	&	4.4	&		&	3	&	0.0288(8)	&	1.36(5)	&	198(2)	&	5.7	&		\\
8	&	1.6196(8)	&	4.4(4)	&	300(5)	&	6.7	&	$f_7$-$f_5$-2$f_1$	&	4	&	0.9478(10)	&	1.01(5)	&	58(3)	&	5.1	&	2$f_{\rm orb}$	\\
\cline{1-6}																							
	\multicolumn{6}{c}{$V$-filter}										&	5	&	14.4616(11)	&	0.98(5)	&	150(3)	&	15.0	&	$f_1$-$f_{4}$-2$f_3$	\\
\cline{1-6}																							
1	&	0.9503(4)	&	13.5(5)	&	258(2)	&	29.0	&		&	6	&	27.8644(12)	&	0.84(5)	&	338(3)	&	18.6	&	3$f_5$-$f_{1}$	\\
2	&	3.5354(5)	&	10.9(5)	&	9(3)	&	19.5	&		&	7	&	27.7637(15)	&	0.68(5)	&	180(4)	&	14.8	&	$f_6$-2$f_{3}$	\\
3	&	0.7564(7)	&	7.1(5)	&	177(4)	&	15.8	&		&	8	&	12.5193(16)	&	0.65(5)	&	175(4)	&	12.2	&	$f_5$-2$f_{4}$	\\
4	&	3.6562(9)	&	5.7(5)	&	103(5)	&	10.3	&	2$f_2$-2$f_1$-2$f_3$	&	9	&	29.6763(21)	&	0.50(5)	&	340(5)	&	10.2	&	$f_7$+2$f_{4}$	\\
5	&	15.7337(14)	&	3.7(5)	&	71(7)	&	7.0	&		&	10	&	14.5538(20)	&	0.51(5)	&	14(5)	&	7.9	&	$f_1$-$f_{4}$	\\
6	&	12.4535(13)	&	3.8(5)	&	26(7)	&	7.9	&	$f_5$-2$f_1$-2$f_2$	&	11	&	27.0191(25)	&	0.41(5)	&	21(6)	&	9.7	&	$f_5$+$f_{8}$	\\
7	&	27.7026(16)	&	3.3(5)	&	4(8)	&	6.1	&	$f_6$-2$f_1$	&	12	&	14.2795(22)	&	0.47(5)	&	208(6)	&	7.0	&	$f_4$+$f_{7}$-$f_{5}$	\\
8	&	11.0764(17)	&	3.0(5)	&	253(9)	&	5.8	&	$f_5$+3$f_4$	&	13	&	15.3411(25)	&	0.42(5)	&	267(6)	&	5.9	&	$f_6$-$f_{8}$	\\
9	&	23.0640(18)	&	2.9(5)	&	26(10)	&	4.3	&	$f_5$+2$f_2$-$f_3$	&	14	&	14.2186(25)	&	0.41(5)	&	67(6)	&	6.1	&	$\sim f_{12}$	\\
10	&	6.3650(16)	&	3.2(5)	&	185(9)	&	4.7	&	$f_1$+$f_4$+$f_6$	&	15	&	23.2011(33)	&	0.31(5)	&	349(8)	&	5.1	&	$f_{11}$-$f_{4}$-2$f_{2}$	\\
\cline{1-6}																							
	\multicolumn{6}{c}{$I$-filter}										&	16	&	28.6083(35)	&	0.30(5)	&	11(9)	&	6.2	&	$2f_{12}$	\\
\cline{1-6}																							
1	&	4.0502(6)	&	8.7(5)	&	125(3)	&	8.4	&		&	17	&	23.4807(35)	&	0.30(5)	&	160(9)	&	5.1	&	$f_{15}$+$f_{5}$-$f_{14}$	\\
2	&	0.6947(8)	&	6.6(5)	&	40(4)	&	7.8	&		&	18	&	25.9232(37)	&	0.28(5)	&	135(9)	&	5.9	&	$f_{11}$+$f_{5}$-$f_{1}$	\\
3	&	3.1096(9)	&	5.7(5)	&	134(5)	&	5.4	&		&	19	&	15.4982(37)	&	0.28(5)	&	241(9)	&	5.1	&	$\sim f_{1}$	\\
\hline																							
\end{tabular}}
\tablefoot{The temporal zeropoints for the computation of the phases for each pass-band were: JD2459325.0 ($B$, $V$, $I$) and JD2459281.0 ($TESS$).}
\end{table}

\end{appendix}
\end{document}